\documentclass[useAMS,usenatbib]{mn2e}

\usepackage{natbib}
\usepackage{ulem}
\usepackage{epsfig}
\usepackage{amsmath}
\usepackage{float}
\usepackage{subfig}
\usepackage{graphicx}
\usepackage{caption}
\def\gae{\mathrel{\hbox{\rlap{\hbox{\lower2pt\hbox{$\sim$}}}\hbox{\raise2pt\hbox{$>$}}}}}
\def\lae{\mathrel{\hbox{\rlap{\hbox{\lower2pt\hbox{$\sim$}}}\hbox{\raise2pt\hbox{$<$}}}}}

\title[Hot Outflows]{Hot Outflows in Galaxy Clusters}

\author[C.~C. Kirkpatrick \& B.~R. McNamara]{C.~C. Kirkpatrick$^{1,2}$ \& B.~R. McNamara$^{1,3,4}$\\
$^{1}$Department of Physics \& Astronomy, University of Waterloo, 200 University Avenue West, Waterloo, ON N2L 3G1, Canada \\
$^{2}$Department of Physics, University of Helsinki, Gustaf H{\"a}llstr{\"o}min katu 2a, FI-00014 Helsinki, Finland \\
$^{3}$Perimeter Institute for Theoretical Physics, 31 Caroline Street North, Waterloo, ON N2L 2Y5, Canada \\
$^{4}$Harvard-Smithsonian Center for Astrophysics, 60 Garden Street, Cambridge, MA 02138, USA}

\begin{document}

\maketitle

\begin{abstract}
The gas-phase metallicity distribution has been analyzed for the hot atmospheres of 29 galaxy clusters using {\it Chandra X-ray Observatory} observations.  All host brightest cluster galaxies (BCGs) with X-ray cavity systems produced by radio AGN.  We find high elemental abundances projected preferentially along the cavities of 16 clusters.  The metal-rich plasma was apparently lifted out of the BCGs with the rising X-ray cavities (bubbles) to altitudes between twenty and several hundred kiloparsecs.  A relationship between the maximum projected altitude of the uplifted gas (the ``iron radius'') and jet power is found with the form $R_{\rm Fe} \propto P_{\rm jet}^{0.45}$.  The estimated outflow rates are typically tens of solar masses per year but exceed $100 ~\rm M_\odot ~yr^{-1}$ in the most powerful AGN.  The outflow rates are 10\% to 20\% of the cooling rates, and thus alone are unable to offset a cooling inflow.  Nevertheless, hot outflows effectively redistribute the cooling gas and may play a significant role at regulating star formation and AGN activity in BCGs and presumably in giant elliptical galaxies.  The metallicity distribution overall can be complex, perhaps due to metal rich gas returning in circulation flows or being blown around in the hot atmospheres.  Roughly 15\% of the work done by the cavities is expended lifting the metal-enriched gas, implying their nuclear black holes have increased in mass by at least $\sim 10^7$ M$_\odot$ to $10^9$ M$_\odot$.  Finally, we show that hot outflows can account for the broad, gas-phase metallicity distribution compared to the stellar light profiles of BCGs, and we consider a possible connection between hot outflows and cold molecular gas flows discovered in recent ALMA observations.
\end{abstract}


\section{Introduction}

Radio active galactic nuclei (AGN) drive shock waves and inflate giant bubbles into the hot,  X-ray atmospheres pervading galaxies and clusters \citep[for reviews see][]{bmc07,bmc12,fab12}.  The enthalpy released as the bubbles rise through hot atmospheres is apparently able to offset radiative cooling \citep{bir04,vr07,moj11} and to regulate the rates at which hot atmospheres condense into stars \citep{bow06, ss06, cro06}.  The rising X-ray bubbles or cavities also drag cool, metal-rich, plasma away from the host galaxy and deposit it on larger scales (Kirkpatrick, McNamara, \& Cavagnolo 2011).  The rate with which this hot, metal-rich gas flows is largely unexplored, and thus may be a significant aspect of radio-mode feedback.

Cluster atmospheres are replete in heavy elements with average metallicities of  approximately 1/3 of the solar value \citep{mus96,mus97,deg01}.  The hot gas became enriched apparently during the early development of clusters, primarily through ejecta from core-collapse supernovae (SNe II) associated with star formation.  The atmosphere immediately surrounding the brightest cluster galaxy (BCG) is chemically enriched above the cluster mean, approaching and sometimes exceeding the solar value \citep{all98,dup00,deg01,tam01}.  This plasma is apparently stripped debris from orbiting galaxies and stellar ejecta from the BCG itself.  Being the dominant stellar structure in the core of a cluster, a BCG enriches cluster cores through stellar winds and type Ia supernovae (SN Ia), a process generating greater quantities of iron compared to SN II \citep{deg04,tam04,dep06}.  A dichotomy therefore exists between the $\alpha$-element enhanced outskirts and the iron-rich core.

Were the iron-rich ejecta produced by a BCG lying in a static atmosphere, the stellar light and atmospheric metal concentrations would track each other.  However, several studies have shown that the radial distribution of iron and other metals in cool core clusters is more broadly distributed than the stellar light profile \citep{reb05,reb06,dav08,ras08}.  Evidently, the metals produced by BCGs are transported to higher altitudes.  Likely agents include bulk and turbulent motions driven by halo mergers and active galactic nuclei (AGN) \citep{gop01,gop03,bru02,omm04,moll07,roe07,pop10}.  A growing body of data implicate AGN.  The Virgo and Hydra A clusters are archetypes.  In M87, bright, X-ray filaments that are metal enriched with respect to the surrounding gas are draped along the radio jets, indicating that the filaments were lifted out by the jets \citep{sim08}.  Similar structure was found in Hydra A but on vastly larger scales exceeding 100 kpc \citep{nuls02,sim09,cck09b,git11}.  Hydra A is experiencing a much more powerful AGN outburst compared to the current outburst in M87, which is why the metal rich gas is seen at such large altitudes.  This phenomenon is observed in rich clusters \citep{san05,dor12} and groups, e.g., HCG 62 \citep{mor06,gu07,git10,raf13}.  A study of ten clusters experiencing radio-AGN activity \citep{cck11} revealed a power law relationship between jet power and the largest projected radius to which metal-enriched gas is observed.  Indications are that persistent AGN activity may be largely responsible for enriching the cluster core.

Here we present an analysis of 29 galaxy clusters drawn from the {\it Chandra X-ray Observatory} archive.  The sample was chosen based on the level of AGN activity exhibited by X-ray cavities and availability of long exposures needed for accurate gas phase metallicity measurements.  In section 2 and 3 we present the entire sample of clusters and discuss preparation of the X-ray data.  Section 4 is our analysis and discussion of the sample.  Our main goals are to further refine the relationship established by \cite{cck11} between jet power and the extent metals are uplifted out of the cluster centre and to test whether or not an AGN can be the driving mechanism behind uplifting all the gas originally associated with the BCG and creating the broad abundance peaks we currently observe.  Sections 5 is the summary of our results.  Throughout this paper we assume a $\Lambda$CDM cosmology with H$_0 = 70$ km s$^{-1}$ Mpc$^{-1}$, $\Omega_{\rm M} = 0.3$, and $\Omega_{\Lambda} = 0.7$.  All uncertainties are quoted at the 67\% confidence level.

\section{Cluster Sample}

Twenty nine clusters, found in Table~\ref{tab:main-hqsamp}, were selected from \citet{raf06} and \citet{df08}, most with clear AGN activity in the form of X-ray cavities.  The sample is divided into two subsamples, dubbed ``High Quality" (HQ) and ``Extended" (EX).  The High Quality sample alone is used to comparing jet power to uplifted metals.  This sample is composed of clusters with Chandra images deep enough to provide accurate gas metallicity measurements projected on scales smaller than the radio/cavity system.   It includes the original 10 clusters with which the trend between the jet power and iron radius was found \citep{cck11}, and an additional six with deep Chandra data that have since become available.  The ``Extended Sample" includes 13 additional systems most of which have X-ray cavity jet power measurements, and central metallicity peaks, but whose Chandra images are too shallow to map the metallicity distribution on fine scales.  The two samples are combined to analyze global metallicity properties and for the analysis of growth rates of super-massive black holes (SMBHs) in BCGs, discussed in Section 4.4.   

\begin{table*}
\centering
\begin{minipage}{140mm}
\caption{The Sample.}
\begin{tabular}{@{}lccccccc@{}}
\hline
~ & ~ & Exp. Time & ~ & $P_{\rm jet}$$^a$ & ~ & $R_{\rm e}$$^c$ & ~ \\
System & Sub-sample & (ks) & $z$ & ($10^{42}$ erg s$^{-1}$) & $M_{\rm B}$$^b$ & (kpc) & $T_{\rm core}$$^d$ \\
\hline
	A133 & HQ & 103.6 & 0.0566 & 620 & -22.64 & 25.0 & 1T \\
	A262 & HQ & 139.4 & 0.016 & 9.7 & -21.07 & 12.3 & 2T \\
	Perseus & HQ & 957 & 0.018 & 150 & -22.65 & --- & 1T \\
	2A 0335 & HQ & 102.4 & 0.035 & 24 & -23.04 & --- & 2T \\
	A478 & HQ & 52.4 & 0.088 & 100 & -23.54 & --- & 1T \\
	MS 0735 & HQ & 476.5 & 0.22 & 6900 & -22.18 & --- & 1T \\
	Hydra A & HQ & 181.9 & 0.055 & 430 & -23.05 & 39.3 & 1T \\
	HCG 62 & HQ & 167 & 0.014 & 3.9 & -19.97 & --- & 1T \\
	A1795 & HQ & 205.8 & 0.063 & 160 & -22.59 & --- & 1T \\
	A1835 & HQ & 193.7 & 0.253 & 1800 & -23.27 & --- & 1T \\
	A2029 & HQ & 97.7 & 0.077 & 87 & -19.87 & 25.1 & 1T \\
	A2052 & HQ & 617.3 & 0.035 & 150 & -21.74 & 14.1 & 2T \\
	A2199 & HQ & 156.5 & 0.030 & 270 & -23.01 & 14.2 & 1T \\
	Sersic 159 & HQ & 107.6 & 0.058 & 780 & -22.34 & --- & 1T \\
	A2597 & HQ & 134.4 & 0.085 & 67 & -22.51 & --- & 2T \\
	M87 & HQ & 480.1 & 0.0043 & 6 & -22.05 & --- & 2T \\
	A85 & EX & 38.4 & 0.055 & 37 & -22.76 & 31.3 & 1T \\
	PKS 0745 & EX & 40.2 & 0.103 & 1700 & -20.53 & --- & 1T \\
	4C 55.16 & EX & 95.8 & 0.241 & 420 & -22.14 & --- & 1T \\
	RBS 797 & EX & 49.7 & 0.35 & 1200 & --- & --- & 1T \\
	A2390 & EX & 110.6 & 0.23 & * & --- & --- & 1T \\
	Zw 2701 & EX & 122.5 & 0.21 & * & -22.25 & --- & 1T \\
	Zw 3146 & EX & 83.3 & 0.29 & * & -22.81 & --- & 1T \\
	MACSJ1423 & EX & 133.6 & 0.54 & 1400 & --- & --- & 1T \\
	MKW 3S & EX & 57.3 & 0.045 & 410 & -22.09 & 18.8 & 1T \\
	Hercules A & EX & 96.8 & 0.15 & 310 & -21.51 & --- & 1T \\
	3C 388 & EX & 38.4 & 0.091 & 200 & -22.98 & --- & 1T \\
	Cygnus A & EX & 225.8 & 0.056 & 1300 & -21.83 & --- & 1T \\
	A4059 & EX & 91.9 & 0.048 & 96 & -22.88 & 23.5 & 1T \\
\hline
\label{tab:main-hqsamp}
\end{tabular}
\begin{itemize}
\item[$^a$] Jet power from \citet{raf06}
\item[$^b$] Absolute magnitudes derived from HyperLeda catalogue
\item[$^c$] Effective radius from HyperLeda catalogue
\item[$^d$] Temperature model used for X-ray core
\end{itemize}
\end{minipage}
\end{table*}

\section{Data Preparation}

The X-ray data were taken with the {\it Chandra X-ray Observatory} using either the ACIS-S3 back-illuminated CCD or the ACIS-I front-illuminated CCD array.  Data preparation was carried out using version 4.1.2 of both \textsc{ciao} and the calibration database.  When present, background flares were eliminated using the \textsc{lc\textunderscore clean} routine.  Over the full energy range (0.3-10 keV), a light curve was extracted from the level 2 event file.  Time intervals with photon count rates varying greater than 3-sigma from the mean were removed.  The corrected exposure time for the combined observations can be found in Table~\ref{tab:main-hqsamp}.  Time-dependent gain and charge transfer inefficiency corrections were applied.  Background subtraction was carried out using blank-sky background files normalized to the source count rate in the $9.5-12$ keV energy band.  Coordinates were reprojected to match the new background file to the source image in order for background spectra to be extracted from the equivalent spatial regions.  Targets with multiple observations were added together to properly identify faint point sources using {\it wavedetect}.  Each point source was confirmed manually and removed from the final level 2 event file.

\subsection{X-ray Spectral Modeling}
\label{subsec:main-spec}

Most observations were taken at a focal plane temperature of $-120^\circ$C, allowing us to create weighted response files using the CIAO tools {\it mkacisrmf} and {\it mkwarf}.  For observations made before 2000 January 29, {\it mkrmf} was used instead.  We modeled all spectra using a single temperature plasma model plus photoelectric absorption unless otherwise noted.  Using a WABS$\times$MEKAL model over the energy range 0.5-7 keV, temperature, abundance, and normalization are allowed to vary.  The column density was set to the values quoted by \citet{dic90}.  Abundance ratios (the fraction of an element of the gas compared to its fraction in the sun by number) were set as in \citet{gre98}, based on solar photospheric data.  These ratios are tied to the abundance of iron, as it is the prominent feature in the X-ray spectrum.

\begin{figure*}
\begin{center}
\begin{minipage}{0.49\linewidth}
\includegraphics[width=\textwidth, trim=0mm 0mm 0mm 0mm, clip]{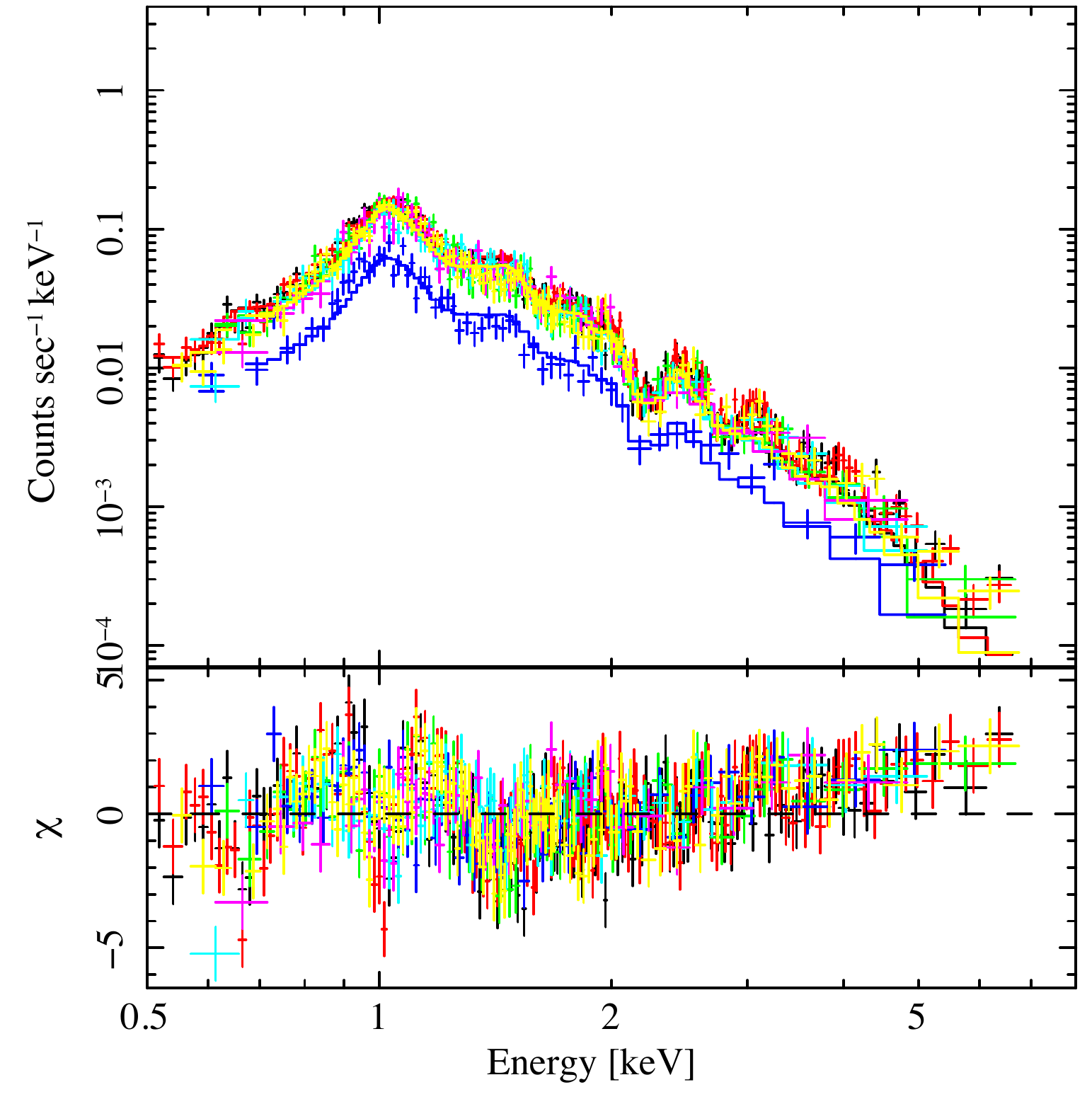}
\end{minipage}
\begin{minipage}{0.49\linewidth}
\includegraphics[width=\textwidth, trim=0mm 0mm 0mm 0mm, clip]{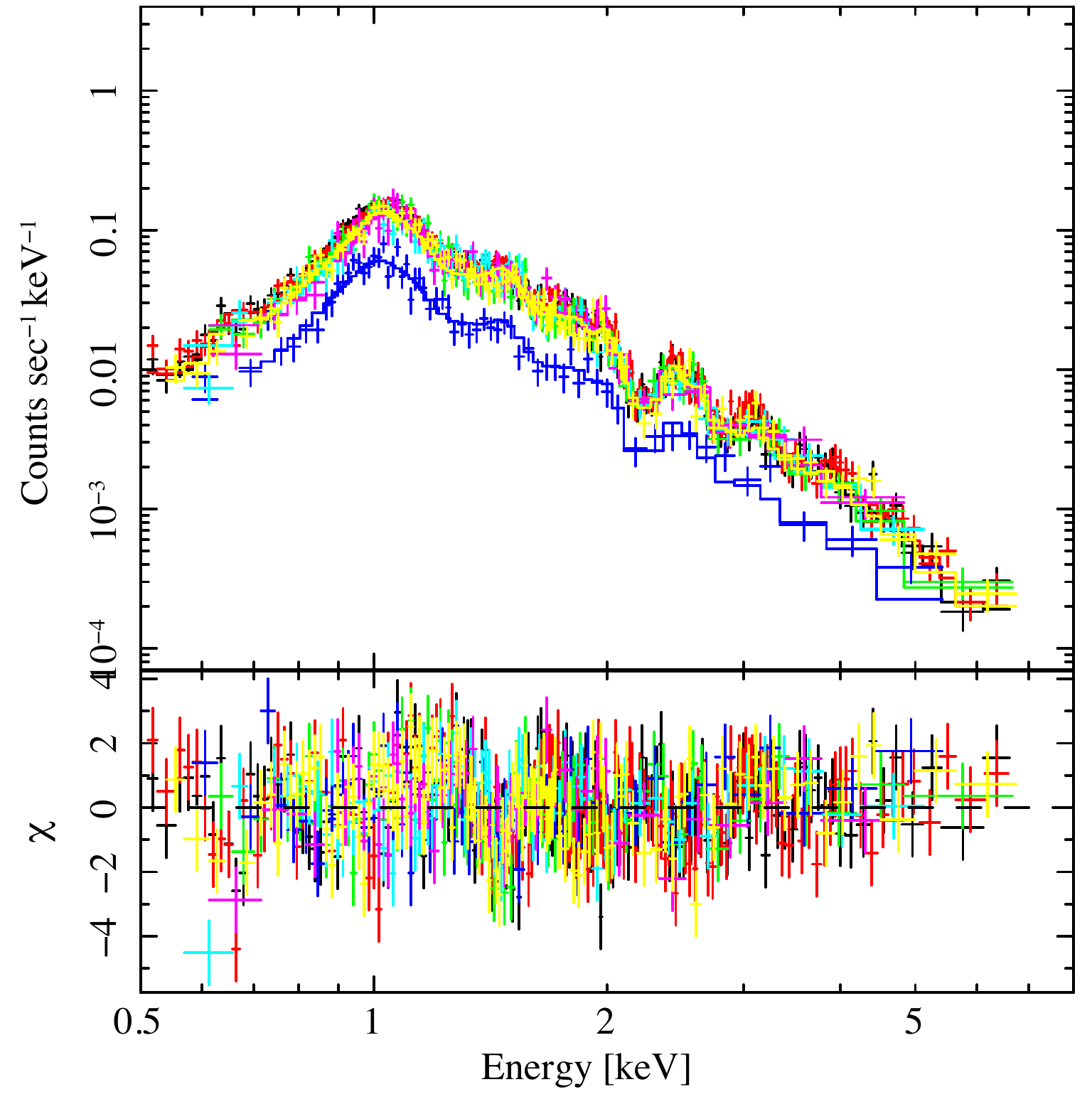}
\end{minipage}
\caption{{\it Left}: A typical, multi-observation X-ray spectrum poorly fit by a single-temperature model. {\it Right}: The same spectrum is better fit at energies above 3 keV when a second temperature component is added.}
\label{fig:main-spec_temp}
\end{center}
\end{figure*}

In cases where single temperature models provided an inadequate fit to the spectrum, multiple temperature model fits were performed eg., \citep{cck09a}.  The failure of the single temperature model at the centre, where the temperature structure becomes complex, usually reveals itself as a  sharp decrease in central metallicity instead of the usual rising metallicity peak \citep{buo00,mol01}.  The single temperature model tends to underestimate emission at energies above 4 keV.  This issue has been noted in the cores of clusters and in some AGN outflow regions.  An example of a typical spectrum is presented in the left panel of Figure~\ref{fig:main-spec_temp}.  The residuals show data above 4 keV systematically above the single temperature model.  When present, the effect was identified and corrected by adding a second temperature component (WABS$\times$(MEKAL+MEKAL)) for the metallicity maps and profile fits.  The second component's temperature and normalization were allowed to vary with the abundance tied between the two component.  The two-temperature fit shown in the right panel of Figure~\ref{fig:main-spec_temp} is shown for example.  We then applied the two-temperature model (2T) to all spectra, including those with acceptable single temperature model fits.  In those cases we found the second temperature component consistent with zero, or an improvement in chi-squared statistics that, when performing an F-test, was shown to be insignificant.  We list in Table~\ref{tab:main-hqsamp} which temperature model is used for each cluster.  When needed, the 2T model is only used in the inner 1 or 2 bins of the profile or  central region of the map, except in M87, where the entire profile along the jet region uses the 2T model.

\subsection{Systematic Errors}

\begin{figure*}
\begin{center}
\begin{minipage}{0.49\linewidth}
\includegraphics[width=\textwidth, trim=0mm 0mm 0mm 0mm, clip]{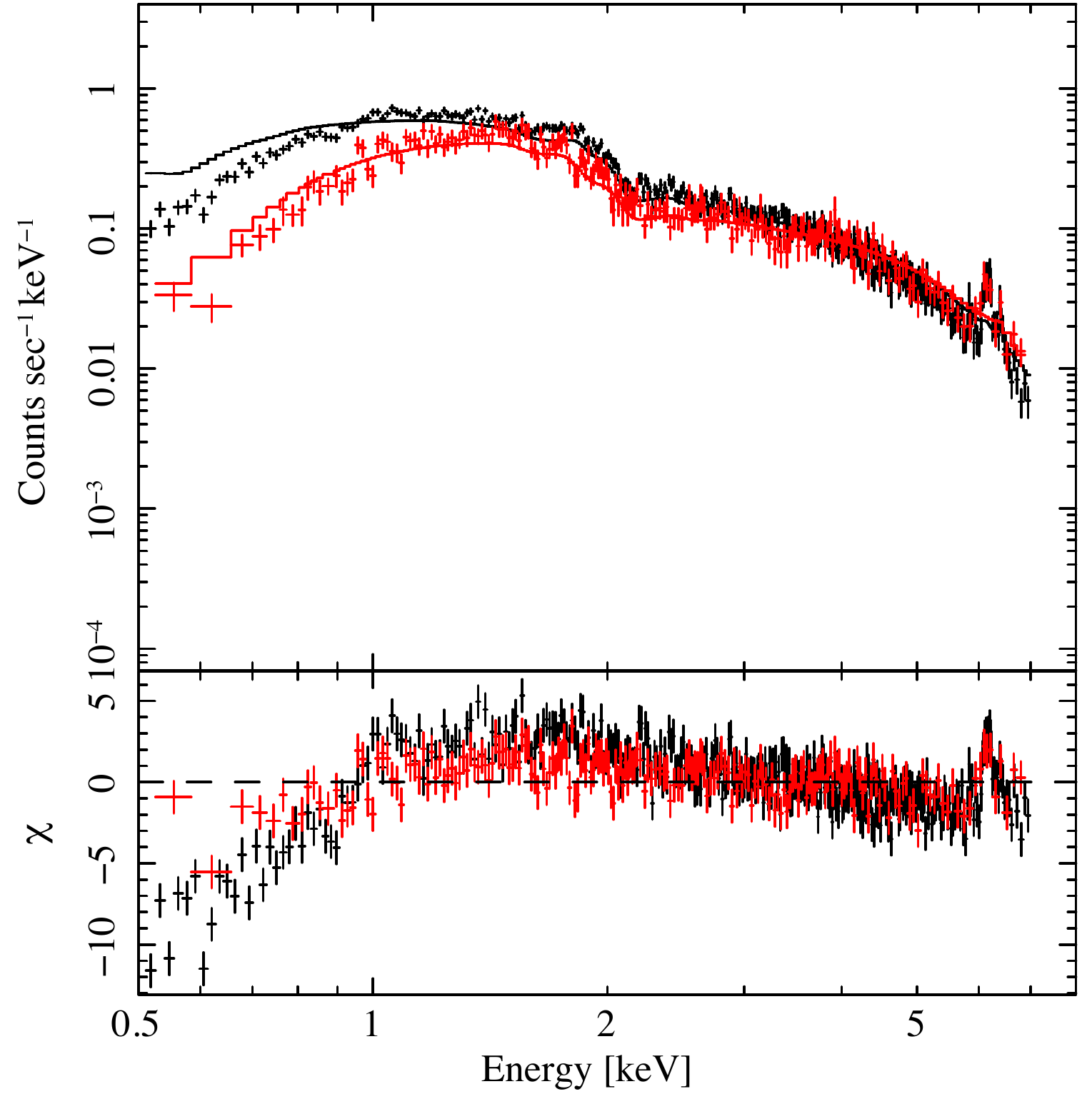}
\end{minipage}
\begin{minipage}{0.49\linewidth}
\includegraphics[width=\textwidth, trim=0mm 0mm 0mm 0mm, clip]{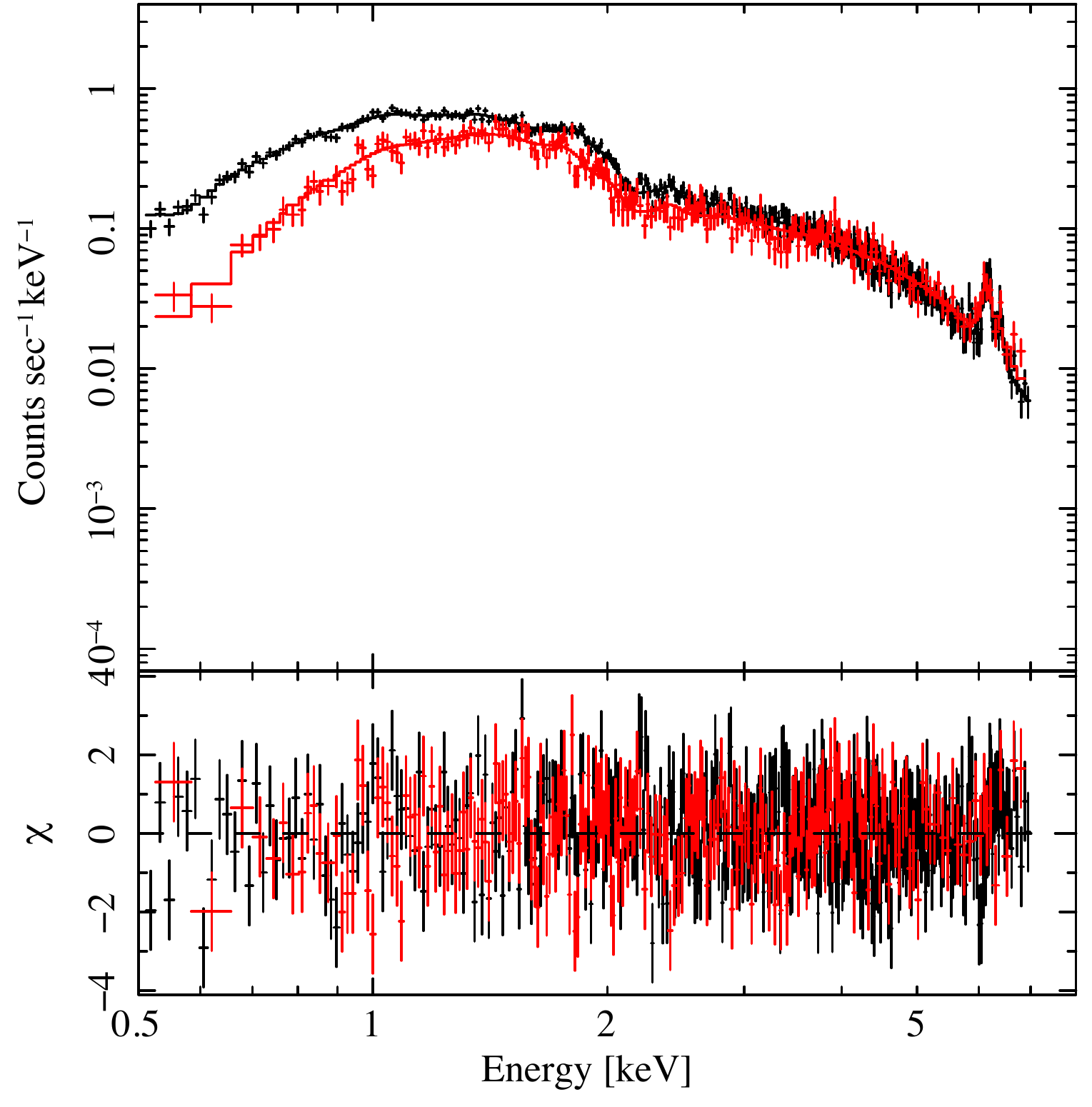}
\end{minipage}
\caption{{\it Left}: A typical, multi-observation X-ray spectrum, where the black and red data are from ACIS-S and ACIS-I, respectively.  It is poorly fit when the average column density is underestimated. {\it Right}: Allowing the column density to vary when large molecular clouds are present along the line of sight improves the fit at energies below 1 keV.}
\label{fig:main-spec_nH}
\end{center}
\end{figure*}

To control for systematic errors, we explored the effects of misestimation of the foreground hydrogen column density on our metal abundance measurements.  We initially assumed a uniform neutral hydrogen column density across the field of view.  However, this assumption is invalid toward fields projected onto Galactic molecular clouds \citep{bre03,bou08}.  The spectrum presented in the left panel of Figure~\ref{fig:main-spec_nH} is an example of a poor spectral fit owing to variations in column density across the field of view.  Based on images from the IRAS all sky survey, we  found that clusters with unacceptably high residuals below 1 keV were all projected toward high Galactic column densities.  The right panel of Figure~\ref{fig:main-spec_nH} represents the improvement in spectral fitting when column density allowed to adjust itself across the field of view.

\section{Analysis \& Discussion}

\subsection{Metallicity Maps}

We present projected metallicity maps for the High Quality sample of clusters in Figure~\ref{fig:main-femaps}.  The spectra were binned using a weighted Voronoi tessellation (WVT) algorithm \citep{cap03,die06}.  We began by defining a minimum signal-to-noise (S/N) of 100 per bin.  The value was increased to between 150 and 200 when possible.  We optimized the analysis to achieve bin sizes that minimize uncertainties but provide sufficient sampling to recover spatial trends in the metallicity distribution on angular scales below that of the radio/cavity system.

\begin{figure*}
\begin{center}
\begin{minipage}{0.32\linewidth}
\includegraphics[width=\textwidth, trim=0mm 0mm 0mm 0mm, clip]{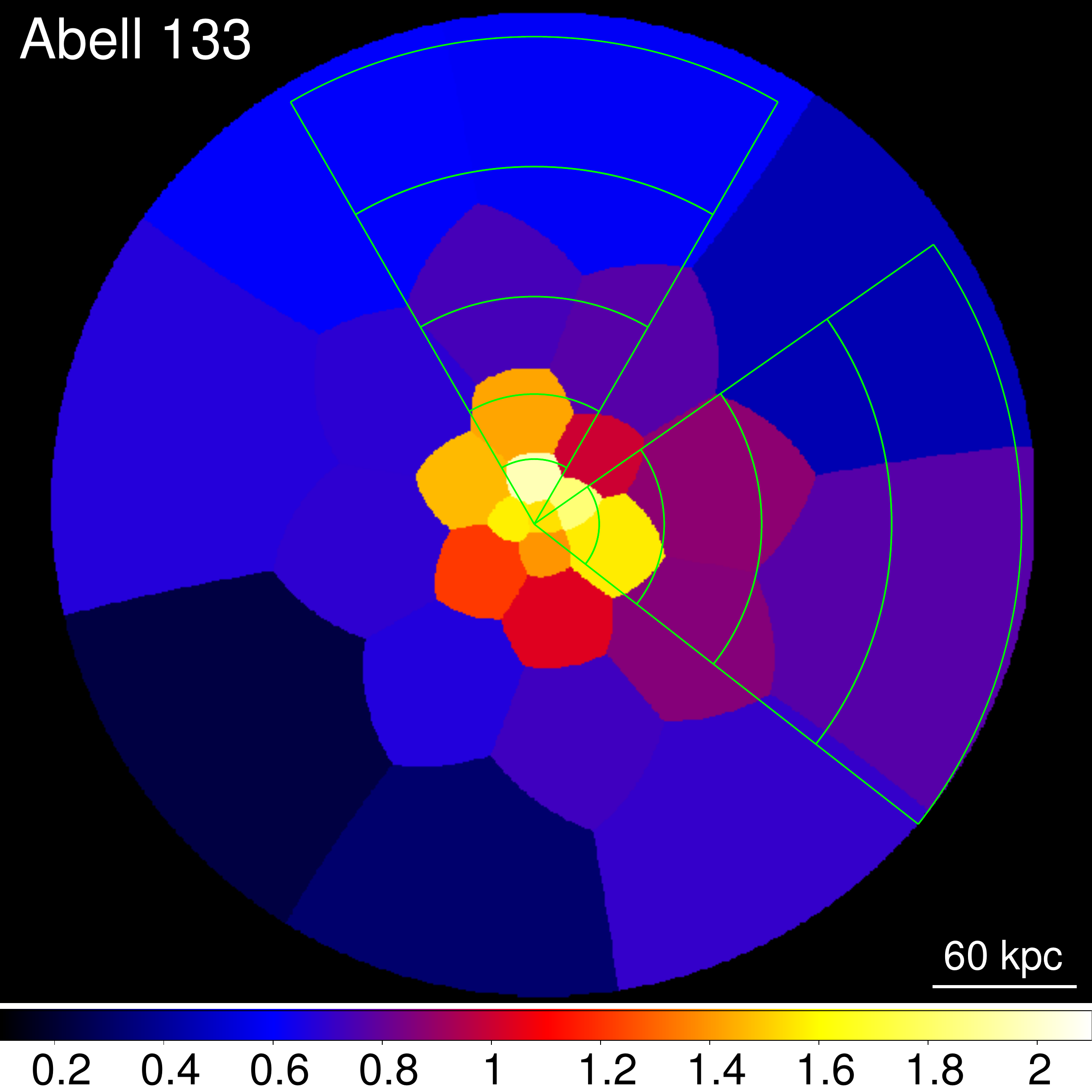}
\end{minipage}
\begin{minipage}{0.32\linewidth}
\includegraphics[width=\textwidth, trim=0mm 0mm 0mm 0mm, clip]{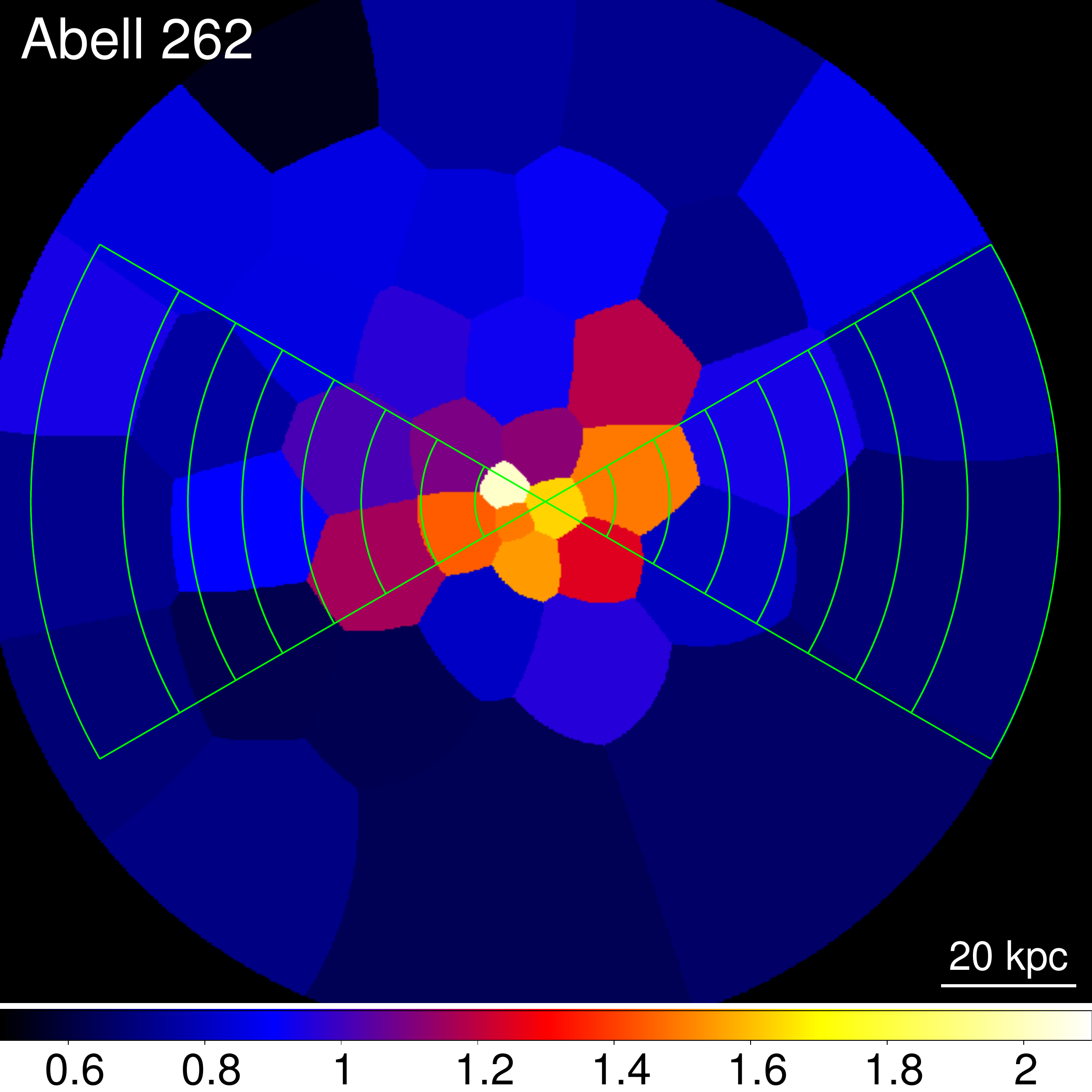}
\end{minipage}
\begin{minipage}{0.32\linewidth}
\includegraphics[width=\textwidth, trim=0mm 0mm 0mm 0mm, clip]{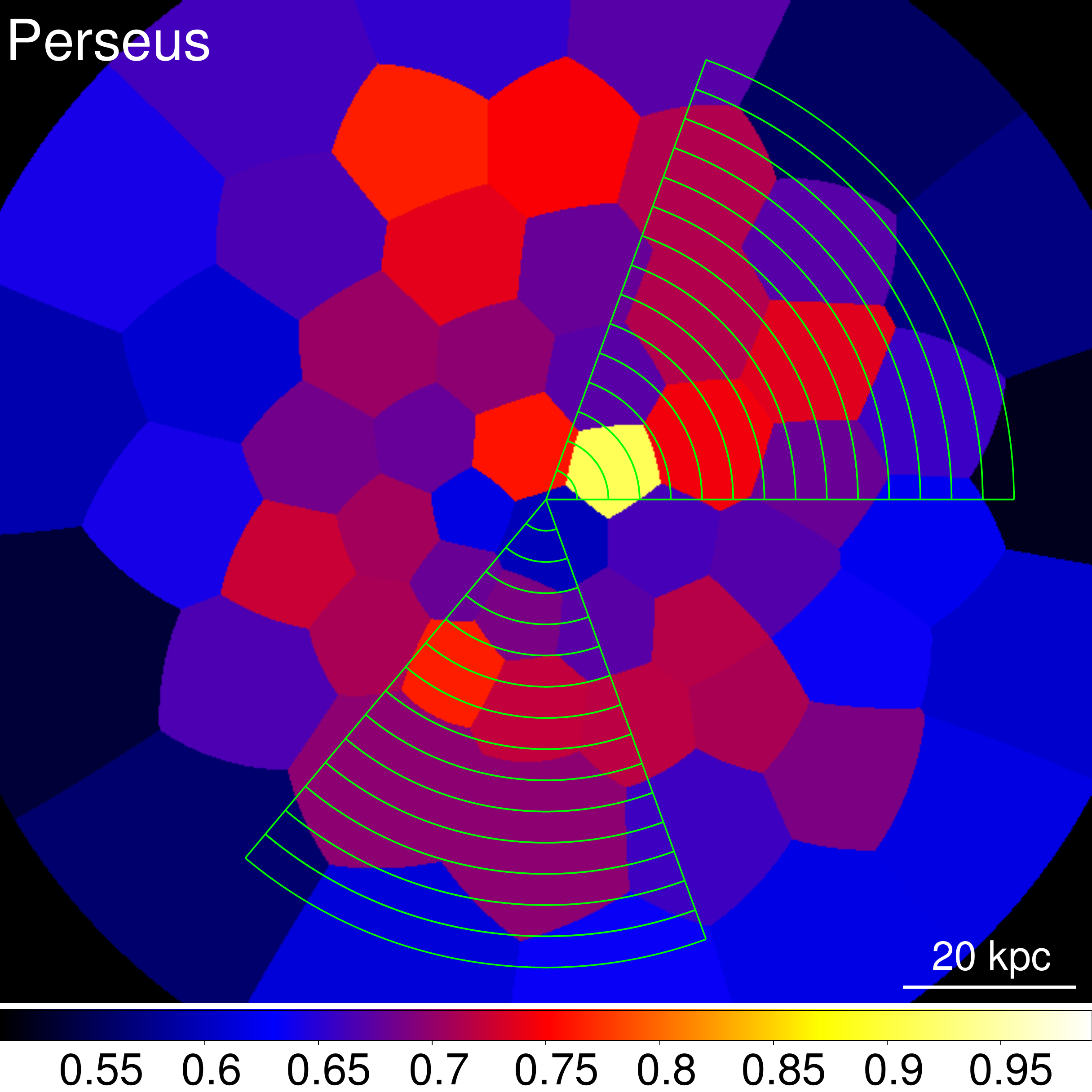}
\end{minipage}

\begin{minipage}{0.32\linewidth}
\includegraphics[width=\textwidth, trim=0mm 0mm 0mm 0mm, clip]{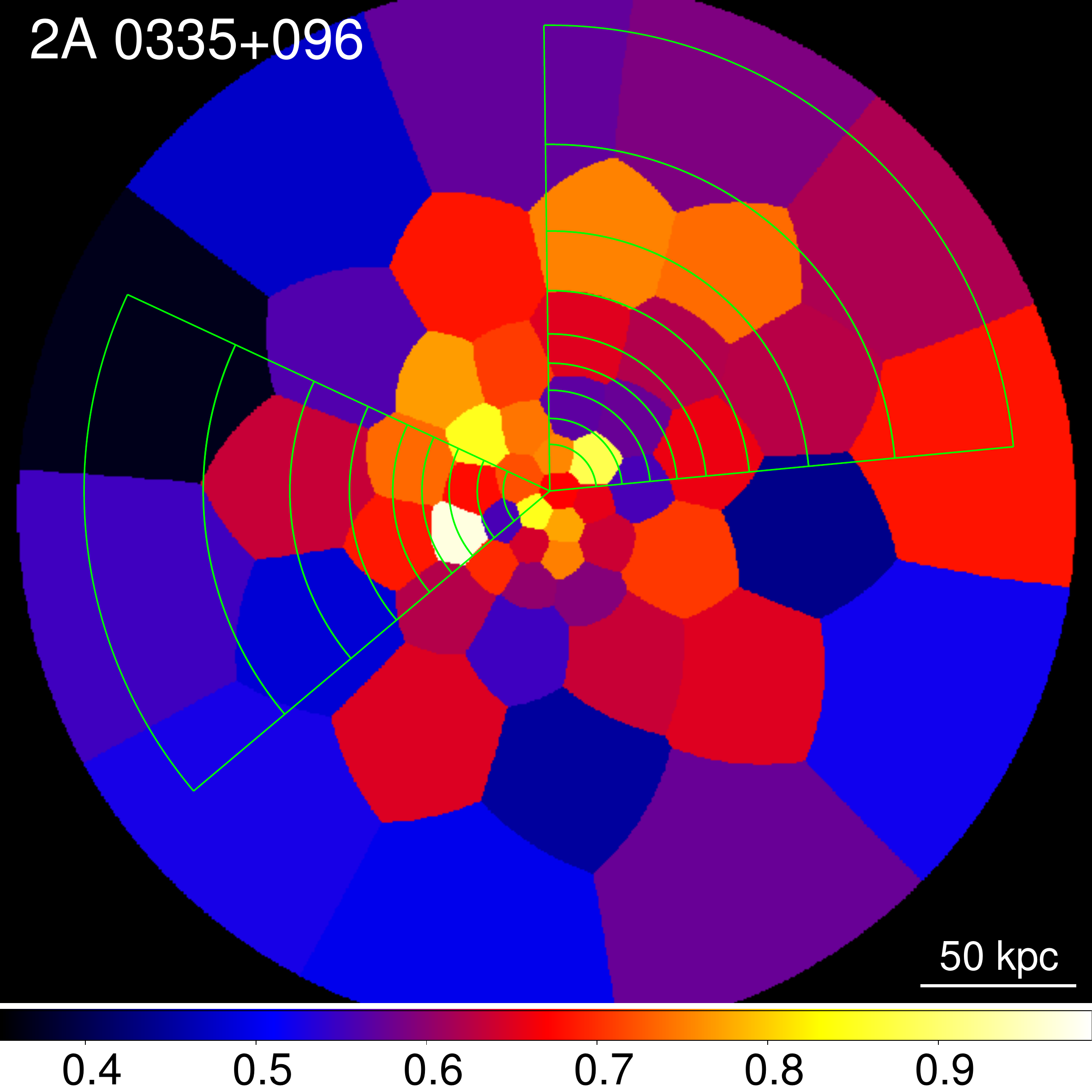}
\end{minipage}
\begin{minipage}{0.32\linewidth}
\includegraphics[width=\textwidth, trim=0mm 0mm 0mm 0mm, clip]{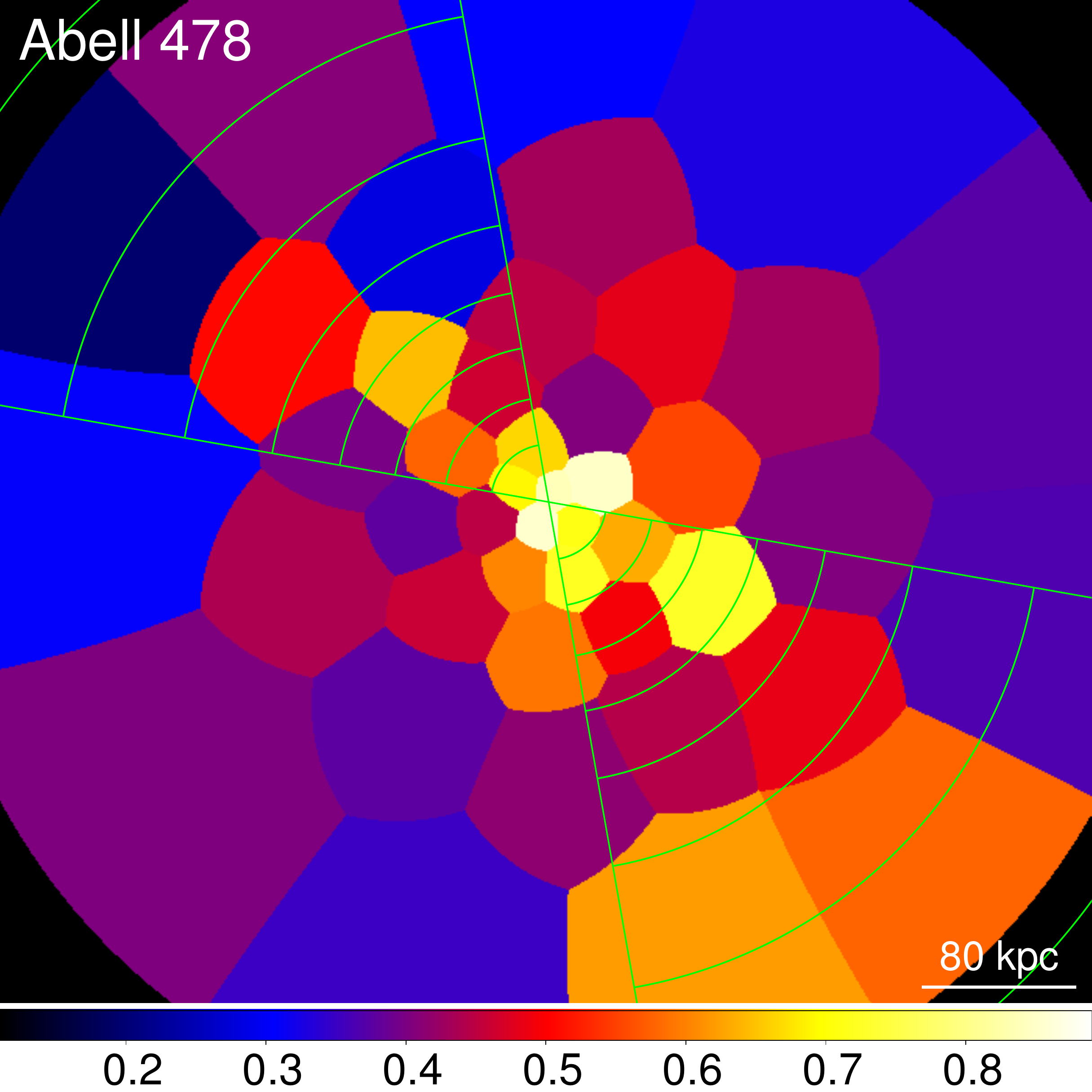}
\end{minipage}
\begin{minipage}{0.32\linewidth}
\includegraphics[width=\textwidth, trim=0mm 0mm 0mm 0mm, clip]{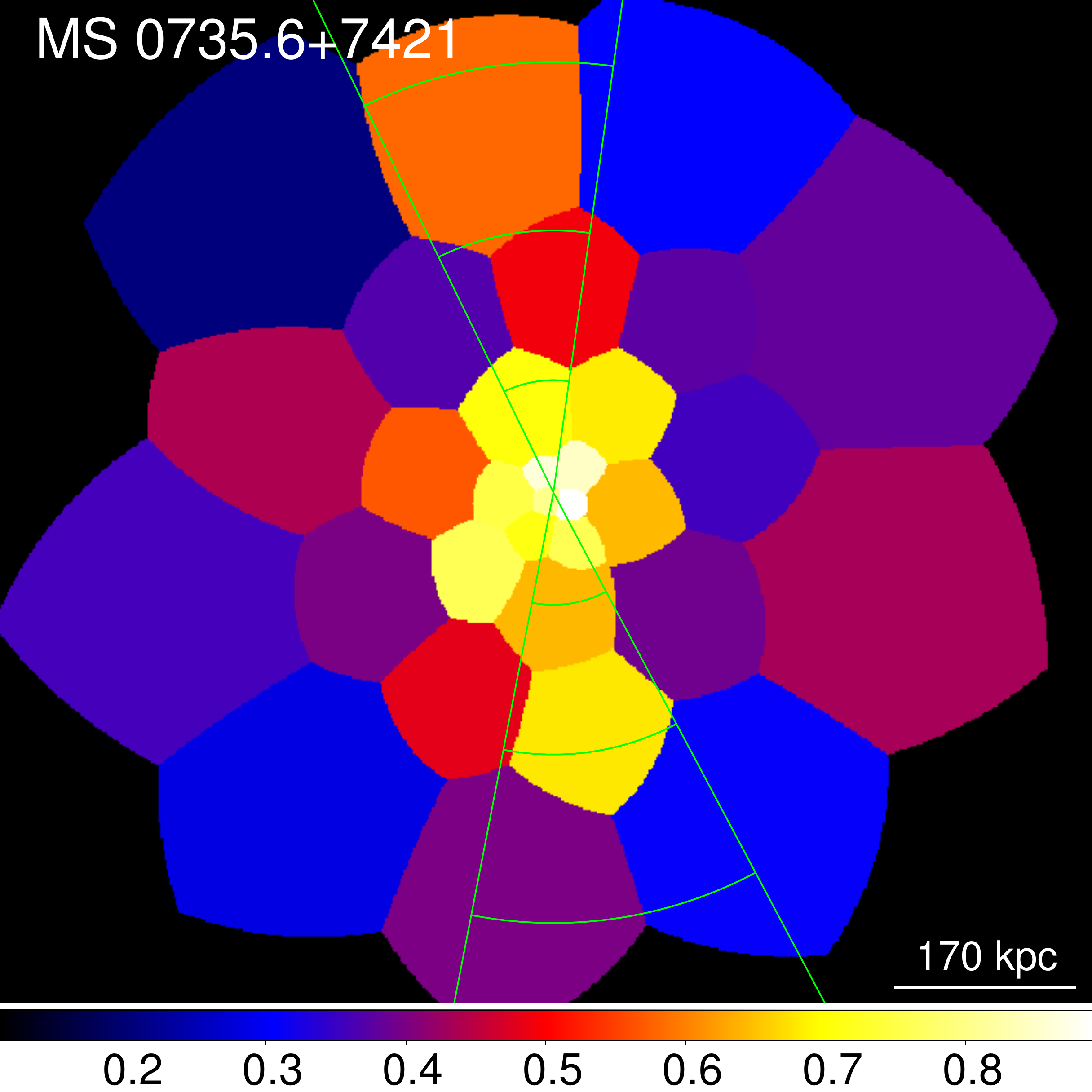}
\end{minipage}

\begin{minipage}{0.32\linewidth}
\includegraphics[width=\textwidth, trim=0mm 0mm 0mm 0mm, clip]{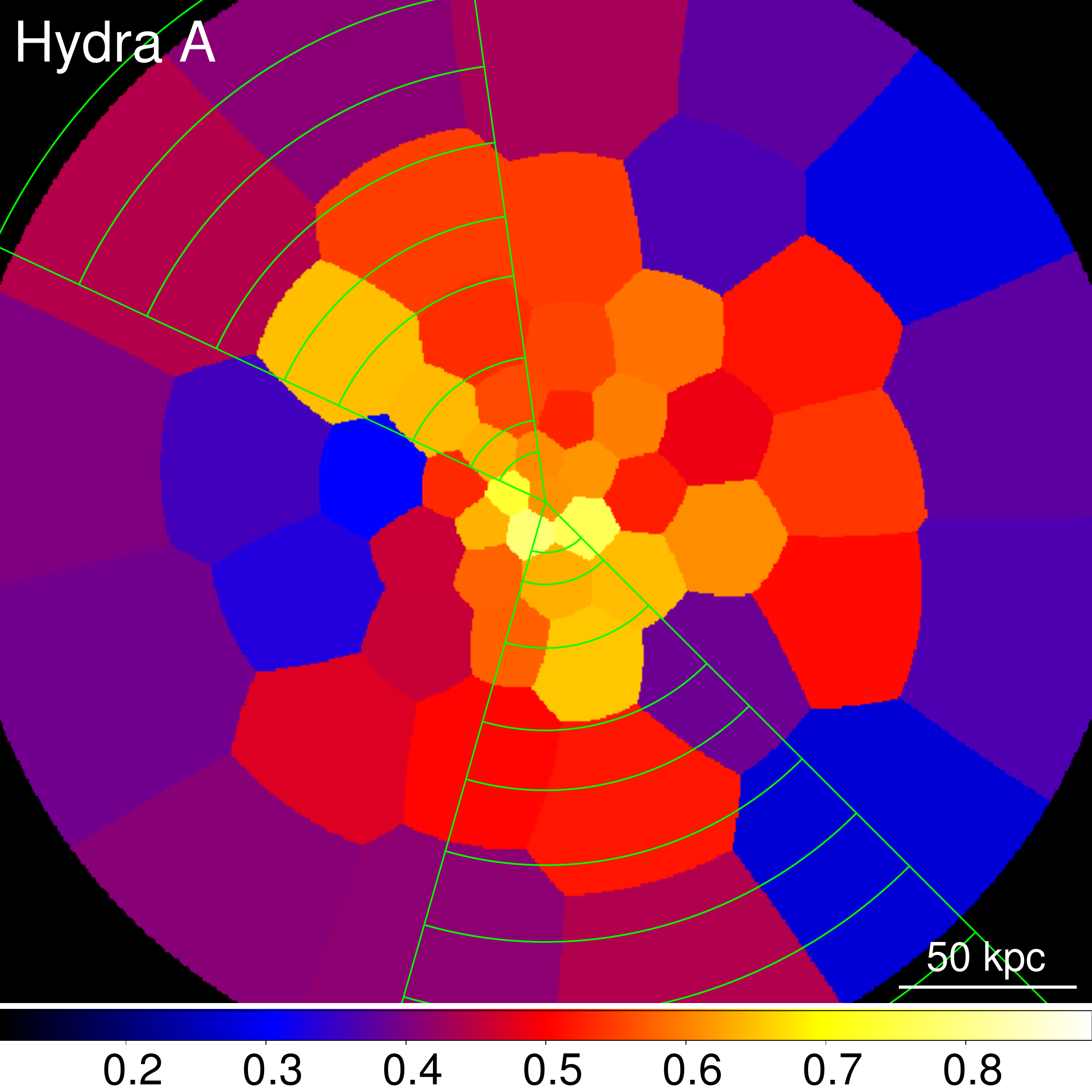}
\end{minipage}
\begin{minipage}{0.32\linewidth}
\includegraphics[width=\textwidth, trim=0mm 0mm 0mm 0mm, clip]{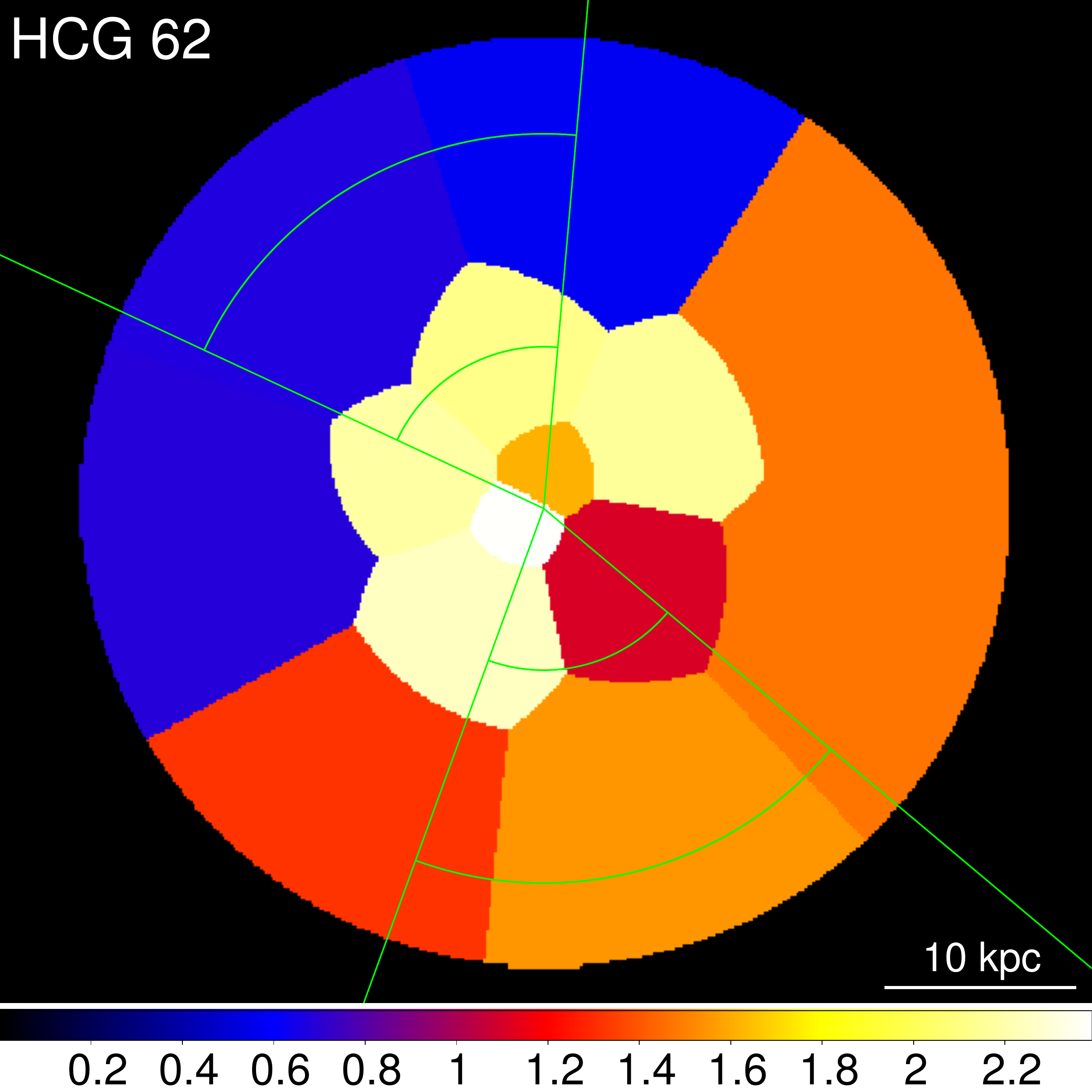}
\end{minipage}
\begin{minipage}{0.32\linewidth}
\includegraphics[width=\textwidth, trim=0mm 0mm 0mm 0mm, clip]{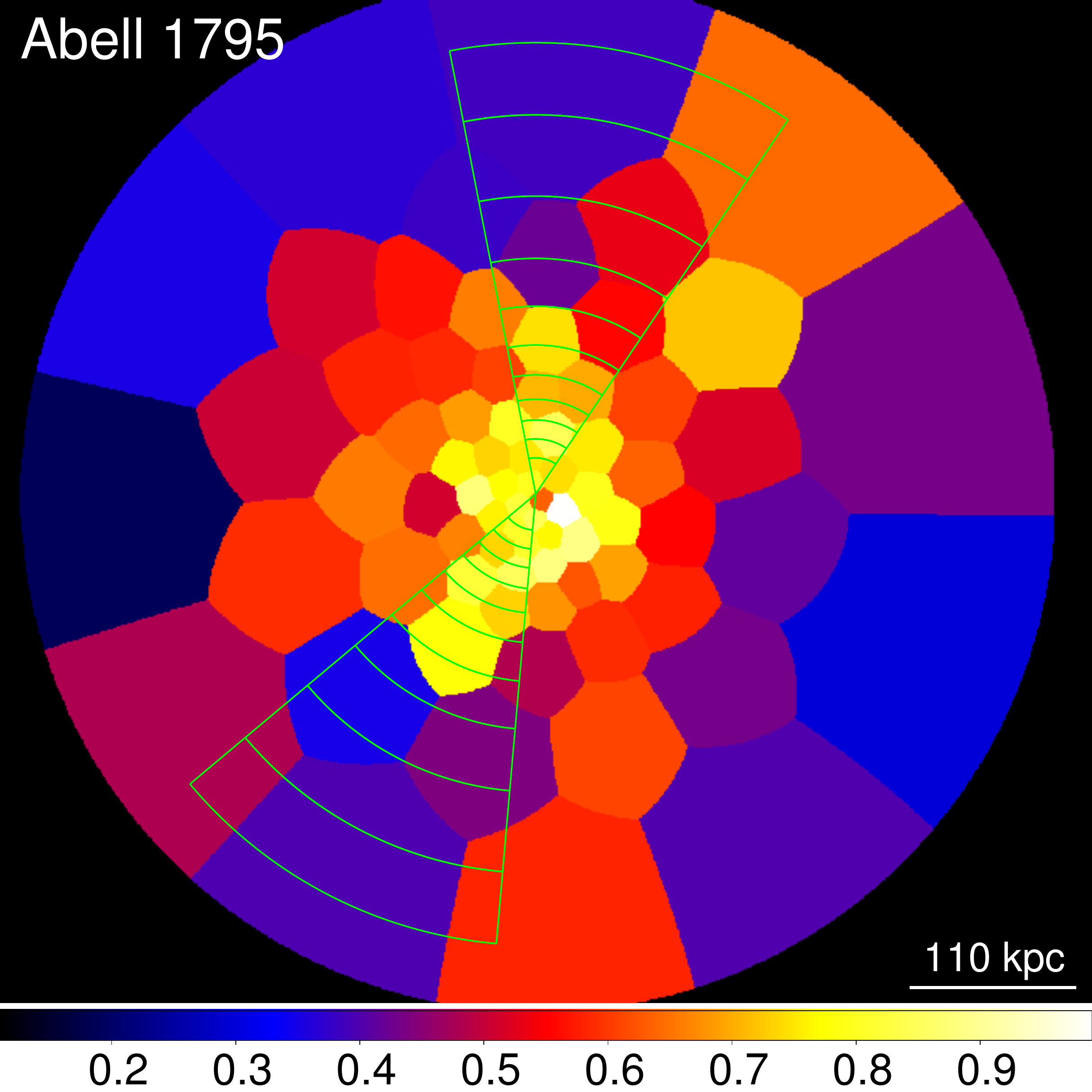}
\end{minipage}

\caption{Metallicity maps for the high quality sample.  The green sectors overlaid are along the axis of the launched X-ray cavities.  These bins are where the spectra are extracted when creating the on-jet metallicity profile.}
\label{fig:main-femaps}
\end{center}
\end{figure*}

\begin{figure*}
\ContinuedFloat
\begin{center}
\begin{minipage}{0.32\linewidth}
\includegraphics[width=\textwidth, trim=0mm 0mm 0mm 0mm, clip]{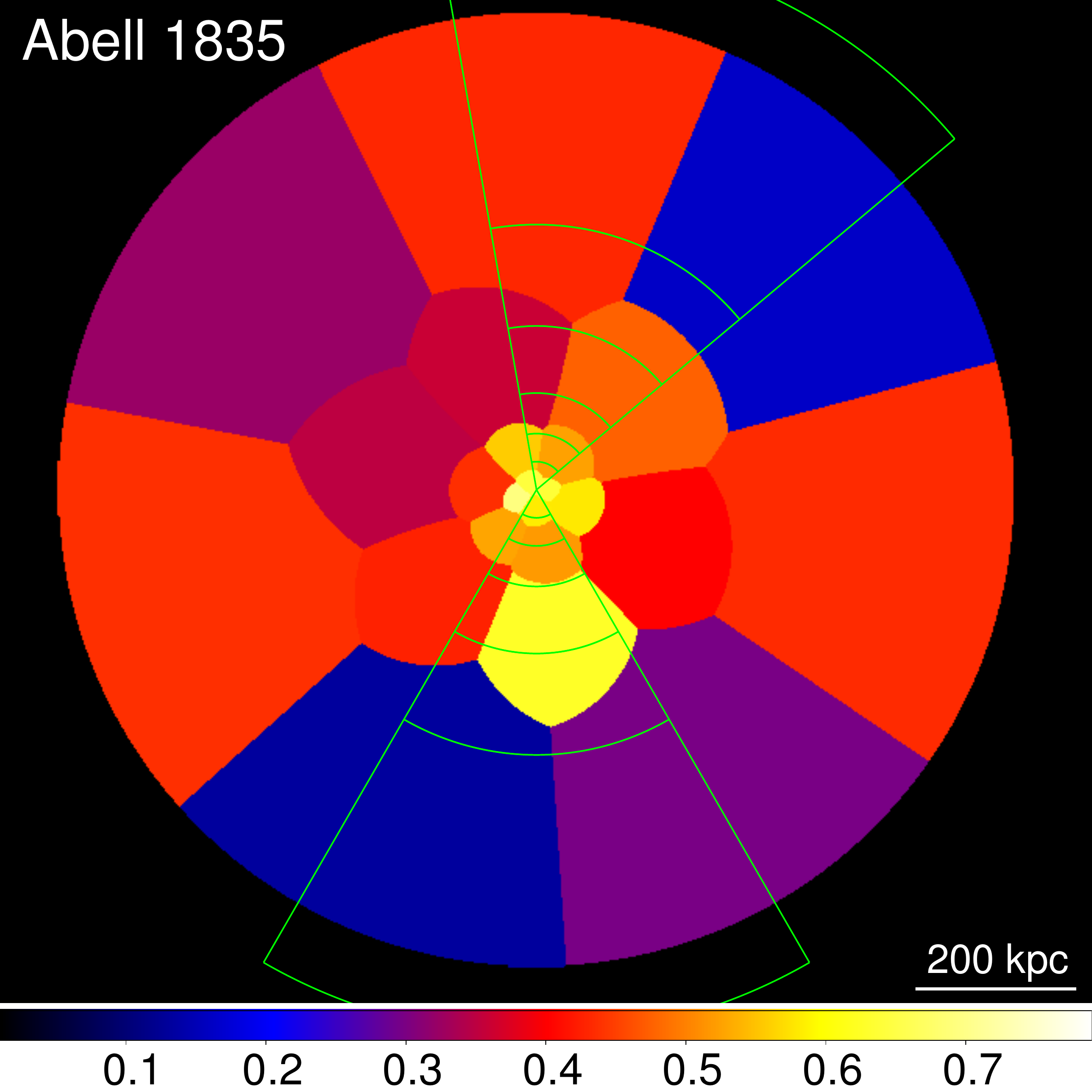}
\end{minipage}
\begin{minipage}{0.32\linewidth}
\includegraphics[width=\textwidth, trim=0mm 0mm 0mm 0mm, clip]{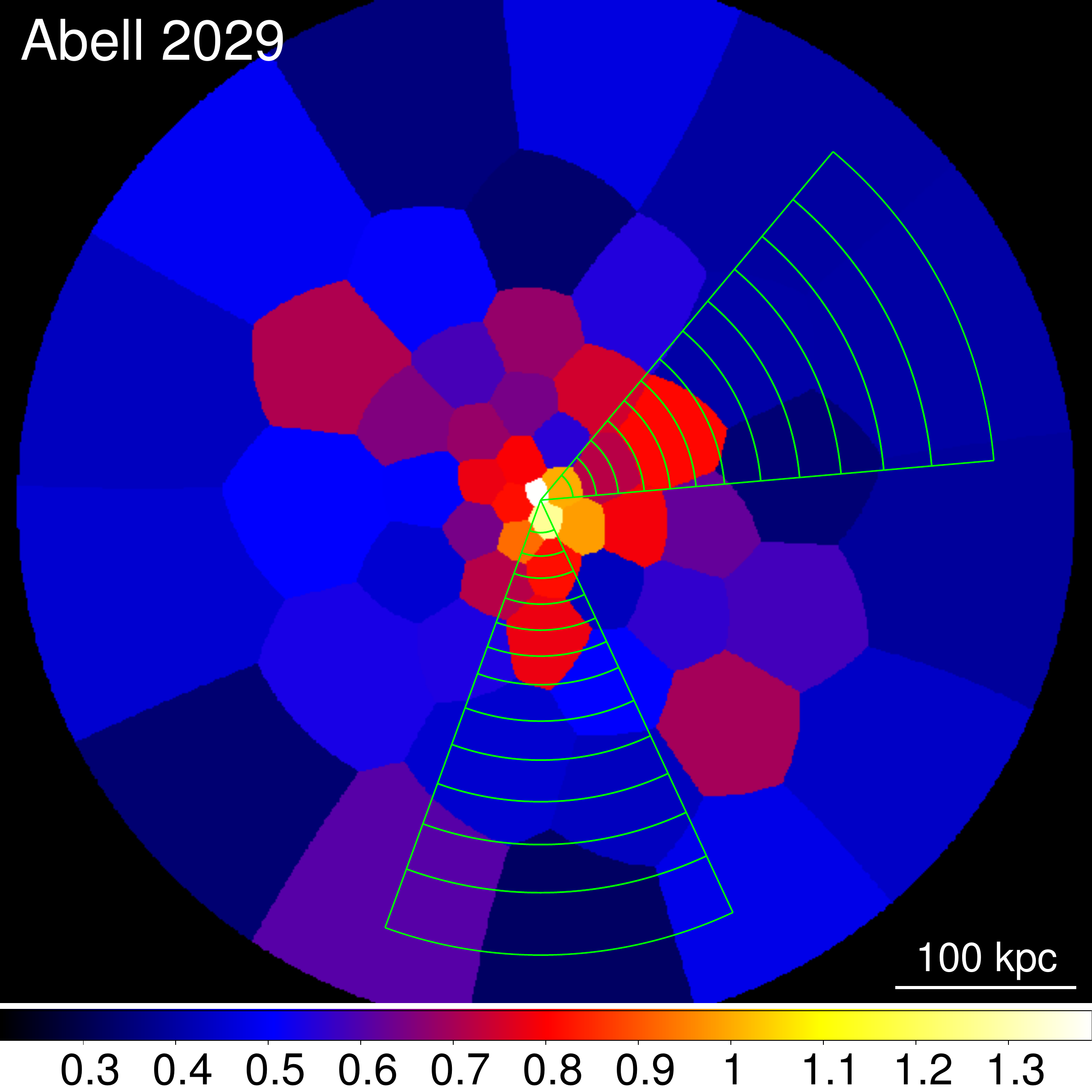}
\end{minipage}
\begin{minipage}{0.32\linewidth}
\includegraphics[width=\textwidth, trim=0mm 0mm 0mm 0mm, clip]{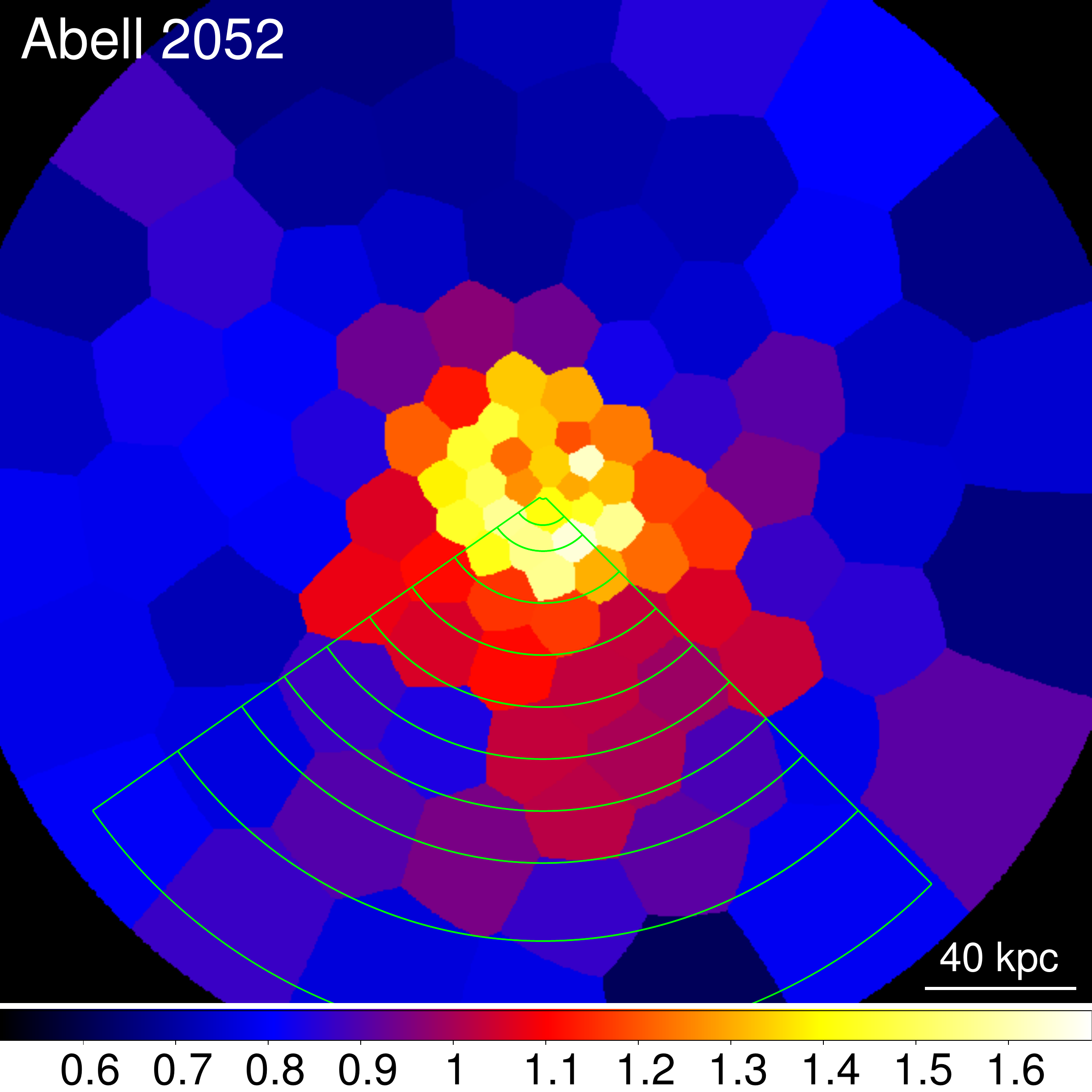}
\end{minipage}

\begin{minipage}{0.32\linewidth}
\includegraphics[width=\textwidth, trim=0mm 0mm 0mm 0mm, clip]{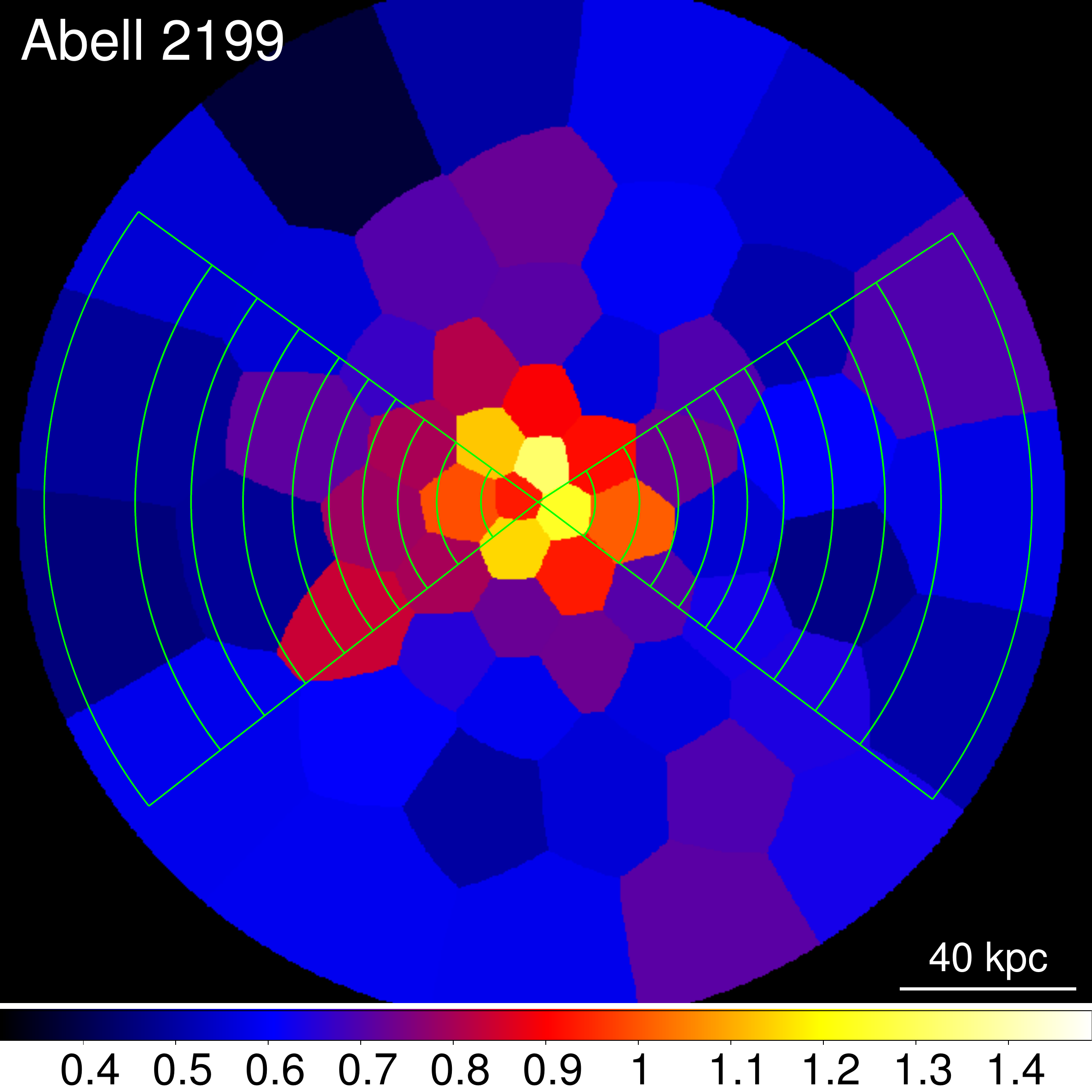}
\end{minipage}
\begin{minipage}{0.32\linewidth}
\includegraphics[width=\textwidth, trim=0mm 0mm 0mm 0mm, clip]{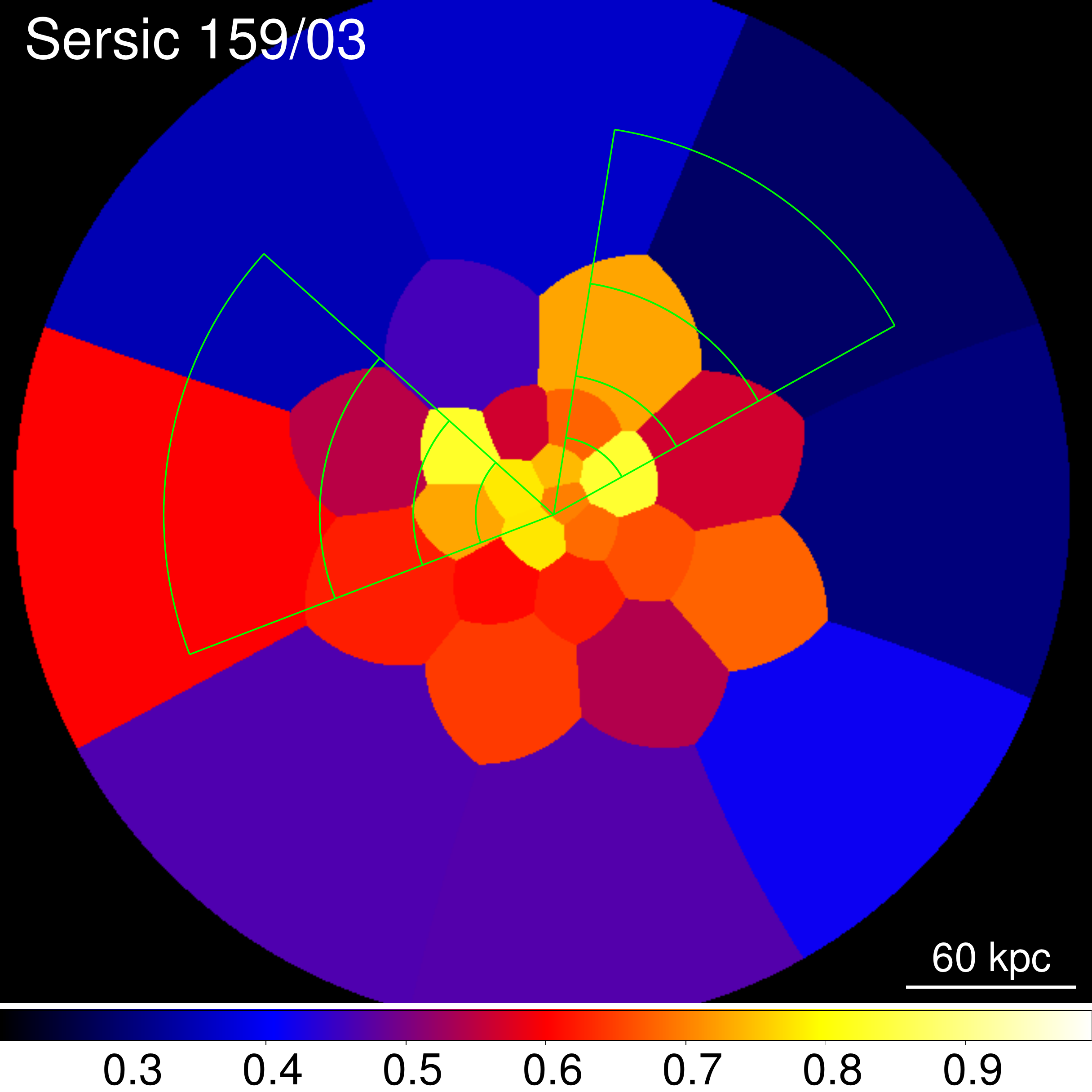}
\end{minipage}
\begin{minipage}{0.32\linewidth}
\includegraphics[width=\textwidth, trim=0mm 0mm 0mm 0mm, clip]{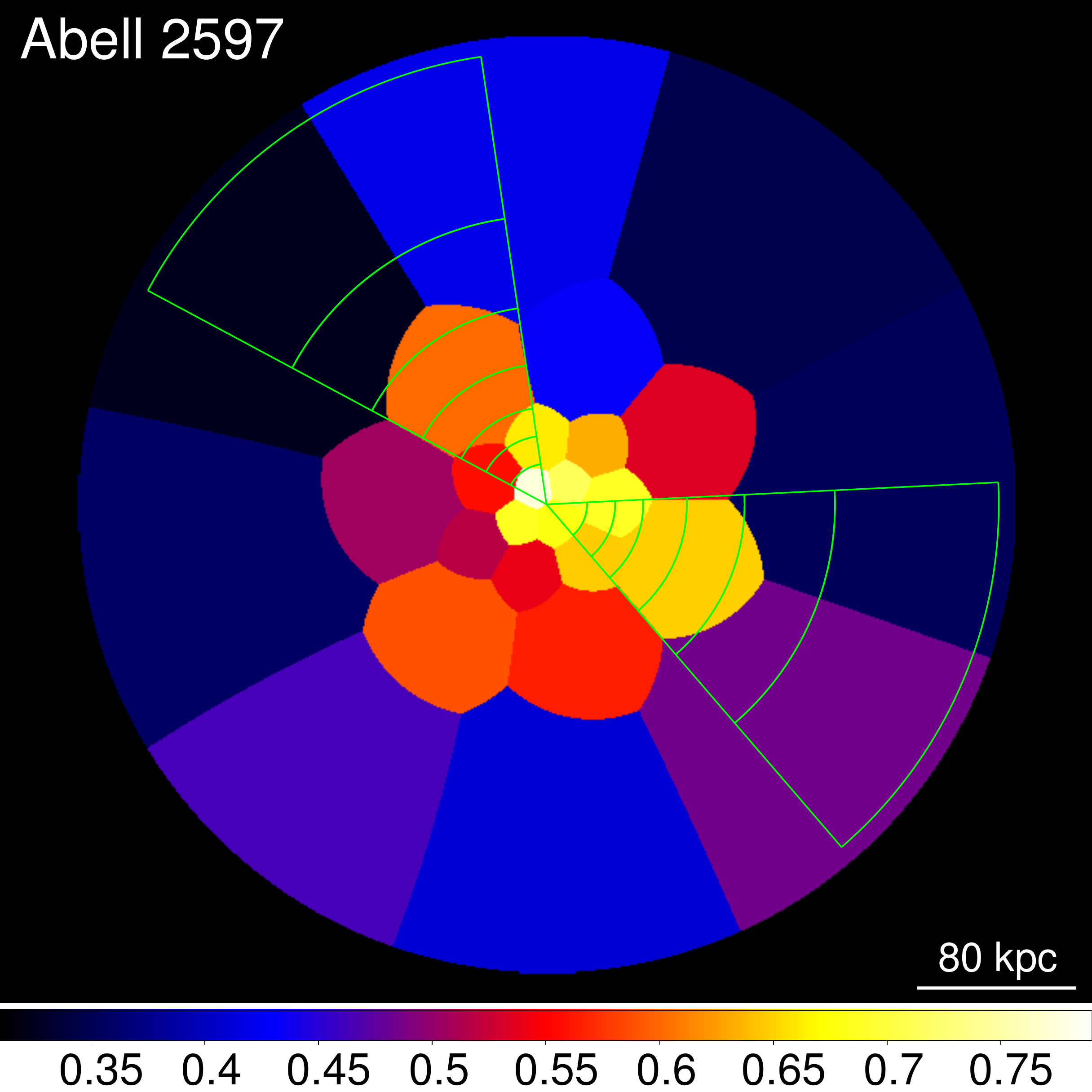}
\end{minipage}

\begin{minipage}{0.32\linewidth}
\includegraphics[width=\textwidth, trim=0mm 0mm 0mm 0mm, clip]{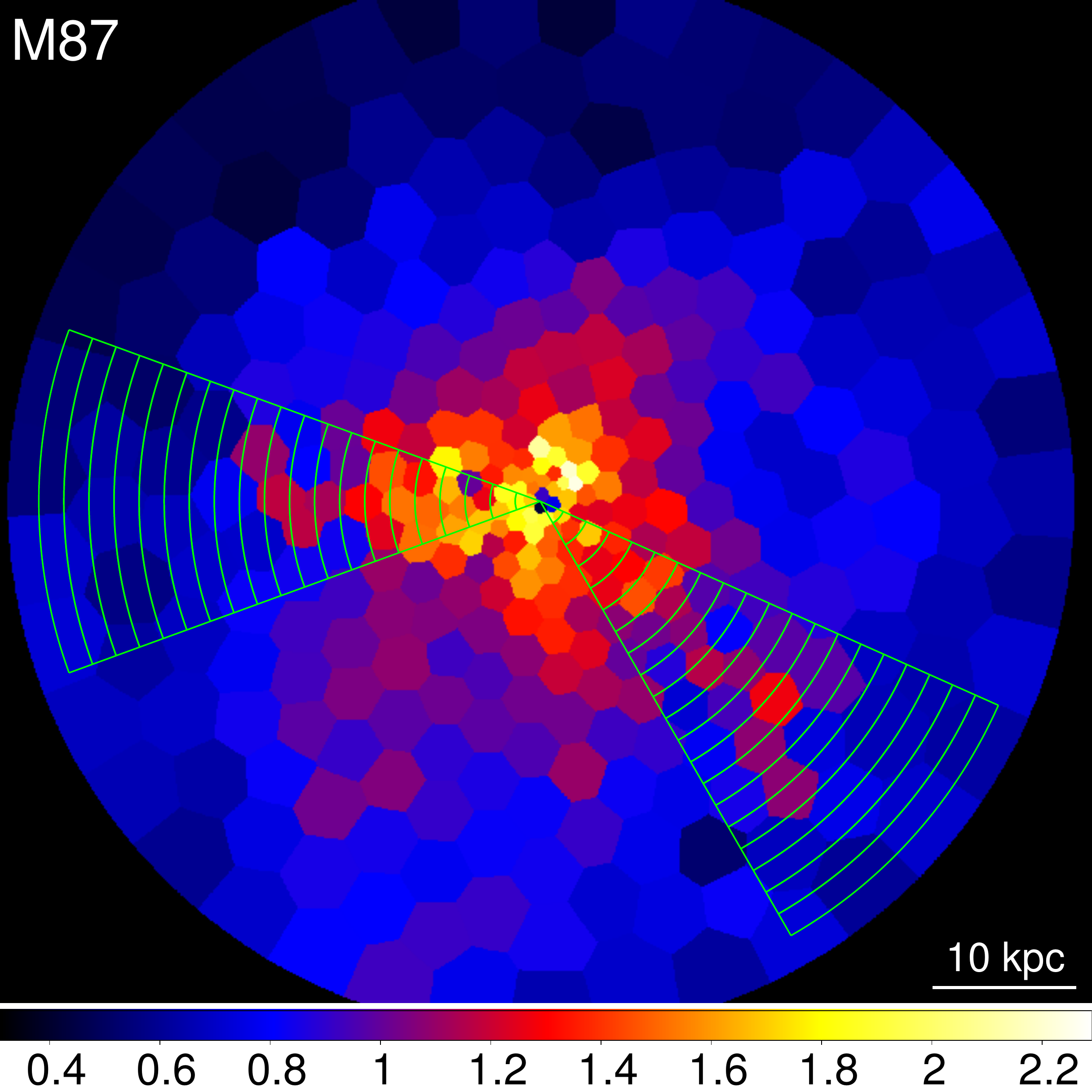}
\end{minipage}

\caption[]{{\it - continued}}
\end{center}
\end{figure*}

The WVT algorithm searches for the highest S/N pixel in the input image, which usually lies near the cluster centre.  The bin is grown by merging with the highest S/N neighbour and the centroid is then recalculated.  The process repeats until the bin reaches the target S/N.  Bins are created similarly until the cluster is mapped at the desired S/N per pixel.  Maps of the more shallowly exposed objects from the Extended Sample were generated at constant spatial resolution rather than a targeted S/N.  We chose bin sizes to insure more than a single bin per quadrant on the sky and S/N ranging between 45 and 75.   

We do not use the metallicity maps to measure the iron radius discussed below. The maps in Figure~\ref{fig:main-femaps} are intended to show how the gas phase metallicity changes with altitude and azimuthal angle.  In each panel of Figure~\ref{fig:main-femaps} we have indicated the spectra extraction regions for abundance profiles along the axis of the cavities system.  The orientation and width of the bins approximates the orientation and width of the cavity systems.

Inspection of the maps reveal plums of high metallicity gas extending from the centre.  Striking examples include Hydra A, Abell 1795, Abell 478, and particularly MS0735, which reveals a prominent ``rust streak" to an altitude approaching 300 kpc along the cavities.  Lacking direct velocity measurements, we assume the metals track outflowing plasma lifted by the bubbles. This assumption may be tested in future with the combination of {\it Astro-H} \citep{tak12} and {\it Athena}.   

Note that high metallicity gas is found preferentially, but not exclusively, along the bubbles.  High metallicity patches are found away from the bubbles as well.  A good example is Hydra A, which shows a high metallicity plume away from the cavities associated with an X-ray filament \citep{cck09b}.  These filaments may be gas returning in a circulation flow \citep{bm06} or gas  being blown around in the dynamic intracluster medium \citep{hbyl06, mor10, ro10, sim12}.

The situation in the Extended Sample, for which we have not shown the metallicity maps, is less clear.  The spatial bins are too large to reveal any convincing evidence for the variation of metallicity along the cavity axis.  Due to these large uncertainties in individual systems, they are not included in the trend analysis below.

\subsection{Updated P$_{\rm jet}$ - R$_{\rm Fe}$ Scaling Relation}
\label{subsec:main-jetab}

With evidence from the metallicity maps for metal-rich plumes, we have estimated the maximum altitude of the metallicity enhancement in the ICM.  The existence and size of the enhancement was found by comparing the metallicity profiles of each cluster along and orthogonal to the X-ray cavities.  The profiles were created using semi-annular bins with opening angles lying between 45 and 90 degrees for most clusters.  One profile enclosed the region along the X-ray cavities and/or where extended radio emission from a jet is found (``on-jet'').  The second profile in the orthogonal direction samples the undisturbed atmosphere (``off-jet'').  We assume the metallicity profile of the undisturbed atmosphere represents the average prior to recent AGN activity.  

To find deviations between the on-jet and off-jet profiles, each semi-annular bin included enough counts to reduce the measurement uncertainties to approximately $10\%$.  This criterion was crucial to preserve spatial resolution because larger radial bins tend to average away small scale metallicity  fluctuations due to the overall metallicity gradient.  This criterion was met by extracting profiles with an average of 9 bins with minimum S/N of 140 for each bin.

Profiles for the High Quality sample are found in Figure~\ref{fig:main-jetab}.  The on-jet profiles are represented by the circular data points.  The off-jet profiles are represented by triangles.  The dotted vertical line represents the location of the ``iron radius".  We define this point by counting bins outward from the centre until we reach the bin where the 1$\sigma$ error bars do not overlap and the regions beyond this radius are either indistinguishable between sectors or the off-jet region is higher in metallicity.

\begin{figure}
\begin{center}
\includegraphics[totalheight=0.45\textwidth]{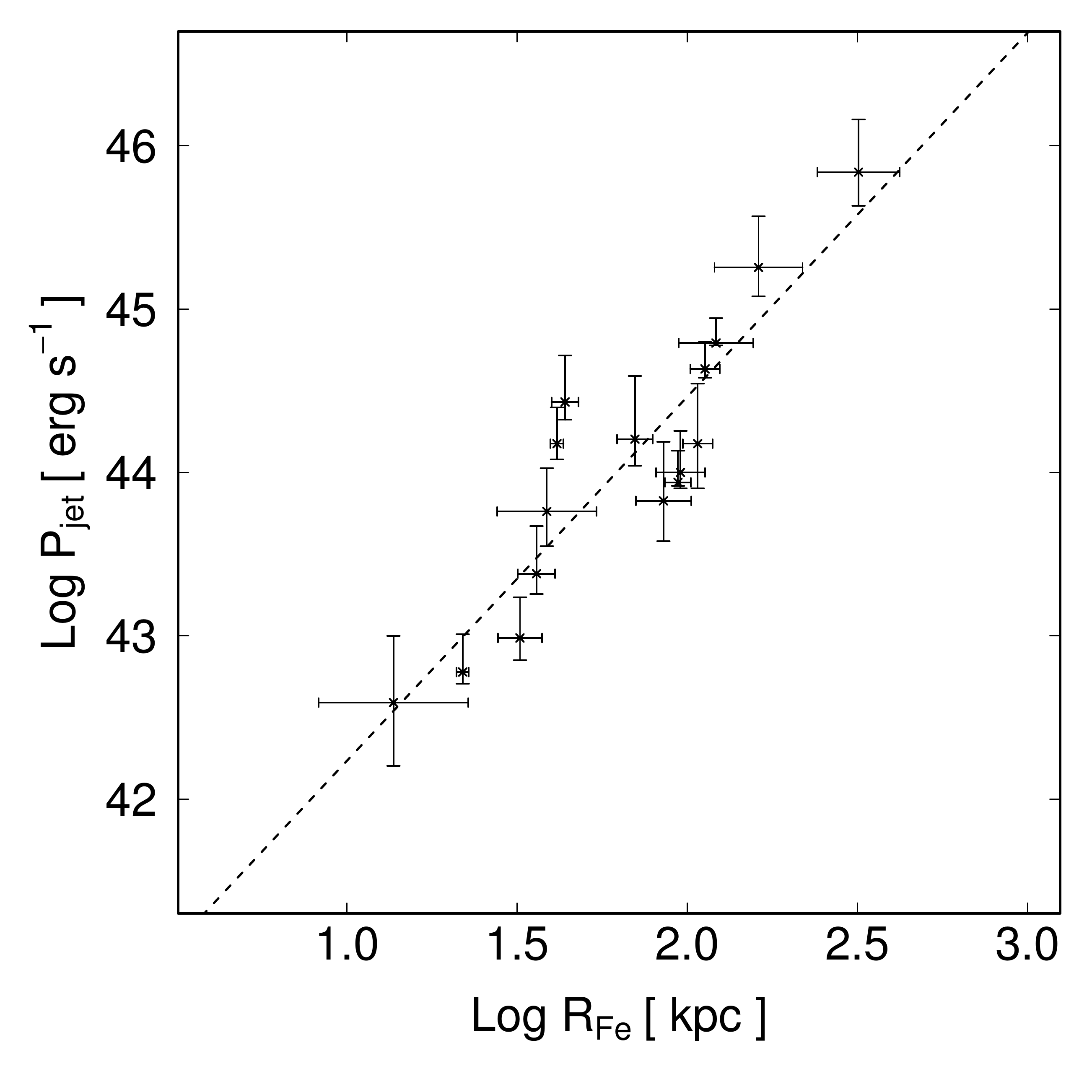}
\caption{Updated jet power vs. iron radius.  This includes six more data points than the previous relationship presented in \citet{cck11}.  The dashed line is the new best-fitting to the data.}
\label{fig:main-jet-dist}
\end{center}
\end{figure}

With a smaller subset of data, a relationship has been found showing the iron radius scales with the AGN jet power \citep{cck11}.  We refined this trend using the entire high quality sample which we expanded here to include additional low-power and high-power systems.  The new jet power vs. iron radius plot is found in Figure~\ref{fig:main-jet-dist}.  The jet powers used here were taken from cavity measurements by \citet{raf06}, except for Sersic 159/03, which were recalculated by including new data from a longer exposure time observation (ObsID 11758.)  The ``error bars'' for $R_{\rm Fe}$ represent the width of the bin in the metallicity profiles (Figure~\ref{fig:main-jetab}).  After performing a least-squares linear regression, the relationship is found to be,
\begin{equation} \label{eqn:main-pjet-rfe}
R_{\rm Fe} = (62 \pm 26) \times P_{\rm jet}^{(0.45 \pm 0.06)} ~(\rm kpc),
\end{equation}
where jet power is in units of $10^{44}$ erg s$^{-1}$.  The standard deviation is approximately 0.87 dex.  Increasing the sample size from 10 to 16 changed the relationship
imperceptibly.  The predicted iron radius is slightly higher, but within the error of the sample the new relationship is indistinguishable from that in \citet{cck11}.

\begin{figure*}
\begin{center}
\begin{minipage}{0.48\linewidth}
\includegraphics[width=\textwidth, trim=0mm 0mm 0mm 0mm, clip]{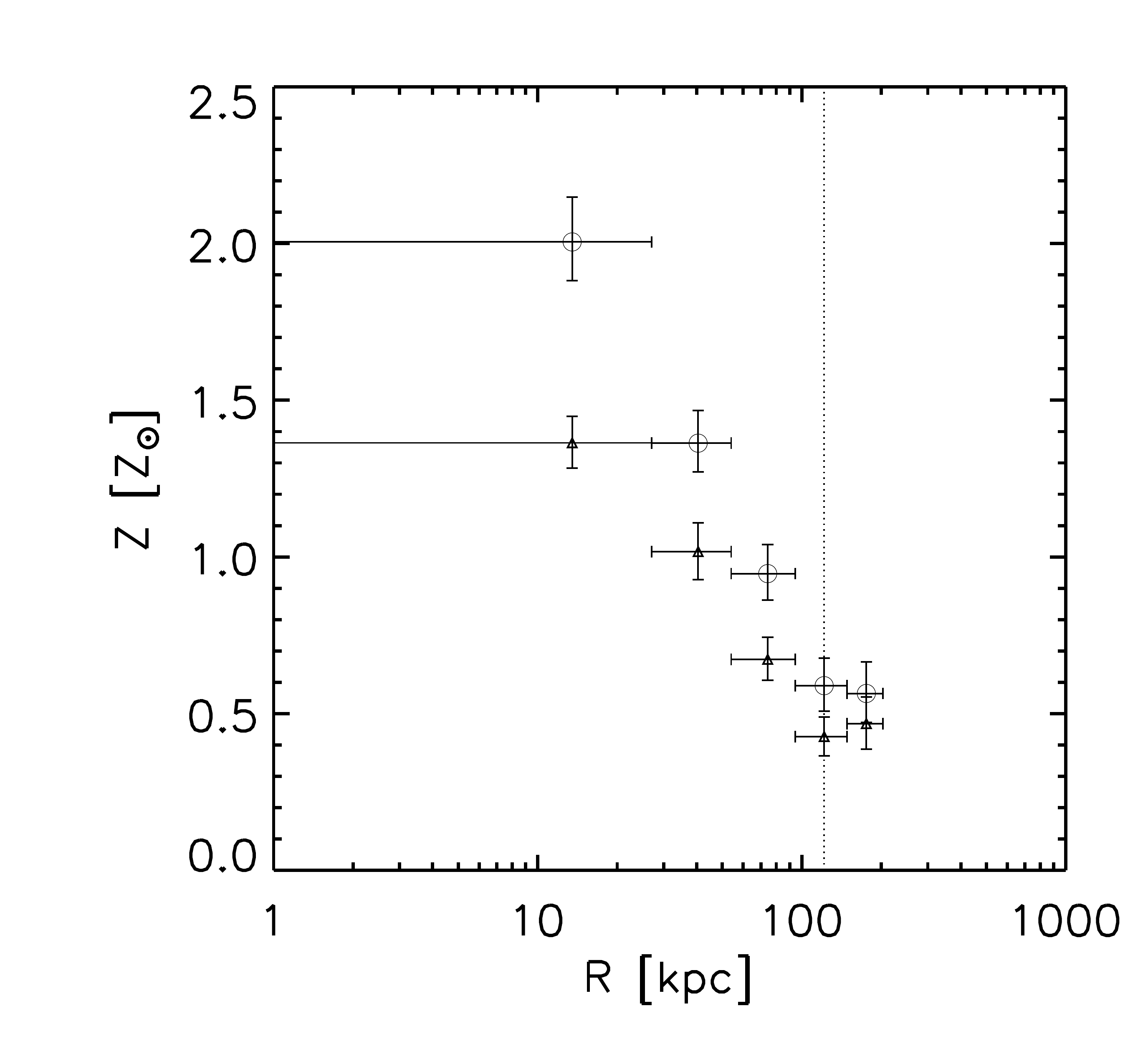}
\end{minipage}
\begin{minipage}{0.48\linewidth}
\includegraphics[width=\textwidth, trim=0mm 0mm 0mm 0mm, clip]{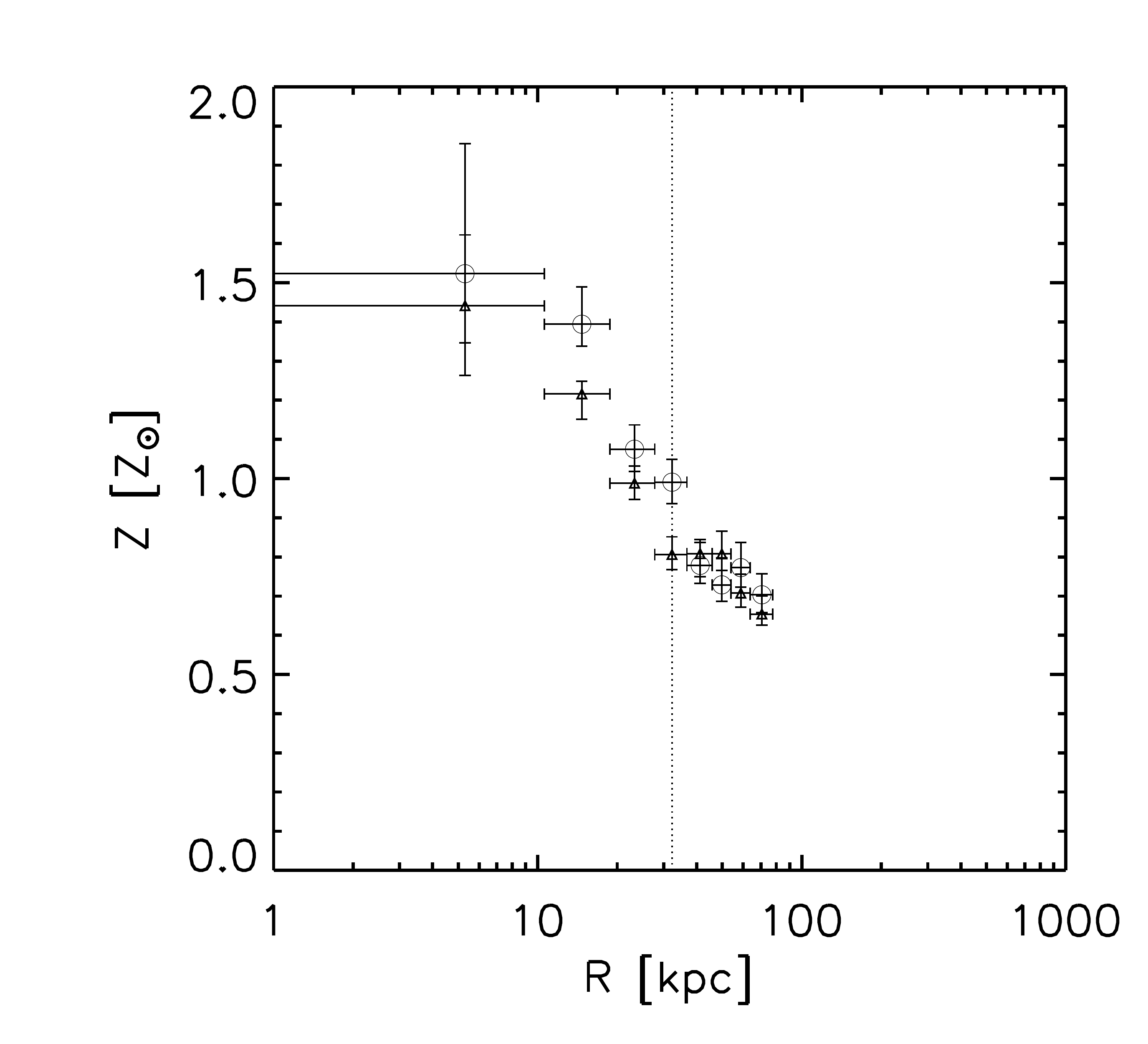}
\end{minipage}

\begin{minipage}{0.48\linewidth}
\includegraphics[width=\textwidth, trim=0mm 0mm 0mm 0mm, clip]{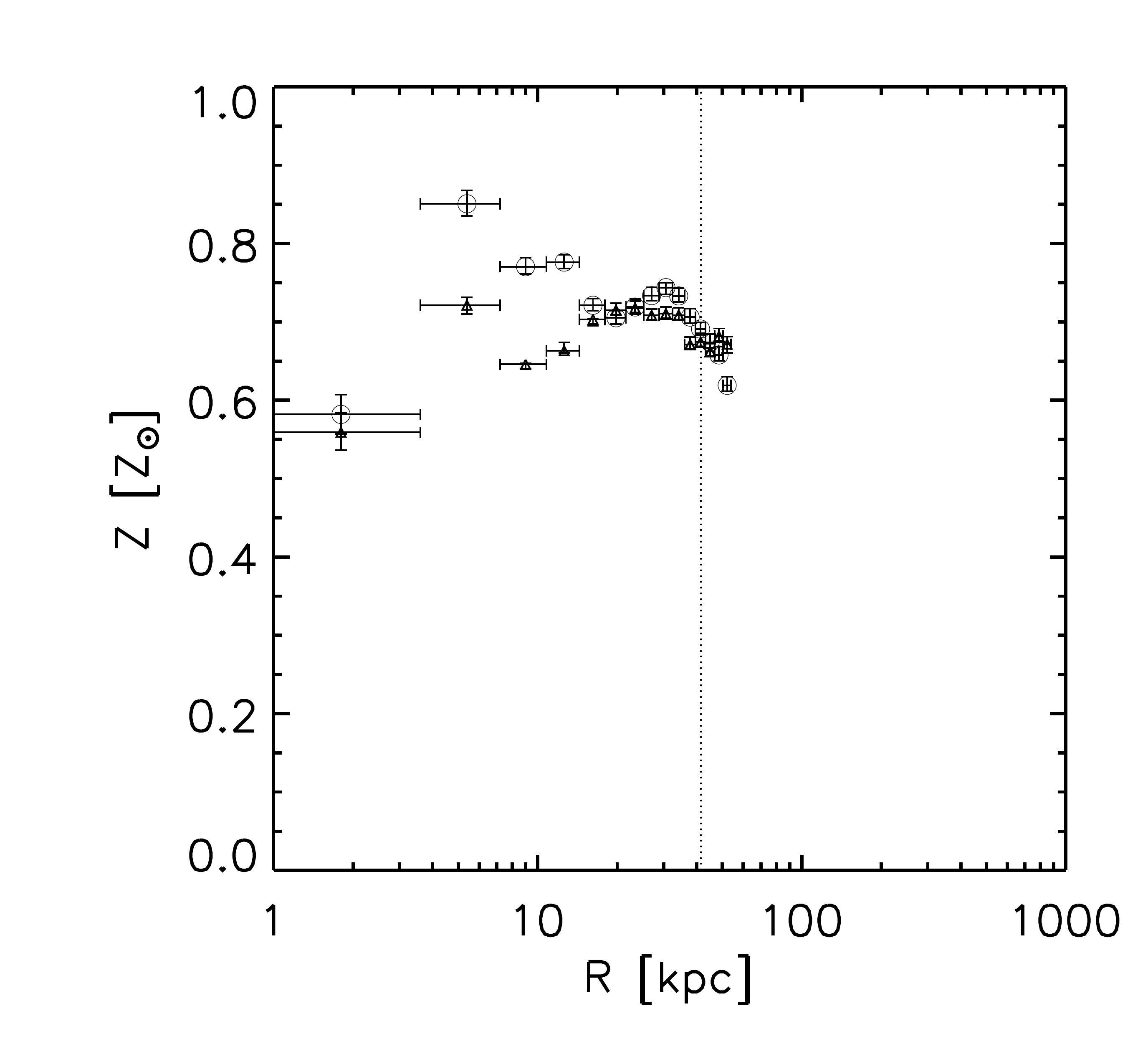}
\end{minipage}
\begin{minipage}{0.48\linewidth}
\includegraphics[width=\textwidth, trim=0mm 0mm 0mm 0mm, clip]{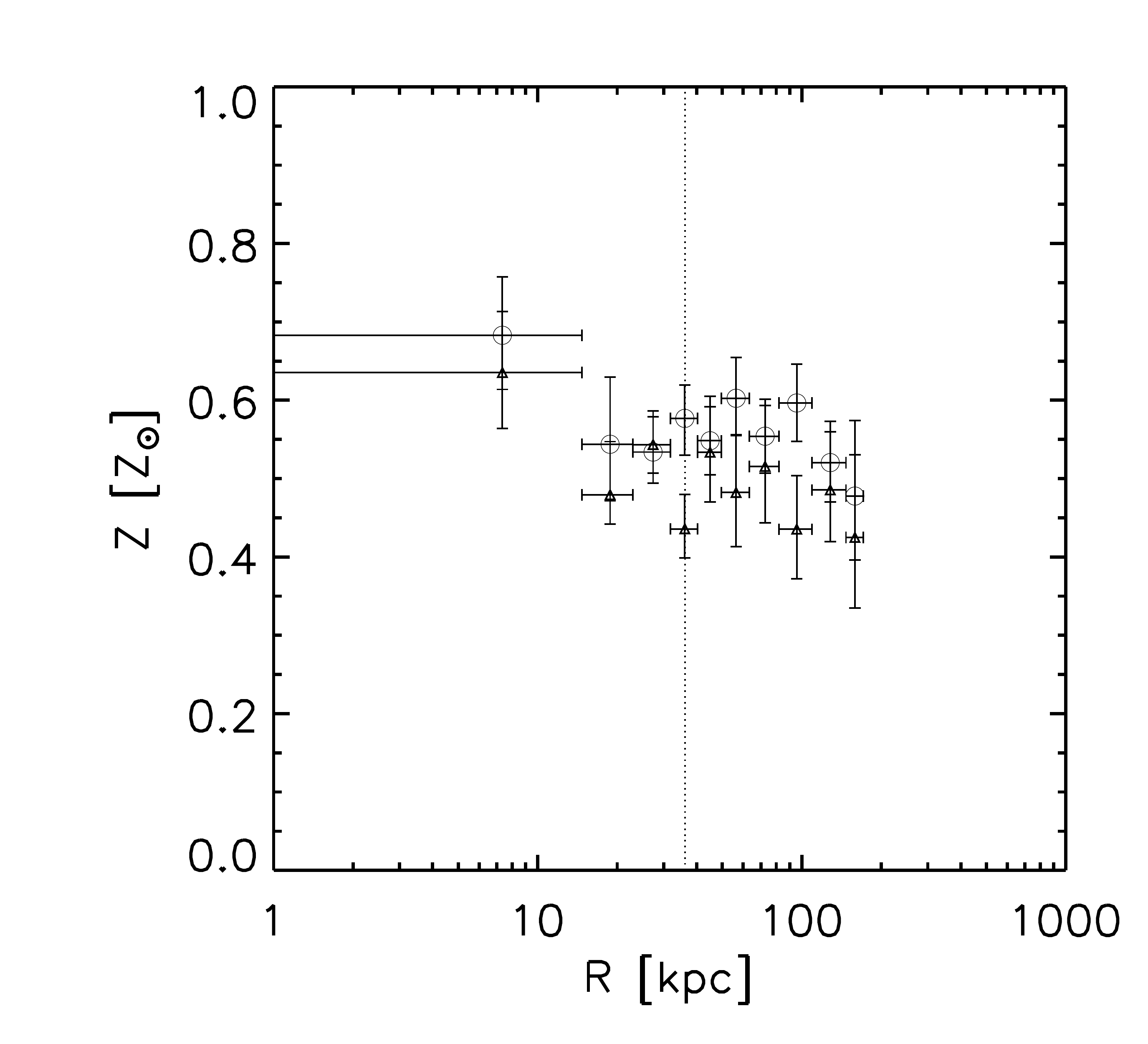}
\end{minipage}

\begin{minipage}{0.48\linewidth}
\includegraphics[width=\textwidth, trim=0mm 0mm 0mm 0mm, clip]{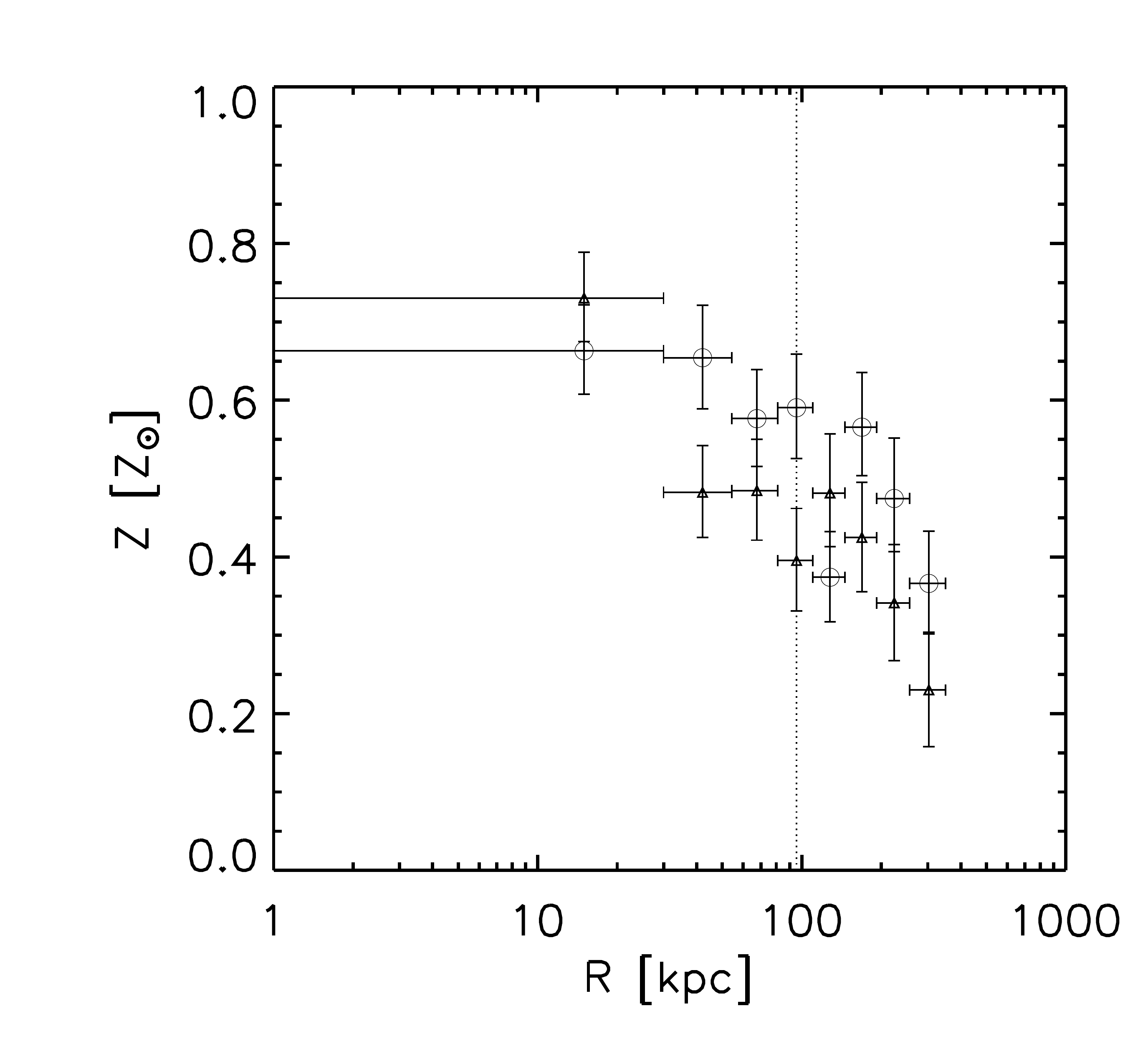}
\end{minipage}
\begin{minipage}{0.48\linewidth}
\includegraphics[width=\textwidth, trim=0mm 0mm 0mm 0mm, clip]{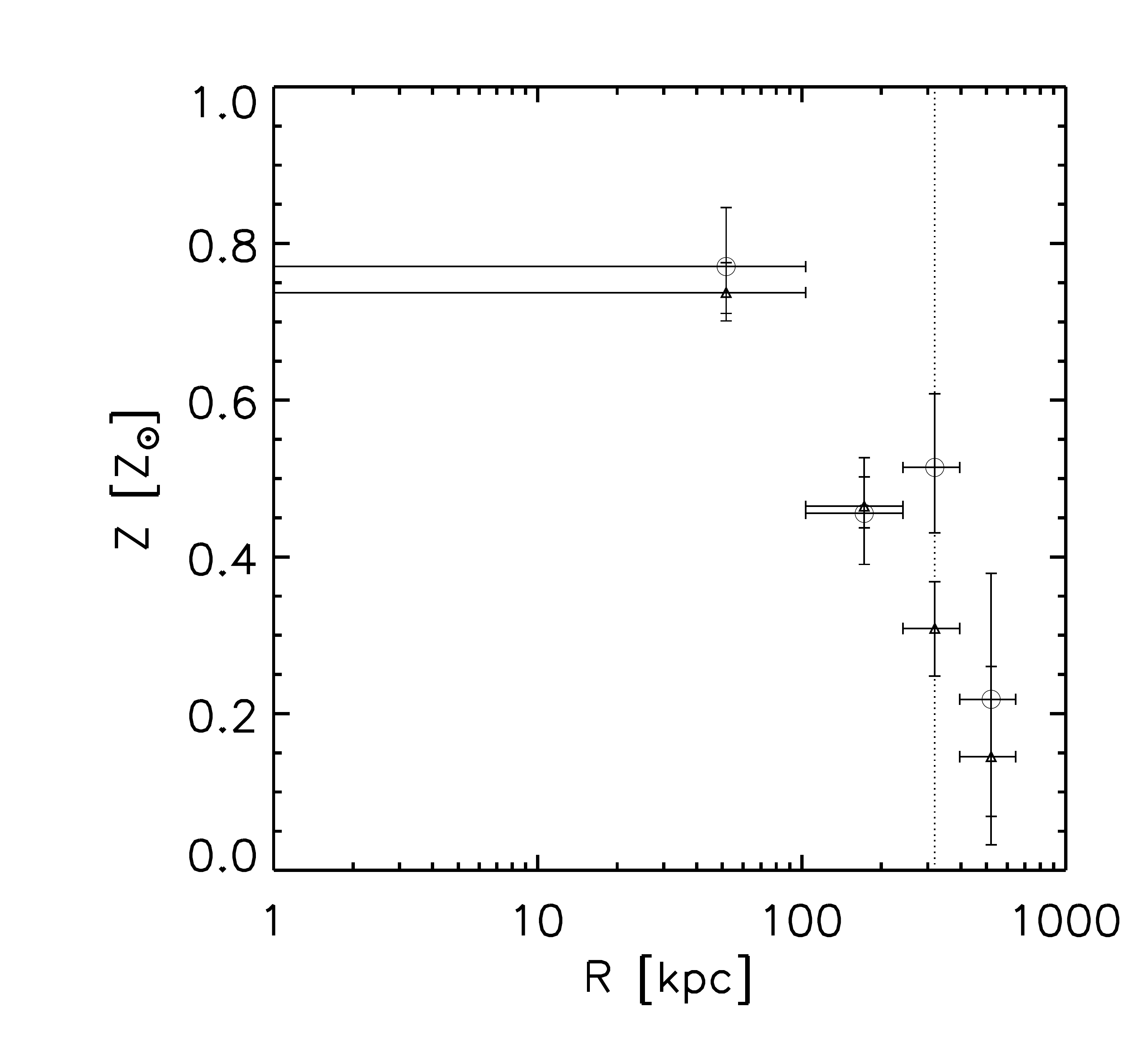}
\end{minipage}
\caption{Metallicity profiles for on-jet regions are represented by circles.  Off-jet profiles are represented by triangles.  The dotted line is the measured iron radius.  Top row: A133 and A262.  Middle row: Perseus and 2A 0335.  Bottom row: A478 and MS 0735.}
\label{fig:main-jetab}
\end{center}
\end{figure*}

\begin{figure*}
\ContinuedFloat
\begin{center}
\begin{minipage}{0.48\linewidth}
\includegraphics[width=\textwidth, trim=0mm 0mm 0mm 0mm, clip]{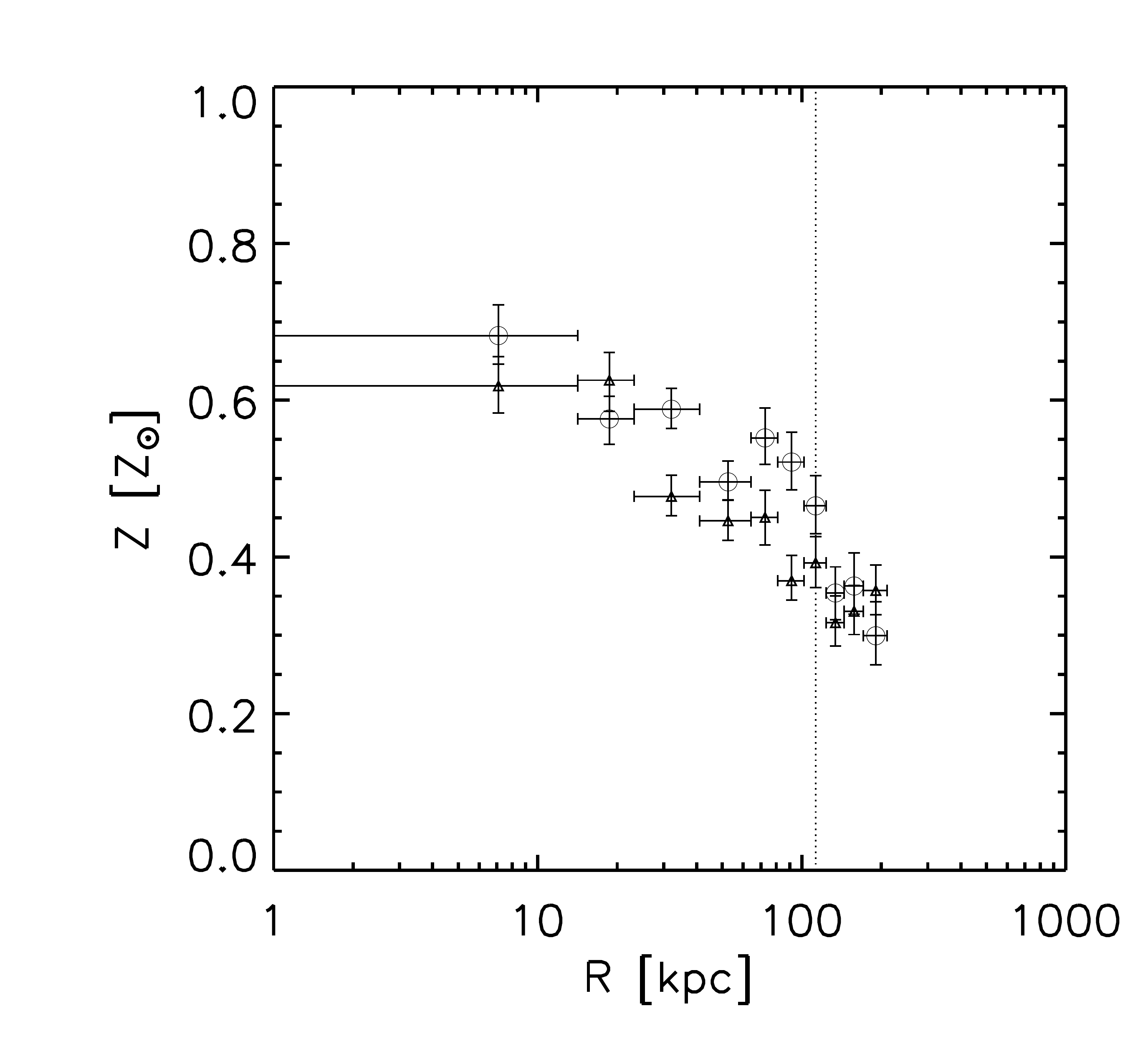}
\end{minipage}
\begin{minipage}{0.48\linewidth}
\includegraphics[width=\textwidth, trim=0mm 0mm 0mm 0mm, clip]{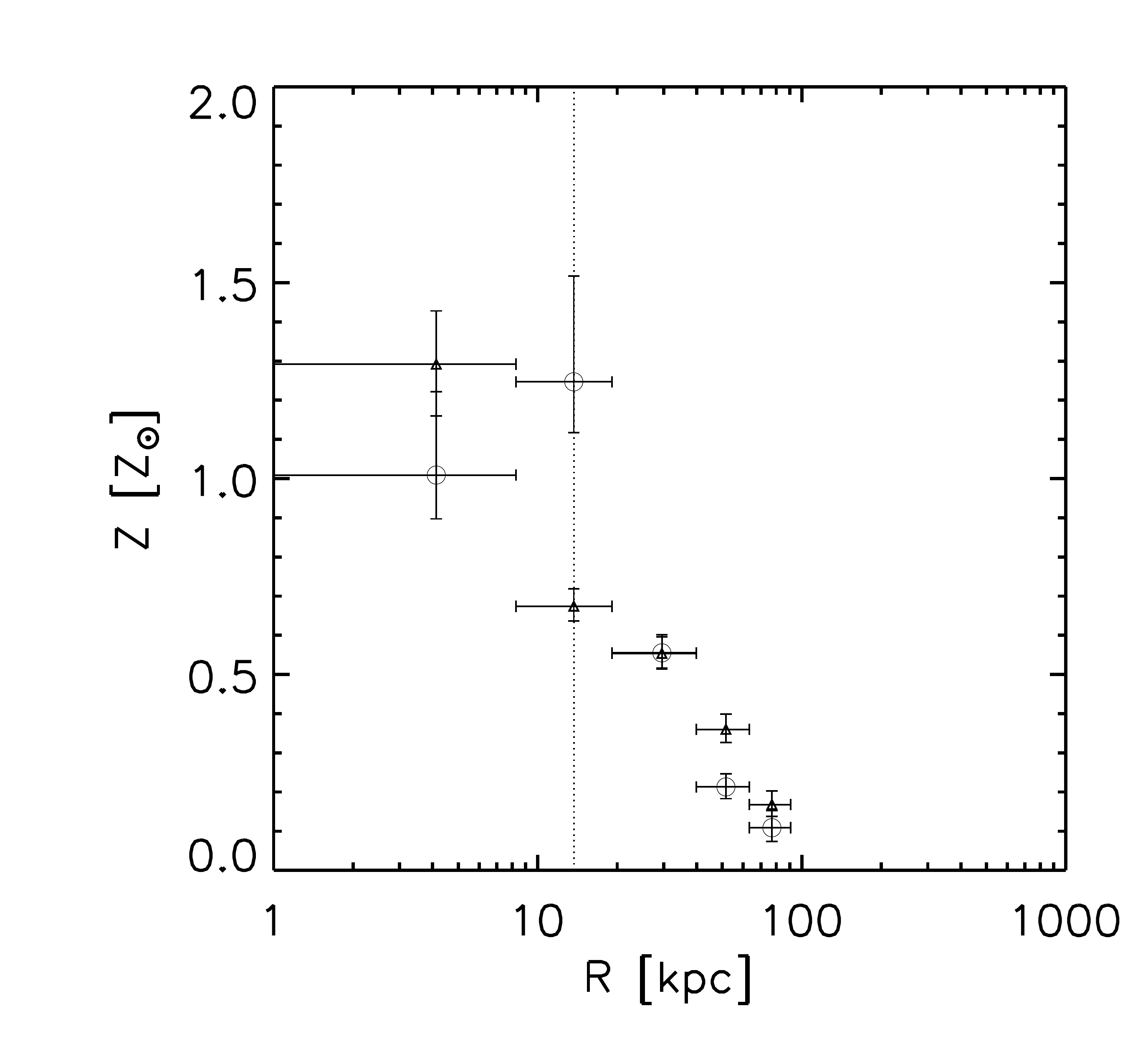}
\end{minipage}

\begin{minipage}{0.48\linewidth}
\includegraphics[width=\textwidth, trim=0mm 0mm 0mm 0mm, clip]{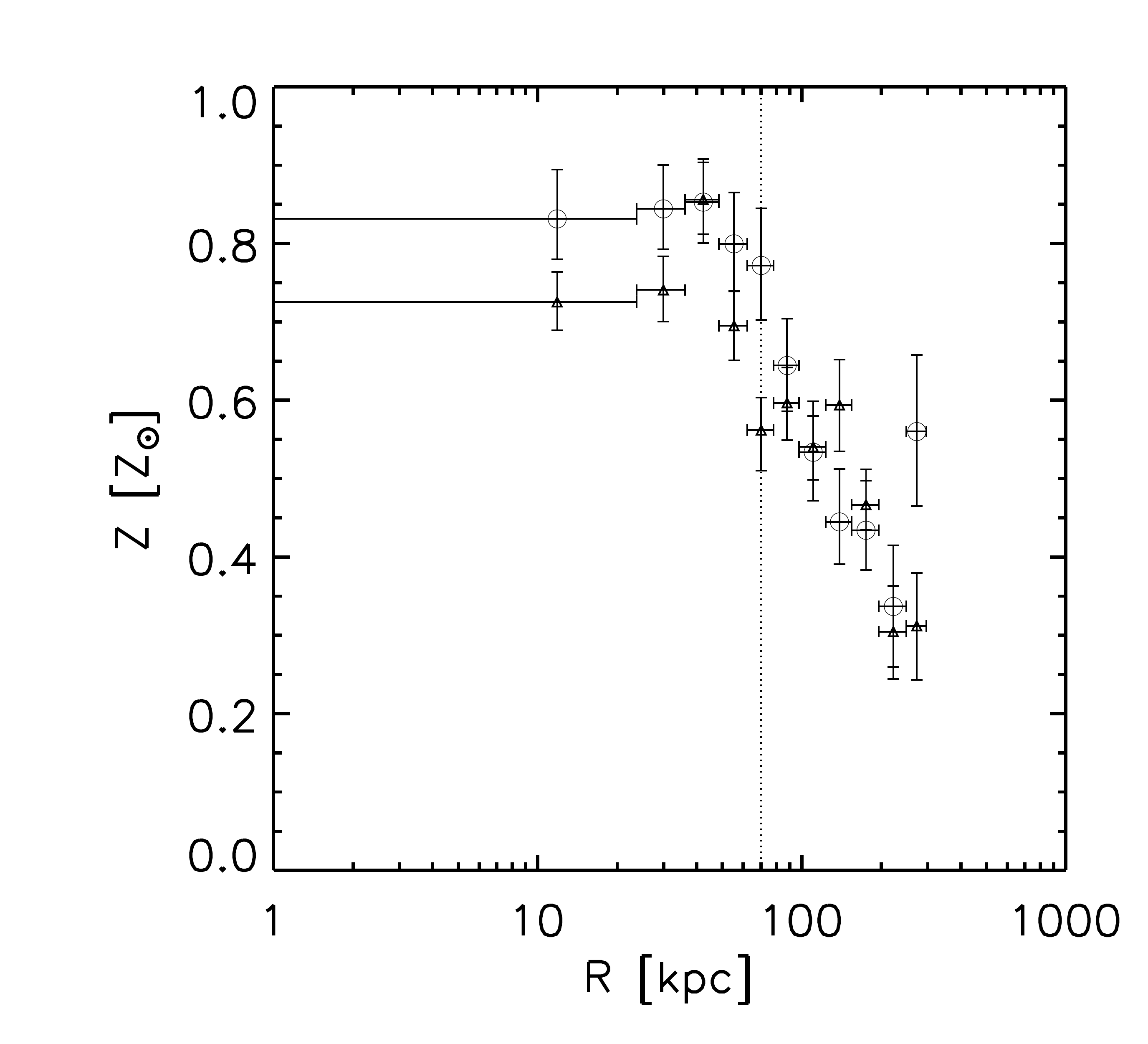}
\end{minipage}
\begin{minipage}{0.48\linewidth}
\includegraphics[width=\textwidth, trim=0mm 0mm 0mm 0mm, clip]{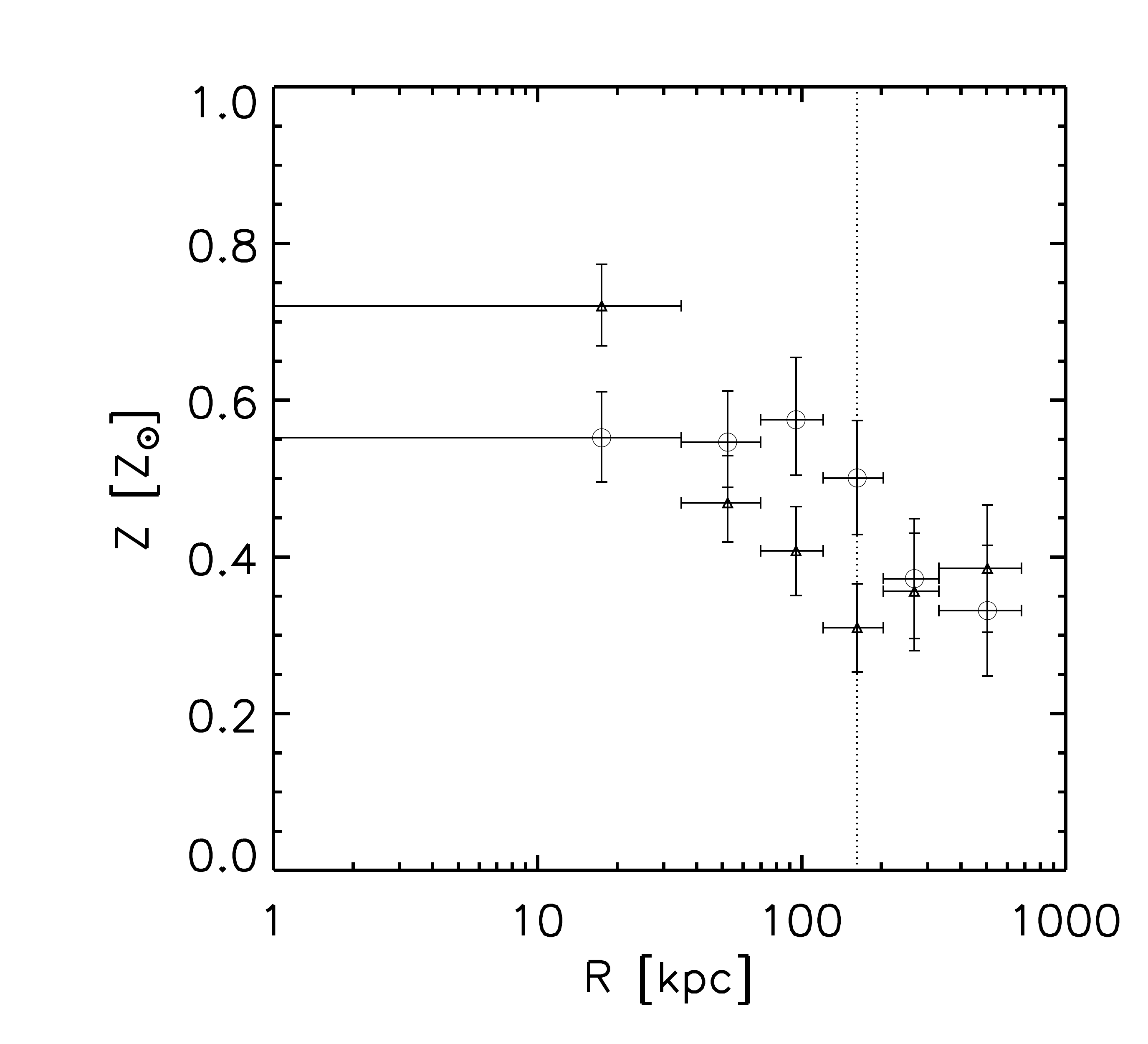}
\end{minipage}

\begin{minipage}{0.48\linewidth}
\includegraphics[width=\textwidth, trim=0mm 0mm 0mm 0mm, clip]{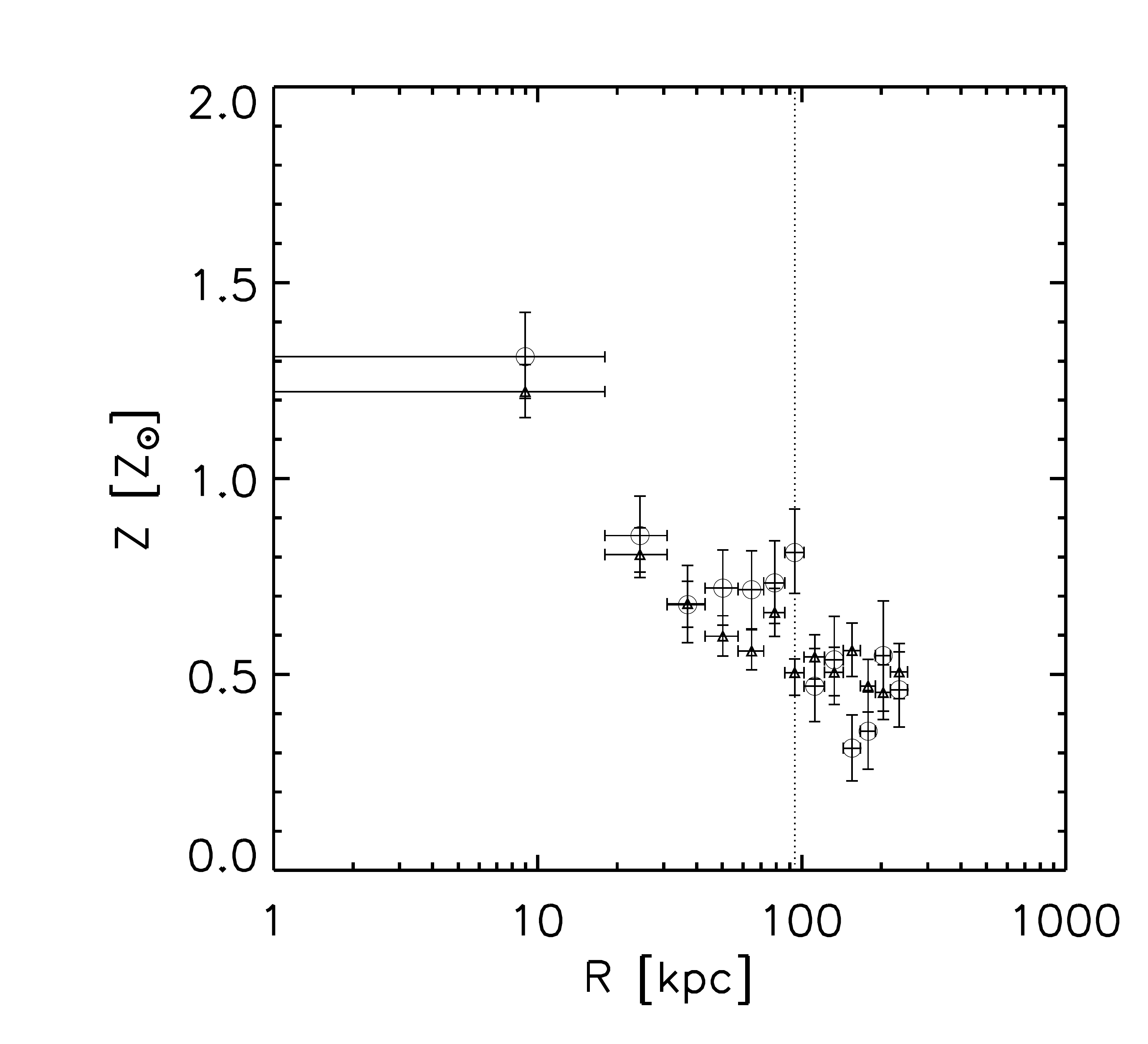}
\end{minipage}
\begin{minipage}{0.48\linewidth}
\includegraphics[width=\textwidth, trim=0mm 0mm 0mm 0mm, clip]{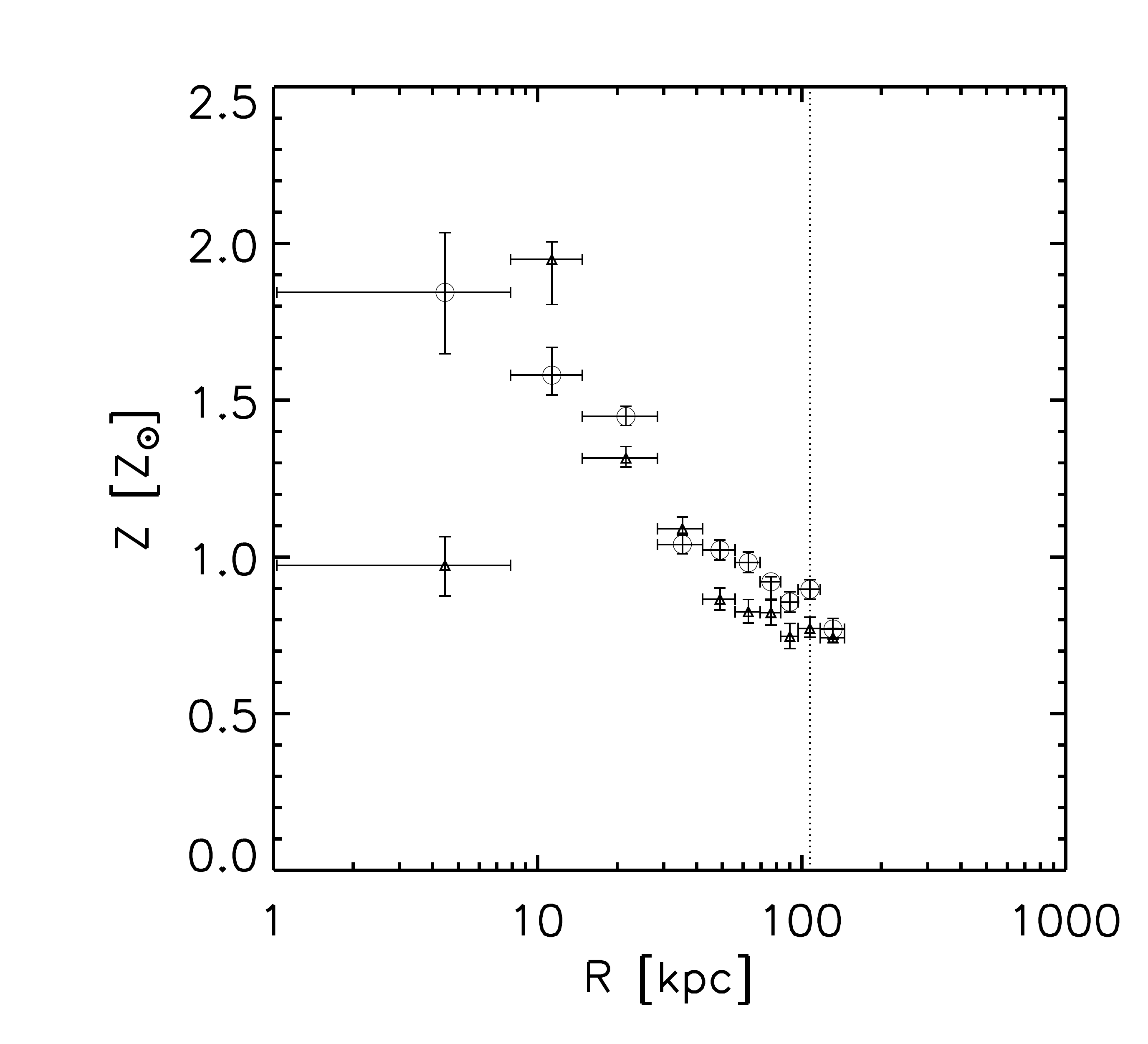}
\end{minipage}
\caption[]{{\it - continued}:  Top row: Hydra A and HCG 62.  Middle row: A1795 and A1835.  Bottom row: A2029 and A2052.}
\end{center}
\end{figure*}

\begin{figure*}
\ContinuedFloat
\begin{center}
\begin{minipage}{0.48\linewidth}
\includegraphics[width=\textwidth, trim=0mm 0mm 0mm 0mm, clip]{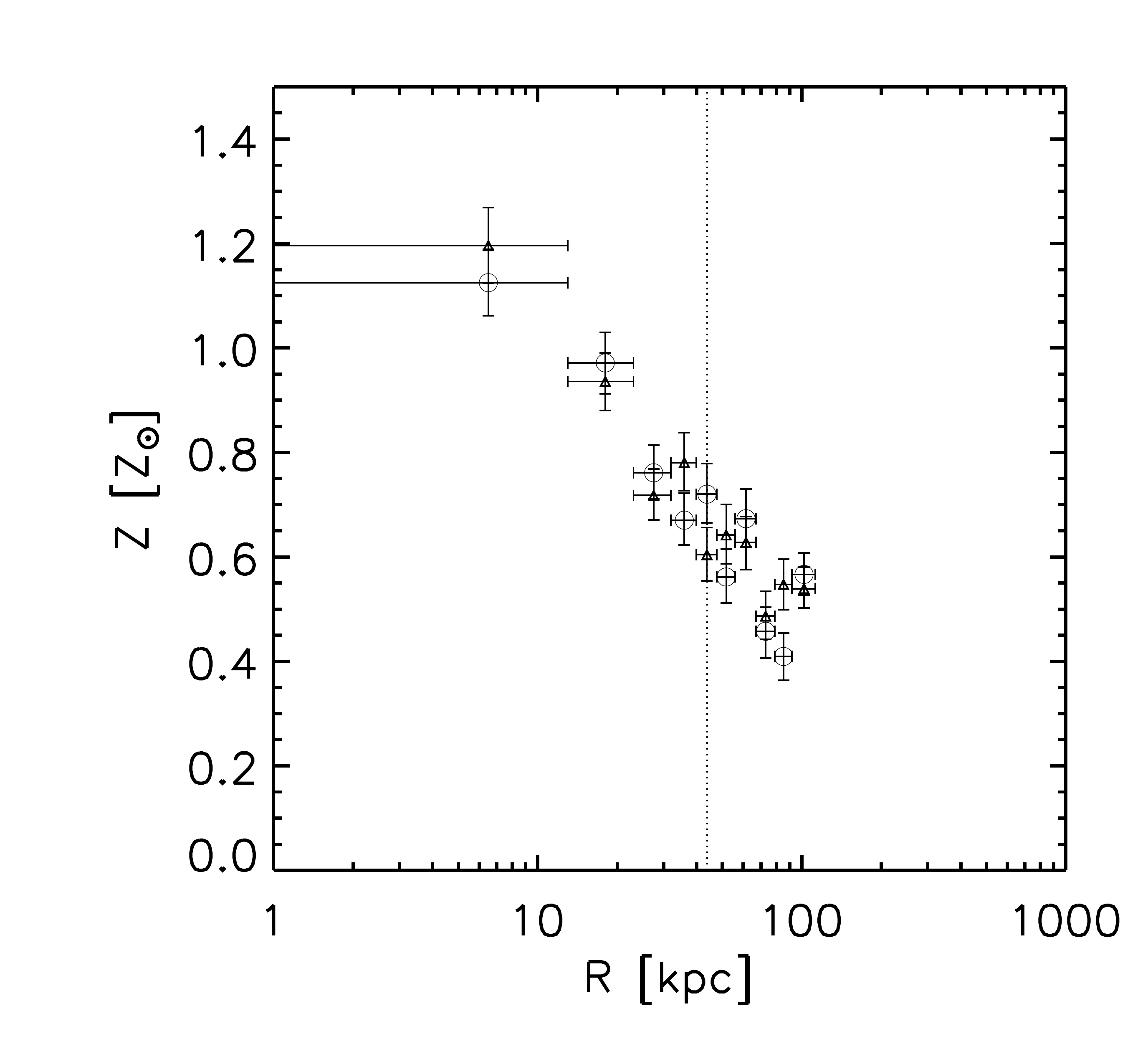}
\end{minipage}
\begin{minipage}{0.48\linewidth}
\includegraphics[width=\textwidth, trim=0mm 0mm 0mm 0mm, clip]{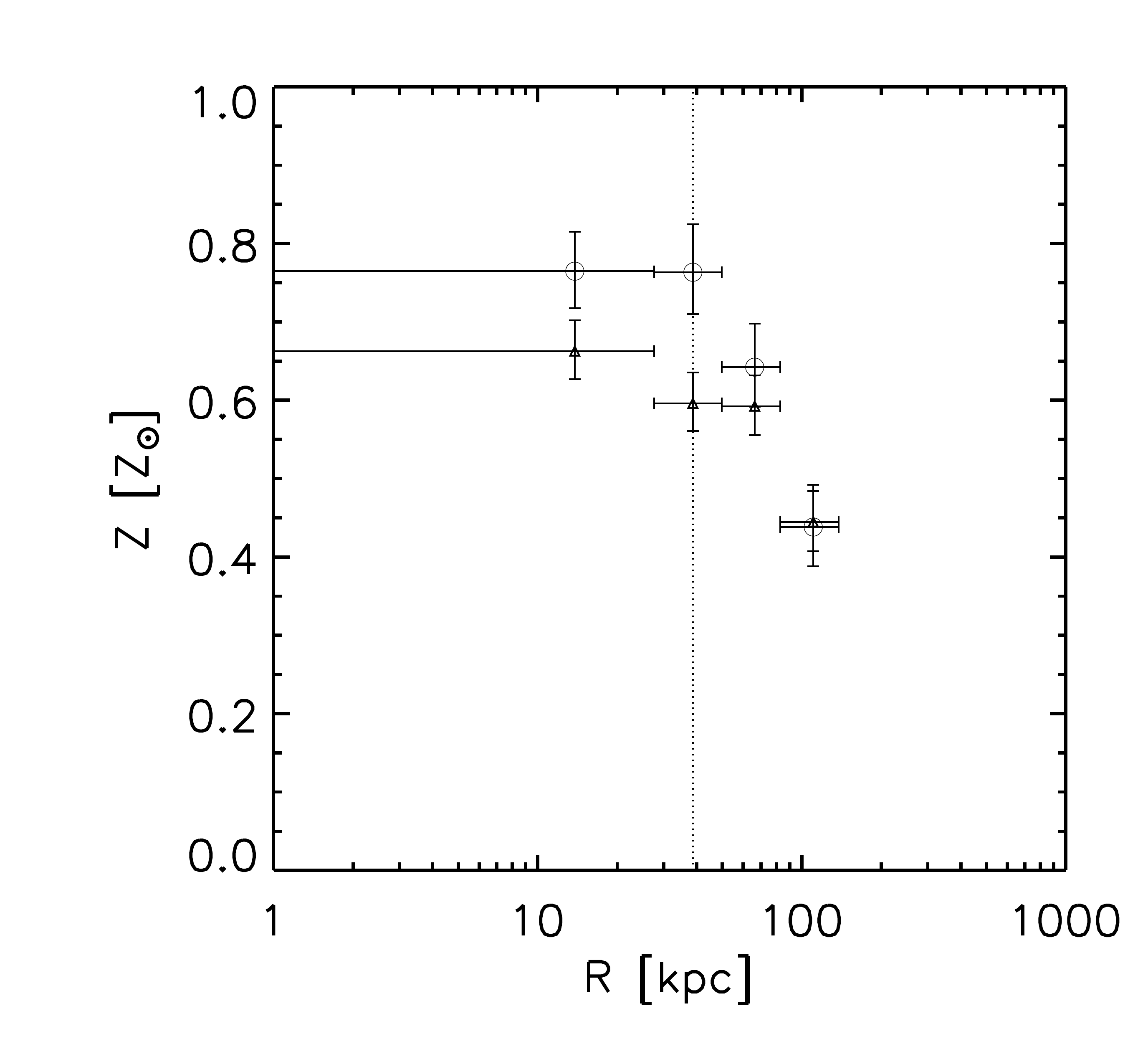}
\end{minipage}

\begin{minipage}{0.48\linewidth}
\includegraphics[width=\textwidth, trim=0mm 0mm 0mm 0mm, clip]{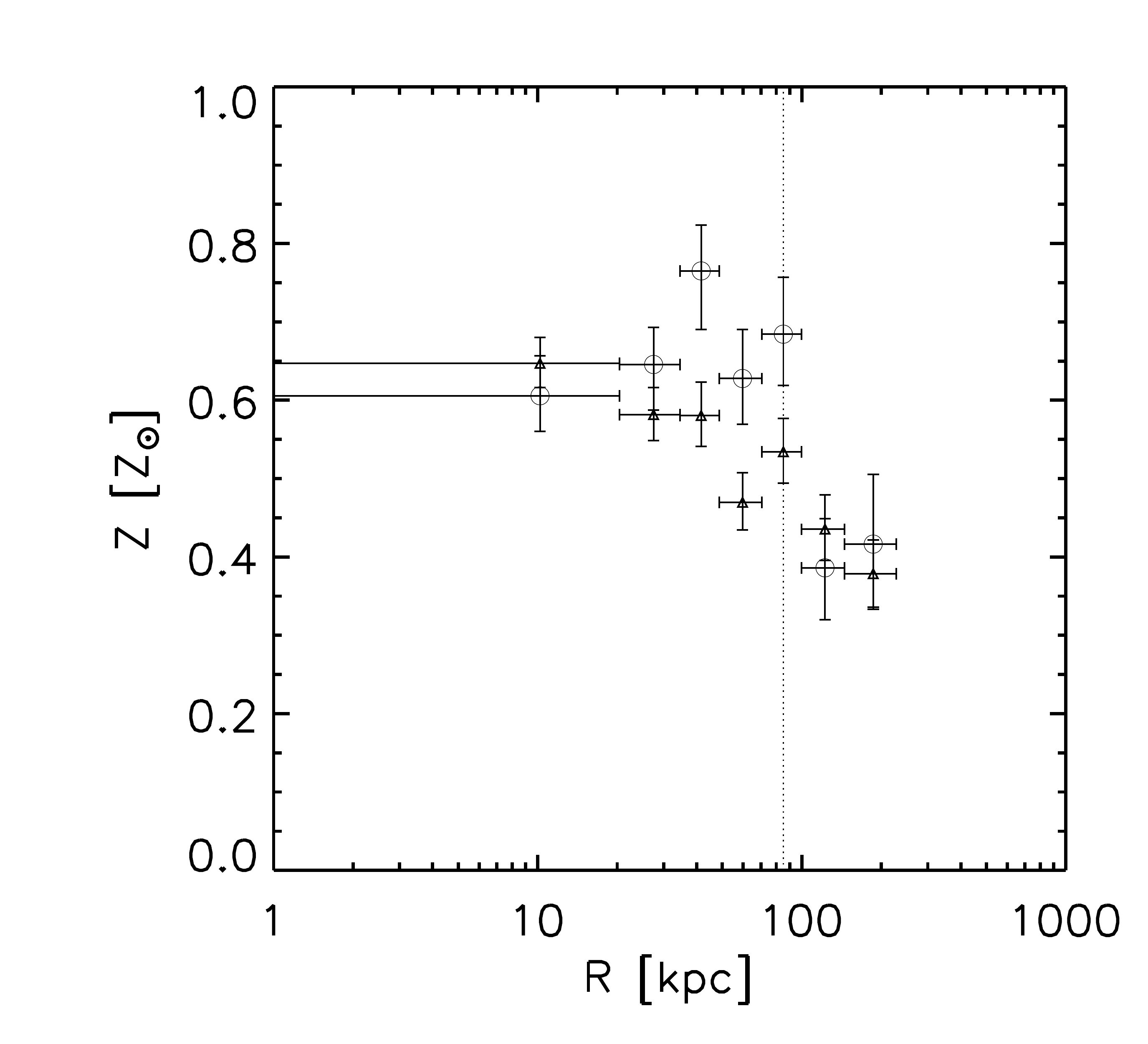}
\end{minipage}
\begin{minipage}{0.48\linewidth}
\includegraphics[width=\textwidth, trim=0mm 0mm 0mm 0mm, clip]{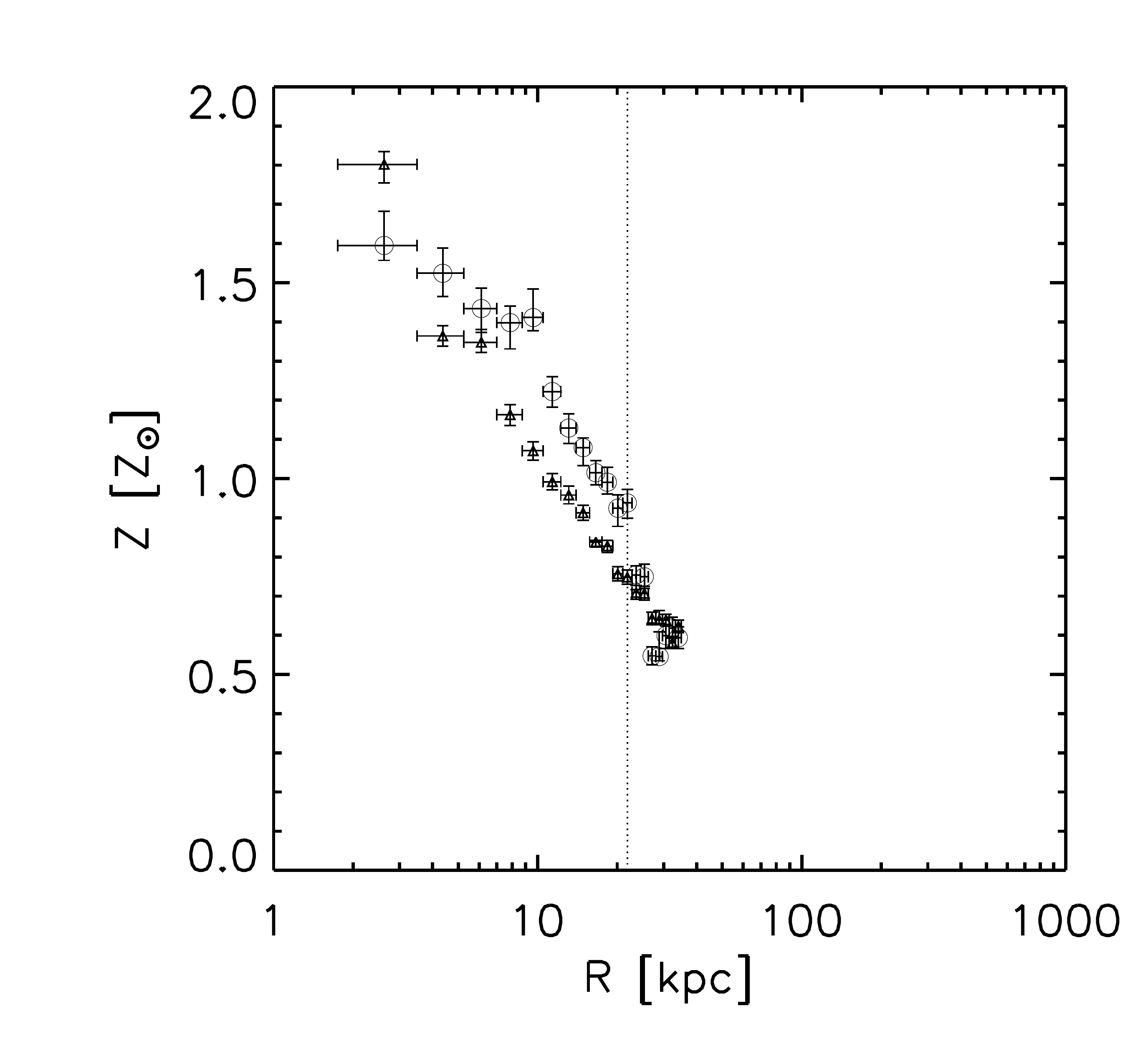}
\end{minipage}
\caption[]{{\it - continued}:  Top row: A2199 and Sersic 159.  Bottom row: A2597 and M87.}
\end{center}
\end{figure*}

We have made no attempt to correct Figure~\ref{fig:main-jet-dist} for projection.  Our $pV$ calculations assume the outbursts lie in the plane of the sky so that the line-of-sight diameter of a cavity is similar to the projected major and minor axis.  It is important to understand what effect this could have on our results and final conclusion.

The iron radii suffer similar uncertainties to the pressure and buoyancy timescales.  The buoyancy time is defined as the time for a cavity to rise at its terminal velocity,  
\begin{equation} \label{eqn:buoy}
t_{\rm buoy} = R\sqrt{SC/2gV},
\end{equation}
where $R$ is the cavities projected distance from the cluster centre, $S$ is the cross section of the cavity, $C$ is the drag coefficient, and $V$ is the cavity volume \citep{chu00,chu01}.  In general, cavities are oriented at oblique angles, which underestimates the altitude and buoyancy time, and overestimates pressure.  In the extreme, metal asymmetries in systems with jets aligned near the line of sight could be foreshortened enough to avoid detection.  On average, the cavities are expected to be oriented 45 degrees or more because cavities projected directly along the line of sight against the nucleus are difficult to detect.  We would then expect the cavity altitude and iron radius to be underestimated by ~30\%.  Any attempt to correct for projection would shift the $P_{\rm jet}-R_{\rm Fe}$ scaling relation to larger distances, but would otherwise leave our conclusions unchanged.  In Figure~\ref{fig:main-jet-dist}, most points on the trend would shift to higher powers and a greater iron radius.  The overall effect is it may take less power to move material to further distances than our relationship implies.

As discussed in \citep{cck11}, qualitative agreement was found between our observed scaling relation and jet simulations.  \citet{mor10} studied the spatial impact jets have on their environment, finding a dependence of $R \propto L_{\rm jet}^{1/3}$, a shallower relationship than what we find in this work.  In the work of \citet{gas11}, with a simulated jet power of $5.7 \times 10^{44}$ erg s$^{-1}$, the gas is lifted to ~200 kpc from the cluster centre.  This is larger than what our scaling relationship predicts by about 1$\sigma$, but the overall agreement is encouraging.  We expect that attempts to match our scaling using simulations will levee restrictive constraints on jet physics.

\subsection{Correlations with Cavity Enthalpy}

\begin{figure}
\begin{center}
\includegraphics[totalheight=0.45\textwidth]{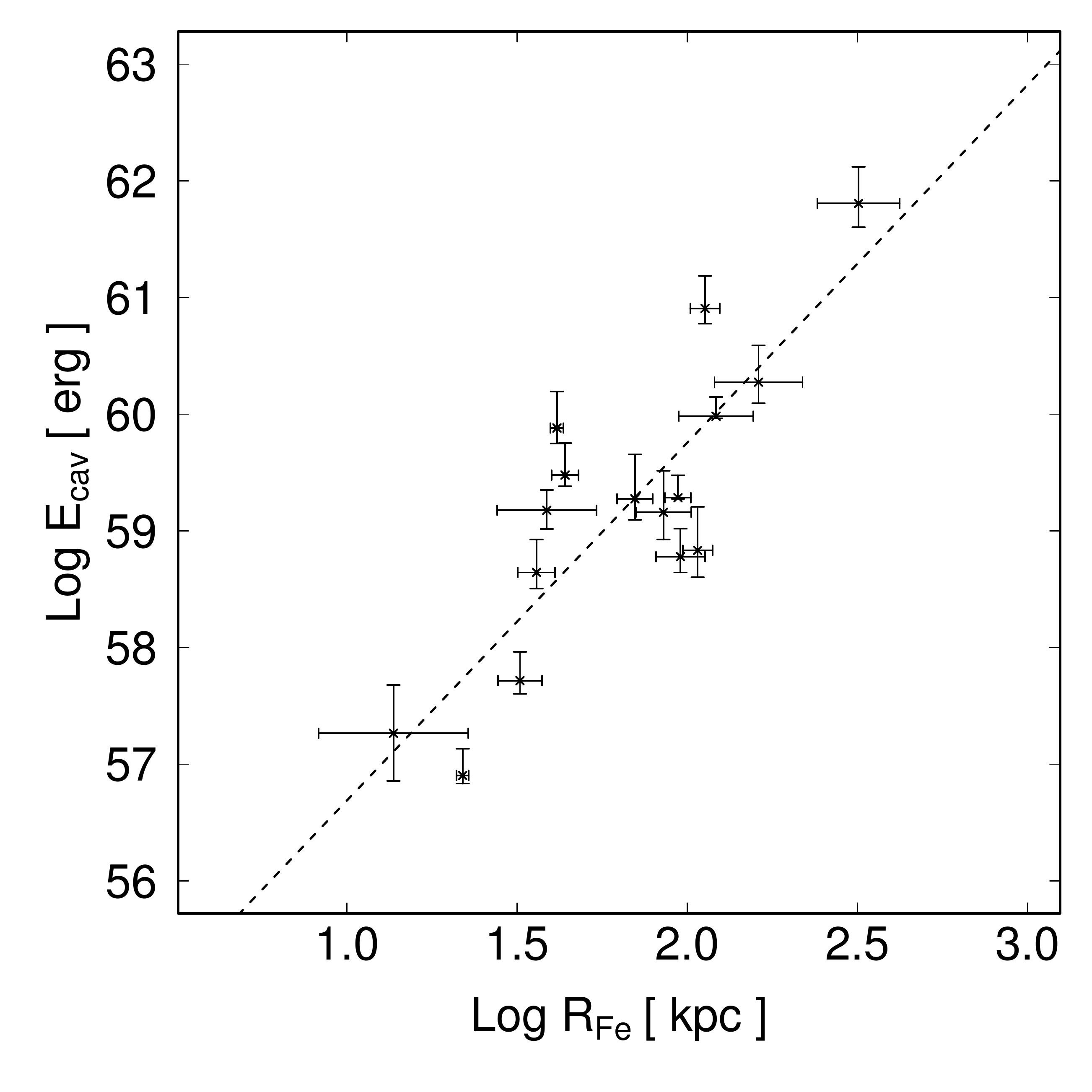}
\caption{The enthalpy of the youngest cavity system for a cluster is plotted against the iron radius.  The best-fitting to the data has slightly more scatter than that in Figure~\ref{fig:main-jet-dist}}
\label{fig:energy-dist}
\end{center}
\end{figure}

The AGN jet powers were calculated from the cavity enthalpy and an estimate of the cavity system's age.  The age can be calculated in several ways \citep{bmc07}, but is generally assumed to be the buoyancy time-scale for the cavity to rise from the nucleus to its projected location.  The power could be significantly higher if the cavities are launched with a substantial initial velocity. Enthalpy is a more direct measurement.  We assume the cavity filling material is relativistic, so the ratio of specific heats is $\gamma = 4/3$, making the enthalpy $4pV$ \citep{raf06}.  But it is unclear in the case of multiple cavity systems what the representative value to use would be.  In Figure~\ref{fig:energy-dist} we plot the enthalpy of the cavity system nearest to the nucleus for each cluster versus the iron radius.  This is a measure of the energy output of the AGN in the past 100 Myr or so.  The dashed line is again the least-squares linear regression best-fitting,
\begin{equation} \label{eqn:ecav-rfe}
R_{\rm Fe} = (57 \pm 30) \times E_{\rm cav}^{(0.33 \pm 0.08)} ~(\rm kpc),
\end{equation}
where the enthalpy is in units of $10^{59}$ erg.  The standard deviation is approximately 1.56 dex, somewhat larger than the power plot.  However, as expected the relationship with increasing iron radius remains.  Hydrodynamical simulations performed by \citet{mor10} show the power, not the total energy, determines the volume an AGN can influence.  Our result cannot confirm this conclusion, but jet power does seem to be more closely correlated.

\begin{figure}
\begin{center}
\includegraphics[totalheight=0.45\textwidth]{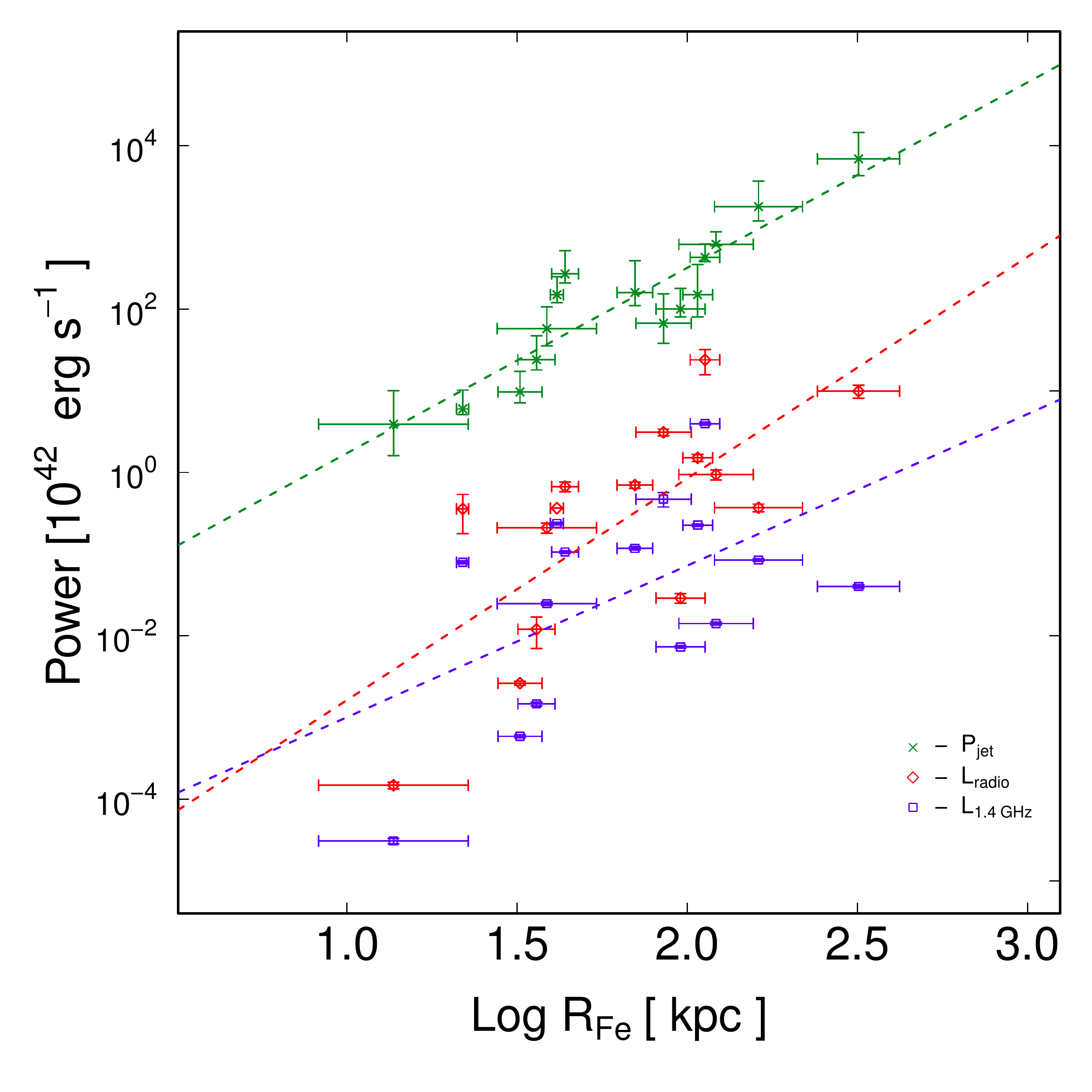}
\caption{Bolometric radio power(red), 1.4 GHz radio power (blue) and AGN jet power (green) have been plotted on the same scale.  The corresponding coloured lines are the correlations best fit.  The jet power, with a correlation coefficient of 0.92, is a stronger indicator of iron radius.  The correlation coefficient with either radio power is 0.74.}
\label{fig:radjet-dist}
\end{center}
\end{figure}

Finally, we consider how radio synchrotron power and iron radius compare to the mechanical jet power.  Using principal component analysis, we can compare how different parameters (jet power, enthalpy, bolometric and 1.4 GHz radio luminosity) correlate with the iron radius.  For jet power, the correlation with iron radius is 0.92.  In comparison, the correlation between enthalpy and iron radius is lower at 0.85.  Finally, the lowest correlation with iron radius is the radio luminosity (bolometric and 1.4 GH), with a value of 0.74.  Figure~\ref{fig:radjet-dist} shows how much tighter the relationship is between iron radius and jet power than the relationship between radio synchrotron power and jet power.  The green crosses represent the jet powers and the red diamonds and blue squares represent the bolometric radio luminosity and 1.4 GHz luminosity, respectively.  This is to be expected to some degree because of the large scatter in the jet power vs radio synchrotron power correlation \citep{bir04, raf06, df08}.  However, Figure~\ref{fig:radjet-dist} shows that the jet power provides the best proxy for the average iron radius.

\subsection{Lifting Hot Gas by Radio Jets}
\label{subsec:main-jet_energy}

Large quantities of plasma are being lifted out of the centres of galaxy clusters, so a significant fraction of the bubble enthalpy must be expended lifting the gas.  This quantity was originally calculated for Hydra A, and the work done lifting the plasma was found to be comparable to the enthalpy of the inner cavity pair \citep{cck09b}.  We have carried out a similar analysis on the High Quality sample.

We assume the outflowing gas is confined to a cylindrical volume with the same approximate radius as the cavity system.  It is unrealistic to believe that the metals are uniformly filling the conical regions shown in Figure~\ref{fig:main-femaps} used to define the average metallicity along the jets.  Because we do not know the filling factor, our best estimate is to assume at most the metals fill a volume with a width no greater than the observed cavity.  We base this off of the highest resolution metallicity maps, such as M87 and Hydra A, where the plumes of metal-rich gas appear to be narrow.  The length of the cylindrical volume is well known, extending from the core to the iron radius.  The width suffers the same uncertainties when measuring cavity sizes.  This alone could further add a factor of 2 to 4 to the mass uncertainties.  We further assume the off-jet profile is the undisturbed, azimuthally averaged profile prior to the AGN outburst.  Finally,  we assume the uplifted metals were originally produced within the half-light radius of the BCG.

The uplifted iron mass is calculated as the sum of the iron mass within each radial bin of the outflow region, minus the metallicity profile of the undisturbed azimuthally averaged profile.  The uplifted iron mass is given as
\begin{equation} \label{eqn:main-mfe_up}
M_{\rm Fe} =  f_{\rm Fe,\odot} \sum\limits_{i=1}^n \rho_i V_i (Z_{{\rm on},i} - Z_{{\rm off},i}),
\end{equation}
where $\rho$ is the density of the gas assuming $n_{\rm e} = 1.2 n_{\rm H}$, $V$ is the volume of the outflow region, $Z_{\rm on}$ and $Z_{\rm off}$ are the metallicity of the gas along the jet and orthogonal to the jet, and $f_{\rm Fe,\odot}$ is the iron mass fraction of the Sun.  Assuming this iron came from the central region of the cluster, we can convert the iron mass values back into total uplifted mass.  This is given as,
\begin{equation} \label{eqn:main-mtot_up}
M_{\rm gas} = \frac{M_{\rm Fe}}{Z_{\circ} f_{\rm Fe,\odot}},
\end{equation}
where $Z_{\circ}$ is the central metallicity value.  The energy that is required to uplift this gas can be estimated by calculating the difference in gravitational potential energy at the cluster centre and its current posiition.  \citet{rey08} gives the energy as
\begin{equation} \label{eqn:main-energy}
\Delta E = \frac{ c_{\rm s}^2 M_{\rm gas}}{\gamma} \ln \left( \frac{\rho_i}{\rho_f} \right),
\end{equation}
where $c_{\rm s}$ is the sound speed, $\gamma = 5/3$ is the ratio of specific heats for the X-ray emitting plasma, and $\rho_i$ and $\rho_f$ are the initial and final gas densities.  The outflow masses, rate, and uplift energy for each cluster in the high quality sample can be found in Table~\ref{tab:outflow_mass-energy}.  These are very rough approximations due to the unknown filling factor of the estimated outflowing volume.  The errors quoted here reflect the propagation of uncertainties due to the measured errors on density and metallicity.  The total gas mass is roughly 3 orders of magnitude greater than the iron mass primarily due to the solar abundance fraction.  Uplift energy as a fraction of the total energy varies from a few percent to as much as 50\%.  Note, the iron mass measured here for Hydra A is smaller than we found earlier in \citet{cck09b} owing to a more conservative estimate of the volume filled by the metals.

\subsection{Creating Broad Abundance Peaks Through AGN Activity}

\begin{table*}
\centering
\begin{minipage}{140mm}
\caption{Jet properties}
\begin{tabular}{@{}lcccc@{}}
\hline
~ & $M_{\rm Fe}$$^a$ & $M_{\rm gas}$$^b$ & $\dot{M}_{\rm outflow}$$^c$ & $E_{\rm uplift}$$^d$ \\
System & ($10^7$ M$_\odot$) & ($10^{10}$ M$_\odot$) & (M$_\odot$ yr$^{-1}$) & ($10^{58}$ erg) \\
\hline
	A133 & $2.74_{-0.69}^{+0.75}$ & $1.00_{-0.25}^{+0.28}$ & $49\pm21$ & $14.5_{-3.9}^{+4.1}$ \\
	A262 & $0.0351_{-0.0110}^{+0.0125}$ & $0.0134_{-0.0421}^{+0.0487}$ & $2.0\pm0.4$ & $0.0852_{-0.0246}^{+0.0279}$ \\
	Perseus & $0.189_{-0.013}^{+0.016}$ & $0.154_{-0.010}^{+0.013}$ & $23\pm12$ & $0.780_{-0.071}^{+0.089}$ \\
	2A 0335 & $0.312_{-0.186}^{+0.241}$ & $0.243_{-0.145}^{+0.188}$ & $36\pm14$ & $0.912_{-0.397}^{+0.462}$ \\
	A478 & $2.98_{-1.03}^{+1.06}$ & $2.26_{-0.78}^{+0.81}$ & $150\pm62$ & $42.7_{-14.5}^{+15.0}$ \\
	MS 0735 & $9.40_{-4.73}^{+5.10}$ & $6.09_{-3.08}^{+3.32}$ & $150\pm81$ & $243_{-123}^{+133}$ \\
	Hydra A & $0.893_{-0.179}^{+0.196}$ & $0.777_{-0.157}^{+0.171}$ & $49\pm19$ & $10.4_{-2.0}^{+2.2}$ \\
	HCG 62 & $0.0256_{-0.0061}^{+0.0122}$ & $0.0122_{-0.0031}^{+0.0060}$ & $3.8\pm0.4$ & $0.0202_{-0.0053}^{+0.0100}$ \\
	A1795 & $1.05_{-0.34}^{+0.36}$ & $0.793_{-0.258}^{+0.270}$ & $77\pm36$ & $6.64_{-2.23}^{+2.22}$ \\
	A1835 & $4.28_{-1.60}^{+1.77}$ & $3.73_{-1.40}^{+1.55}$ & $170\pm86$ & $161_{-55}^{+58}$ \\
	A2029 & $0.891_{-0.392}^{+0.416}$ & $0.472_{-0.208}^{+0.220}$ & $49\pm15$ & $13.1_{-4.6}^{+4.7}$ \\
	A2052 & $0.121_{-0.022}^{+0.023}$ & $0.0545_{-0.0100}^{+0.0104}$ & $2.2\pm0.8$ & $0.374_{-0.082}^{+0.084}$ \\
	A2199 & $0.0610_{-0.0501}^{+0.0509}$ & $0.0313_{-0.0257}^{+0.0261}$ & $4.6\pm1.9$ & $0.319_{-0.211}^{+0.219}$ \\
	Sersic 159 & $0.634_{-0.243}^{+0.277}$ & $0.501_{-0.193}^{+0.220}$ & $73\pm30$ & $2.37_{-0.92}^{+1.04}$ \\
	A2597 & $1.44_{-0.408}^{+0.384}$ & $1.34_{-0.383}^{+0.360}$ & $98\pm50$ & $14.5_{-3.9}^{+4.1}$ \\
	M87 & $0.0382_{-0.0039}^{+0.0039}$ & $0.0289_{-0.0043}^{+0.0043}$ & $0.92\pm0.18$ & $0.284_{-0.038}^{0.038}$ \\
\hline
\label{tab:outflow_mass-energy}
\end{tabular}
\begin{itemize}
\item[$^a$] Mass of uplifted iron
\item[$^b$] Total mass of uplifted gas
\item[$^c$] Outflow rate of the gas
\item[$^d$] Energy required to raise gas from centre to its observed location
\end{itemize}
\end{minipage}
\end{table*}

AGN activity has been shown to be capable of redistributing gas in galaxy clusters \citep{sim08,sim09,cck09b,cck11,git11}.  Because this appears to be an efficient mechanism, there is reason to believe AGN could in principle be responsible for the broadened abundance peak observed in cool-core clusters.  We have tested the feasibility of AGN broadened abundance peaks by calculating the amount of energy required to move the entire mass of the abundance peak from the centre of the cluster to its current position using the same methods as described in the previous section.

\begin{figure*}
\begin{center}
\begin{minipage}{0.49\linewidth}
\includegraphics[width=\textwidth, trim=0mm 0mm 0mm 0mm, clip]{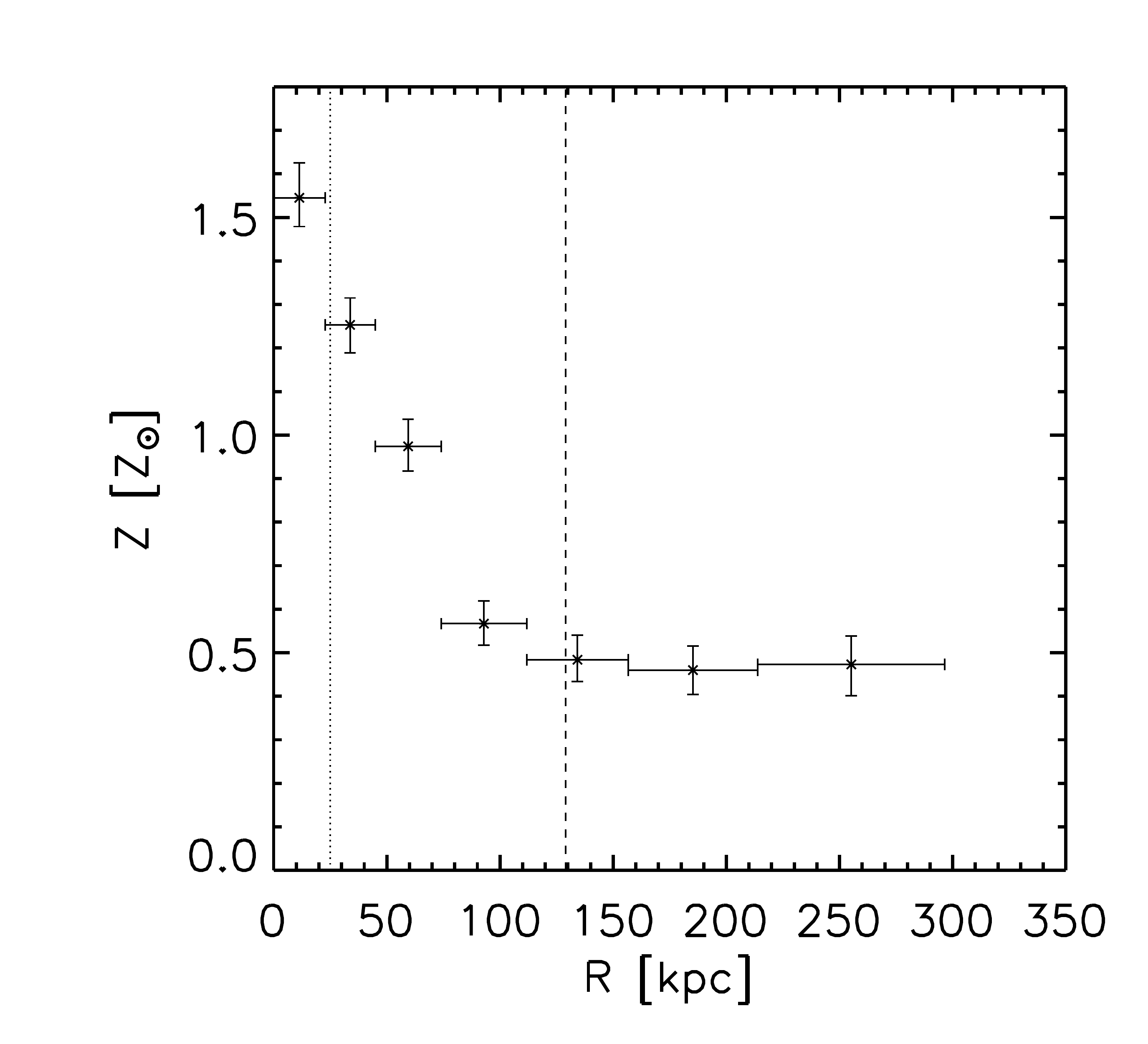}
\end{minipage}
\begin{minipage}{0.49\linewidth}
\includegraphics[width=\textwidth, trim=0mm 0mm 0mm 0mm, clip]{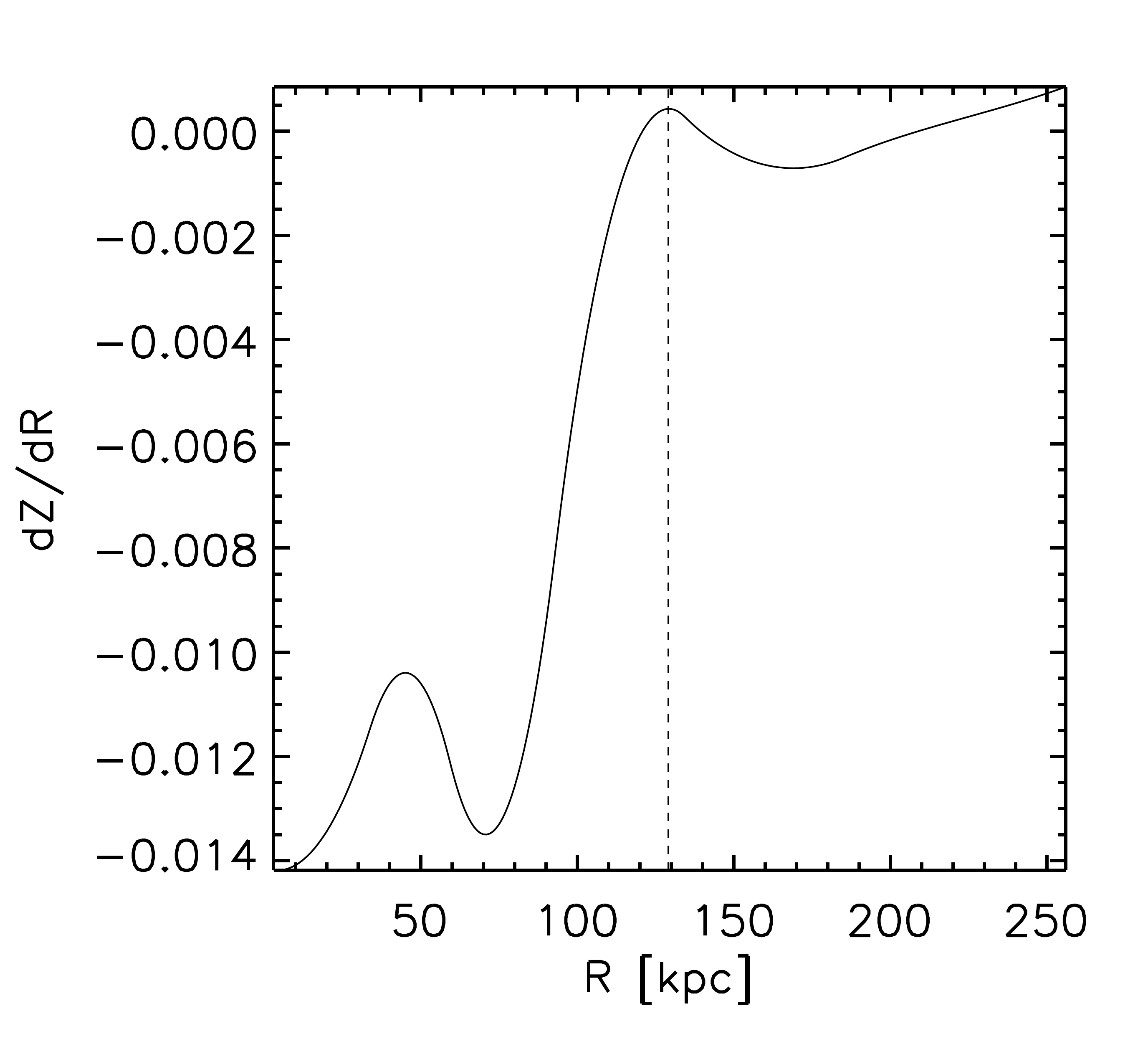}
\end{minipage}
\caption{{\it Left}: The abundance profile of Abell 133.  Profiles such as this were created for all clusters in the sample.  The dotted line is the half-light radius of the BCG.  The dashed line is the abundance peak radius {\it Right}: The derivative of the spline fit to the abundance profile.  The abundance peak radius was derived here at the point where the slope approaches zero and remains mostly flat.}
\label{fig:peak_example}
\end{center}
\end{figure*}

Azimuthally averaged abundance profiles have been created for the entire sample with 6 to 8 bins per profile, spanning the field of view of the CCD.  The large bin size allows for lower uncertainties and the spatial resolution to distinguish the abundance peak from the background.  The profiles were fit using the method described in section~\ref{subsec:main-spec}.  Only three clusters (2A 0335, A262, and A2052) were fitted using a two-temperature model in the central region.  The left panel of Figure~\ref{fig:peak_example} is an ordinary example of the resulting profiles.  Targets such as Perseus and M87, where the field of view only covers a small part of the cluster centre, have been excluded from this analysis.  The dotted vertical lines indicate the effective radius of the BCG from the HyperLeda catalogue\footnote{http://leda.univ-lyon1.fr}.  If material were not diffused outward, then the abundance peak should lie within this radius \citep{deg01,deg04,reb05,reb06,dav08}.  

We first must define peak and background metallicity.  We started by fitting a spline to each abundance profile.  Taking the derivative of the spline fit, we searched for an inflection point in slope where the profile is transitioning to the background value.  We choose this point as the turnover in slope closest to zero, where the slope is rapidly changing interior to this point and remaining near zero exterior to this point.  We refer to this turnover point as the significant abundance peak radius ($R_{\rm Z}$).  The right column of Figure~\ref{fig:peak_example} is the change in slope plotted against radius.  The dashed line represents $R_{\rm Z}$.

\begin{figure}
\begin{center}
\includegraphics[totalheight=0.45\textwidth]{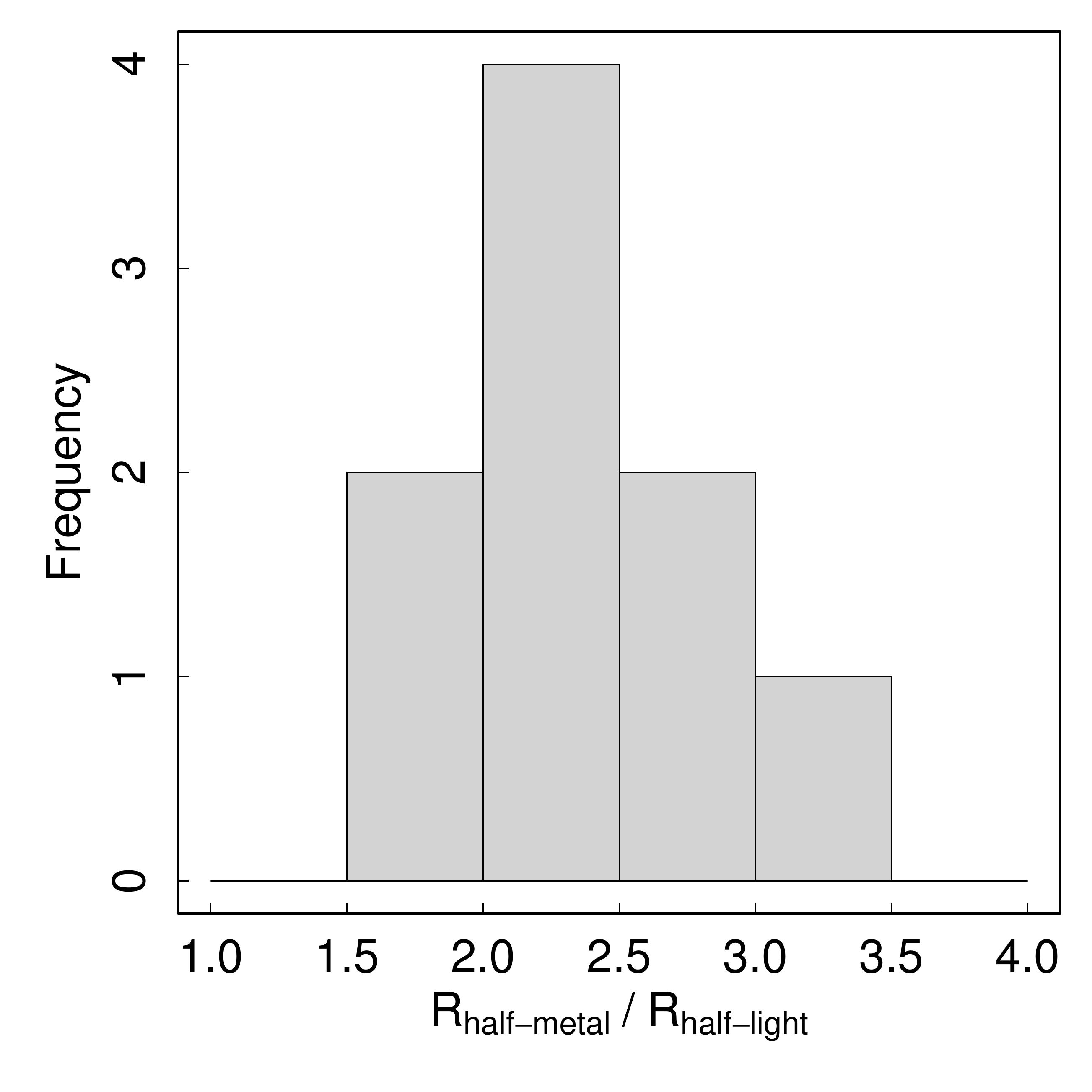}
\caption{The fraction of half-metal radius to half-light radius.  Half of the iron produced within the BCG is extended many times beyond its effective radius.  The small subset of clusters used here are from the objects in Table 1 and Table 2 where effective radii measurements were available in the HyperLeda catalogue.}
\label{fig:hmr_hlr}
\end{center}
\end{figure}

To test whether our defined metallicity peak is more extended than the light, we have calculated the half-metal radius.  The half-metal radius is the radius at which half the iron within $R_{\rm Z}$, by mass, is located.  In Figure~\ref{fig:hmr_hlr} we present a histogram of the half-metal to half-light radius fraction.  For the nine clusters where we have an effective radius measurement, we find that the half-metal radius is 1.5 to 3.5 time more extended than the light.  If there was no broadening of the abundance peak, these values would be expected to be similar.

With a peak radius defined, we can now estimate the approximate metallicity background.  We use the average metallicity value of the profile points outside of $R_{\rm Z}$ to be the subtracted background.  The modified version of equation~\ref{eqn:main-mfe_up} is now
\begin{equation} \label{eqn:main-mfe_up-mod}
M_{\rm Fe} = f_{\rm Fe,\odot} \sum\limits_{i=1}^n \rho_i V_i (Z_i - Z_{\rm back}),
\end{equation}
where V is the volume of the spherical shells, $Z$ is the total metallicity, and $Z_{\rm back}$ is the background metallicity.  For the purpose of this analysis, the iron mass given here is assumed to have all originated from the BCG.  Following the same assumptions we made for the mass uplifted by a single jet, the iron mass has been converted to a total gas mass using equation~\ref{eqn:main-mtot_up}.  The final step is to calculate the energy to displace the total mass of gas from the centre to its current observed location using equation~\ref{eqn:main-energy}.  All values calculated here are presented in Table~\ref{tab:main-blackhole}.  The average background metallicity over the sample is 0.422 $Z_\odot$ and most clusters in the sample have displaced on the order of $10^8$ $M_\odot$ of iron.  The energy to displace the iron ranges as low as $10^{56}$ erg for the galaxy group HCG 62 and as high as $10^{60}$ erg for massive clusters such as A1835 and MS 0735.

In order to confirm that a single BCG can plausibly be responsible for the magnitude and width of a cluster abundance peak, we have determined the time-scales over which the material is created by SNe Ia, and how much the SMBHs must grow.  If enrichment times are too long, this would indicate that some of the material is coming from another source (i.e., stripping), meaning a different mechanism besides AGN must be invoked to explain this phenomenon.  Conversely, time-scales that are extremely short would suggest our abundance peak size calculations are grossly underestimated.  To calculate enrichment time, the relationship used by \citet{boh04} takes into account two components, SNe Ia and stellar mass loss, that enrich the surrounding ICM with metals.  The enrichment time to create all the metals in the abundance peak is given as
\begin{equation} \label{eqn:main-enrich}
t_{\rm enr} =  \left ( 10^{-12} S \eta_{\rm Fe} + 2.5\times10^{-11} \gamma_{\rm Fe}\right ) ^{-1} \frac{M_{\rm Fe} L_{{\rm B}\odot}}{L_{\rm B}},
\end{equation}
where $S=0.15$ SNU is the supernova rate \citep{cap99}, $\eta_{\rm Fe}=0.7$ M$_\odot$ is the SN Ia iron yield, $\gamma_{\rm Fe}=2.8\times10^{-3}$ is stellar mass loss fraction of iron, and $L_{\rm B}$ is the total $B$-band luminosity referenced from Table~\ref{tab:main-hqsamp}.  Most clusters in the sample have little to no star formation in the BCG.  An insignificant fraction of the total iron mass of the abundance peak is expected to originate from SN II \citep{cck09b}.  For most clusters in our sample, SN II only contributes at most 1\% of the peak iron mass.  However, A1835 stands out as exhibiting a high star formation rate between 100 and 180 M$_\odot$ yr$^{-1}$ \citep{bmc06}.  At a star formation rate of over 100 M$_\odot$ yr$^{-1}$, the contribution of SN II would increase to over 10\%.  To account for this, equation~\ref{eqn:main-enrich} is modified to
\begin{multline}
t_{\rm enr} = M_{\rm Fe}   [  (10^{-12} S_{\rm Ia} \eta_{\rm Fe} +\\ 2.5\times10^{-11} \gamma_{\rm Fe}  ) \frac{L_{\rm B}}{L_{{\rm B}\odot}} + {\rm SFR} \gamma_{\rm Fe, II} S_{\rm II}  ]^{-1},
\end{multline}
where ${\rm SFR}$ is the star formation rate, $\gamma_{\rm Fe,II} = 0.07$ M$_\odot$ is the SN II iron yield, and $S_{\rm II} = 0.01$ SN M$_{\odot}^{-1}$.  The enrichment time-scale for each cluster in our sample can be found in Table~\ref{tab:main-blackhole}.  The enrichment times range from a few hundred Myr to a few Gyr.  This is shorter than cluster ages so it is plausible for all the material to have been produced within the BCGs.

\begin{table*}
\centering
\begin{minipage}{140mm}
\caption{Properties of metal-enriched gas in galaxy clusters abundance peaks.}
\begin{tabular}{@{}lcccccc@{}}
\hline
~ & L$_{\rm B}$$^a$ & Z$_{\rm back}$$^b$ & E$^c$ & M$_{\rm Fe}$$^d$ & $\Delta$M$_{\rm BH}$$^e$ & t$_{\rm enr}$$^f$ \\
System & 10$^{11}$ L$_\odot$ & Z$_\odot$ & 10$^{58}$ erg & 10$^7$ M$_\odot$ & 10$^7$ M$_\odot$ & Gyr \\
\hline
	A133 & 1.77 & 0.467 & 233$_{-77}^{+24}$ & 52.4$_{-7.7}^{+2.3}$ & 8.38$_{-2.78}^{+0.86}$ & 2.71$_{-0.40}^{+0.12}$ \\
	A262 & 0.417 & 0.481 & 3.95$_{-0.97}^{+0.33}$ & 1.63$_{-0.18}^{+0.08}$ & 0.142$_{-0.035}^{+0.012}$ & 0.358$_{-0.039}^{+0.018}$ \\
	2A 0335 & 2.56 & 0.534 & 46.2$_{-33.1}^{+79.0}$ & 7.86$_{-3.70}^{+4.54}$ & 1.66$_{-1.19}^{+2.84}$ & 0.281$_{-0.132}^{+0.161}$ \\
	A478 & 4.06 & 0.345 & 1850$_{-1260}^{+1740}$ & 108$_{-51}^{+52}$ & 66.6$_{-45.5}^{+62.4}$ & 2.43$_{-1.14}^{+1.17}$ \\
	MS 0735 & 1.16 & 0.346 & 1210$_{-725}^{+1050}$ & 96.8$_{-33.6}^{+40.7}$ & 43.6$_{-26.2}^{+37.6}$ & 7.63$_{-2.64}^{+3.17}$ \\
	Hydra A & 2.58 & 0.345 & 160$_{-97.9}^{+178}$ & 19.2$_{-7.4}^{+10.1}$ & 5.75$_{-3.52}^{+6.41}$ & 0.680$_{-0.260}^{+0.360}$ \\
	HCG 62 & 0.151 & 0.333 & 0.189$_{-0.104}^{+0.669}$ & 0.188$_{-0.055}^{+0.267}$ & 0.010$_{-0.007}^{+0.021}$ & 0.114$_{-0.033}^{+0.161}$ \\
	A1795 & 1.69 & 0.349 & 1170$_{-476}^{+550}$ & 95.7$_{-22.5}^{+19.6}$ & 42.0$_{-17.1}^{+19.8}$ & 5.18$_{-1.22}^{+1.06}$ \\
	A1835 & 3.16 & 0.324 & 5738$_{-3353}^{+5350}$ & 172$_{-62}^{+77}$ & 206$_{-120}^{+193}$ & 4.98$_{-1.79}^{+2.22}$ \\
	A2029 & 1.64 & 0.455 & 574$_{-337}^{+426}$ & 66.2$_{-23.6}^{+22.1}$ & 20.6$_{-12.1}^{+15.4}$ & 3.69$_{-1.31}^{+1.24}$ \\
	A2052 & 0.773 & 0.526 & 69.5$_{-26.0}^{+40.5}$ & 17.8$_{-3.4}^{+4.6}$ & 2.50$_{-0.94}^{+1.44}$ & 2.11$_{-0.41}^{+0.54}$ \\
	A2199 & 2.49 & 0.492 & 47.1$_{-2.1}^{+32.8}$ & 11.7$_{-0.2}^{+2.7}$ & 1.69$_{-0.08}^{+1.18}$ & 0.430$_{-0.009}^{+0.098}$ \\
	Sersic 159 & 1.34 & 0.456 & 36.4$_{-2.7}^{+2.9}$ & 10.1$_{-0.3}^{+0.4}$ & 1.31$_{-0.10}^{+0.12}$ & 0.686$_{-0.023}^{+0.025}$ \\
	A2597 & 1.57 & 0.439 & 103$_{-6}^{+6}$ & 15.4$_{-0.4}^{+0.4}$ & 3.72$_{-0.20}^{+0.21}$ & 0.896$_{-0.024}^{+0.026}$ \\
	A85 & 1.98 & 0.470 & 221$_{-22}^{+24}$ & 41.1$_{-1.8}^{+1.9}$ & 7.94$_{-0.78}^{+0.85}$ & 1.90$_{-0.08}^{+0.09}$ \\
	PKS 0745 & 0.254 & 0.494 & 22.4$_{-3.1}^{+3.3}$ & 8.25$_{-0.45}^{+0.47}$ & 0.805$_{-0.112}^{+0.119}$ & 2.97$_{-0.16}^{+0.17}$ \\
	4C 55.16 & 1.12 & 0.473 & 386$_{-67}^{+74}$ & 71.4$_{-5.5}^{+6.0}$ & 13.9$_{-2.4}^{+2.7}$ & 5.83$_{-0.45}^{+0.49}$ \\
	RBS 797 & --- & 0.351 & 1660$_{-377}^{+409}$ & 122$_{-12}^{+13}$ & 59.6$_{-13.5}^{+14.7}$ & --- \\
	A2390 & --- & 0.366 & 251$_{-53}^{+54}$ & 61.8$_{-5.0}^{+4.0}$ & 9.02$_{-1.89}^{+1.93}$ & --- \\
	Zw2701 & 1.24 & 0.418 & 60.1$_{-18.1}^{+24.3}$ & 56.3$_{-4.1}^{+4.6}$ & 2.16$_{-0.65}^{+0.87}$ & 4.18$_{-0.30}^{+0.34}$ \\
	Zw3146 & 2.07 & 0.367 & 738$_{-134}^{+144}$ & 71.4$_{-5.7}^{+6.0}$ & 26.5$_{-4.8}^{+5.2}$ & 3.16$_{-0.25}^{+0.27}$ \\
	MACSJ1423 & --- & 0.416 & 855$_{-176}^{+206}$ & 80.4$_{-7.9}^{+8.5}$ & 30.8$_{-6.3}^{+7.4}$ & --- \\
	MKW 3S & 1.07 & 0.435 & 150$_{-14}^{+15}$ & 31.8$_{-1.5}^{+1.5}$ & 5.39$_{-0.51}^{+0.53}$ & 2.73$_{-0.13}^{+0.13}$ \\
	Hercules A & 0.625 & 0.296 & 562$_{-103}^{+111}$ & 69.9$_{-6.0}^{+6.6}$ & 28.3$_{-5.2}^{+5.6}$ & 10.2$_{-0.9}^{+0.9}$ \\
	Cygnus A & 0.840 & 0.391 & 276$_{-160}^{+163}$ & 61.7$_{-17.7}^{+10.1}$ & 9.93$_{-5.75}^{+5.87}$ & 6.71$_{-1.92}^{+1.11}$ \\
	A4059 & 2.21 & 0.607 & 40.4$_{-4.6}^{+5.0}$ & 14.7$_{-0.6}^{+0.7}$ & 2.04$_{-0.23}^{+0.25}$ & 0.609$_{-0.027}^{+0.029}$ \\
\hline
\label{tab:main-blackhole}
\end{tabular}
\begin{itemize}
\item[$^a$] $B$-band luminosity derived from HyperLeda catalogue
\item[$^b$] Background metallicity
\item[$^c$] Estimated uplift energy + cavity/shock energy
\item[$^d$] Abundance peak iron mass
\item[$^e$] Change in black hole mass needed to account for total output of energy
\item[$^f$] Enrichment time to create the observed iron mass
\end{itemize}
\end{minipage}
\end{table*}

\begin{figure}
\begin{center}
\includegraphics[totalheight=0.45\textwidth]{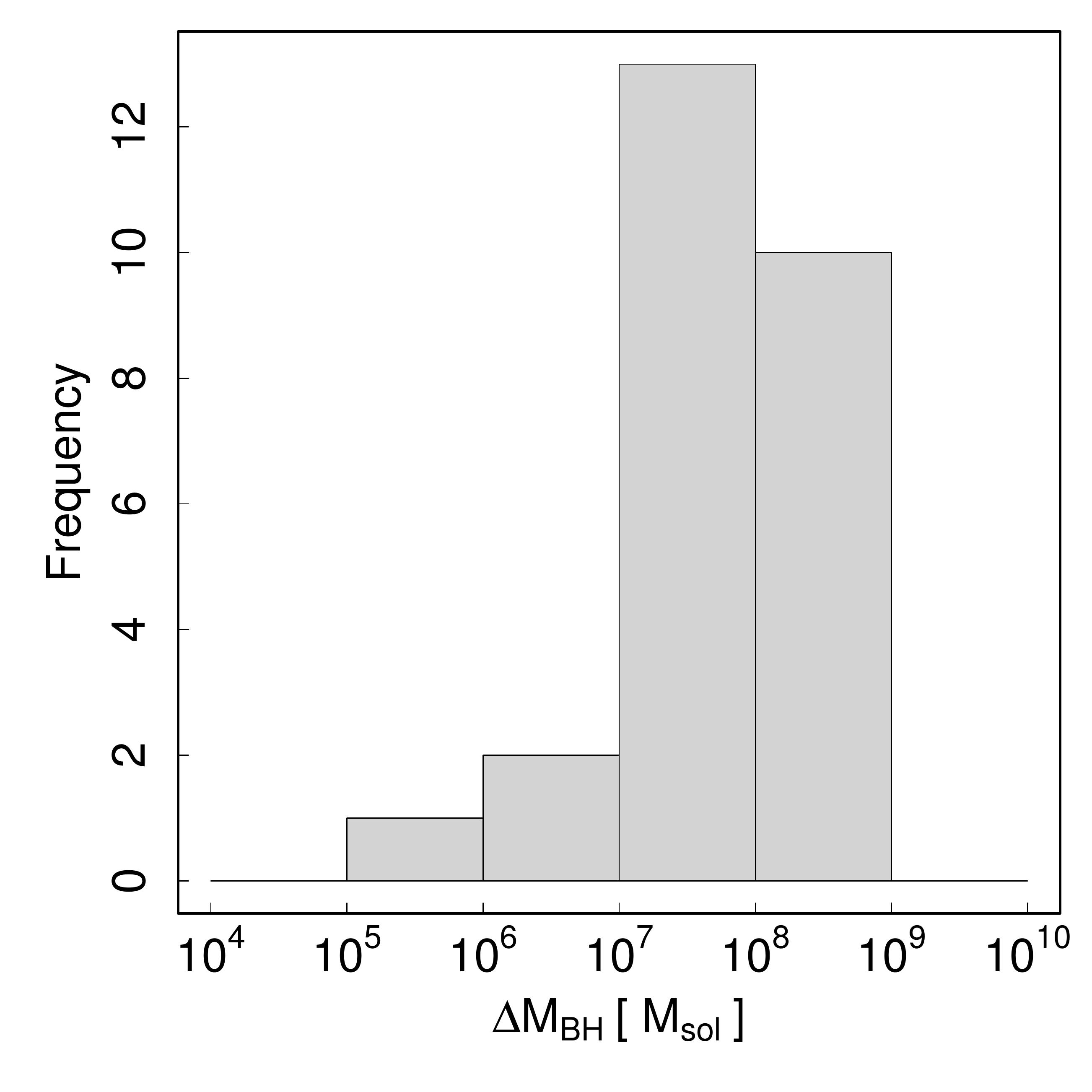}
\caption{Histogram of the change in mass of the SMBH would have to take place to account for all the energy used in uplifting all the material in the abundance peak.}
\label{fig:mbh_hist}
\end{center}
\end{figure}

To determine if enough energy is available, we measure the change in SMBH mass required to output the total energy that went into uplifting all the metal-rich gas to form the abundance peak.  The black hole grows by
\begin{equation} \label{eqn:main-bh-m}
\Delta M_{\rm BH} = (1 - \epsilon) M_{acc},
\end{equation}
where $\epsilon$ is the efficiency depending on the spin of the black hole and $M_{acc}$ is the accretion mass.  The black hole spin is important here because it determines the radius of the last stable orbit, which in turn determines the binding energy that can be released.  This efficiency can range from approximately $\epsilon=0.06$ for a non-rotating Schwarzschild black hole to as much $\epsilon=0.4$ for a maximal spinning Kerr black hole \citep{kin02}.  For this analysis we have adopted a standard value of $\epsilon=0.1$.  Accretion mass can be written in terms of the energy released, giving
\begin{equation} \label{eqn:main-bh-e}
\Delta M_{\rm BH} = (1 - \epsilon) \frac{E}{\epsilon c^2},
\end{equation}
where E is the uplift energy, given from equation~\ref{eqn:main-energy}, multiplied by an additional efficiency factor.  The extra efficiency factor is to account for the additional energy put into cavity expansion and shocks during an AGN induced outflow.  We know this for each cluster by taking the uplift energy of a single outburst from Table~\ref{tab:outflow_mass-energy} and comparing it to the total energy output during the lifetime of the outflow.  The average age of outflows in our sample is approximately $10^8$ yr and an average uplift-to-total energy fraction of 14\%.  The histogram of black hole mass growth are presented in Figure~\ref{fig:mbh_hist} and values reported in Table~\ref{tab:main-blackhole}.  Again, the errors here reflect the propagation of uncertainties due to the measured errors on density and metallicity.  It is plausible that all of these SMBHs could have grown by this amount.  The range of total black hole growth in our sample are consistent with previous estimations of SMBH masses \citep{raf06}.  The largest outburst in our sample, MS 0735, is on the same order growth of what \citet{bmc09} found to be the maximum growth the SMBH must undergo to create the observed cavity system.  When considering the spin of a black hole, the values reported in Table~\ref{tab:main-blackhole} would be reduced.  Less accretion mass is needed in a spinning black hole to maintain the same output, resulting in less overall growth.  At the maximal spin rate, the change in black hole mass for every cluster in our sample would be reduced by an order of magnitude.

It may be surprising at first glance that MS0735, the system with the most energetic outburst, does not have the largest estimated SMBH growth.  However, the mass and overall distribution of iron in a cluster core encodes the power output history of the SMBH. The integrated power of the AGN over time will generally exceed the instantaneous power.  The estimated SMBH growth found for A1835 is the largest because its abundance peak has the largest iron mass and therefore the largest expended enthalpy.

Radio AGN outflows may be responsible for the broadly distributed abundance peaks in cool core clusters.  Across our sample, it takes on average 2 Gyr to produce all the iron enriching the abundance peaks through the process of SN Ia explosions, stellar mass loss, and core-collapse supernovae associated with star formation.  The change in black hole mass associated with the energy required to uplift gas ranges from $10^5$ - $10^8$ M$_\odot$.  This range is less than the expected masses of SMBH in large BCGs and is consistent with previous predictions of expected growth.  Adding black hole spin to the calculation lowers the threshold for the amount of accretion material need for equivalent energy output.  We believe because of these easily achievable standards, AGN are the prime mechanism for redistributing all the metal enriched gas associated with the BCG.

\subsection{Regulating Star Formation by Hot and Cold Molecular Outflows}

The hot outflow rates estimated here are comparable to or exceed the star formation rates found in BCGs \citep{raf06}.  Therefore, outflows of hot gas may play a role in offsetting cooling and regulating star formation.  

\begin{figure}
\begin{center}
\includegraphics[totalheight=0.45\textwidth]{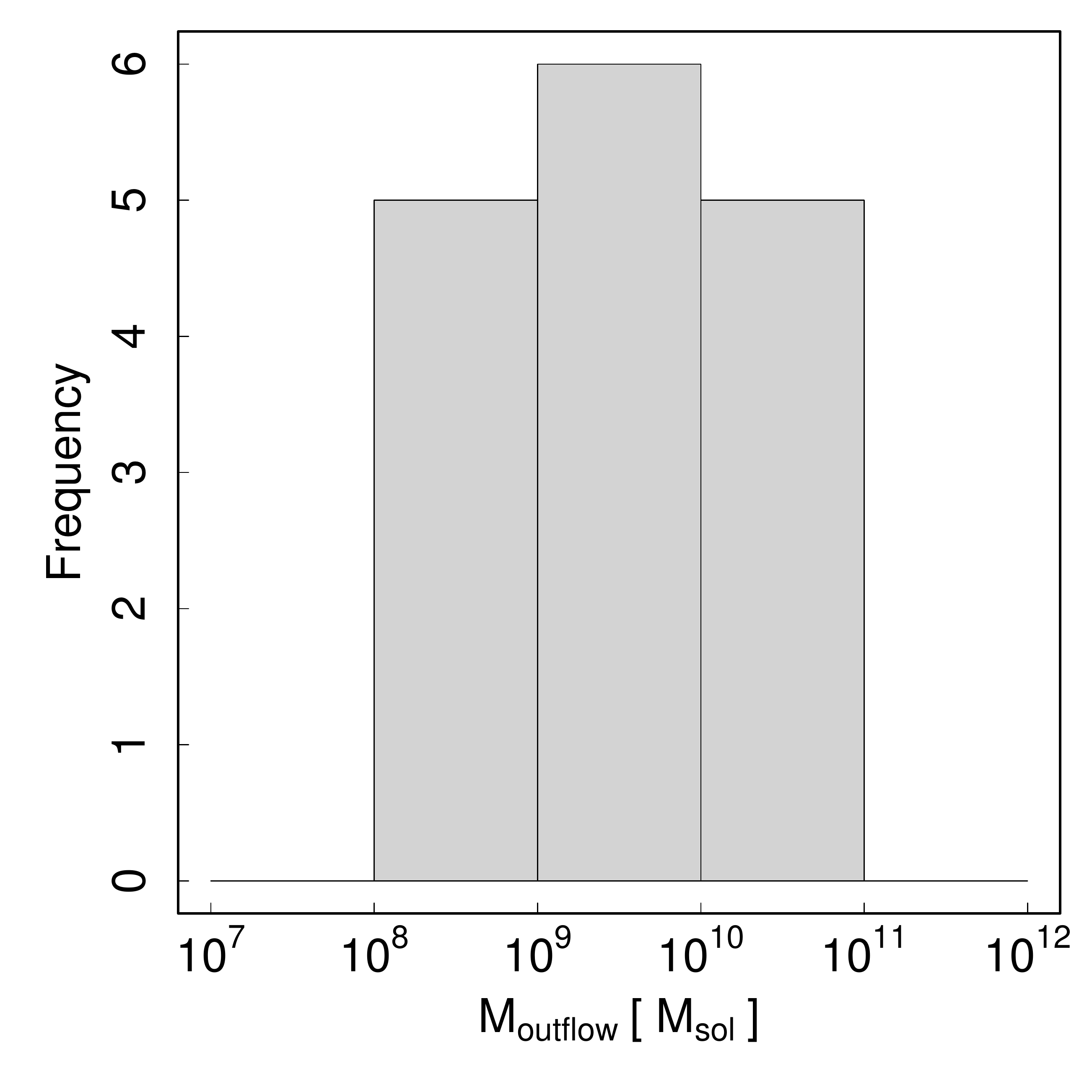}
\caption{Distribution in total mass outflowing from each cluster in the high quality sample.}
\label{fig:outflow_hist}
\end{center}
\end{figure}

\begin{figure}
\begin{center}
\includegraphics[totalheight=0.45\textwidth]{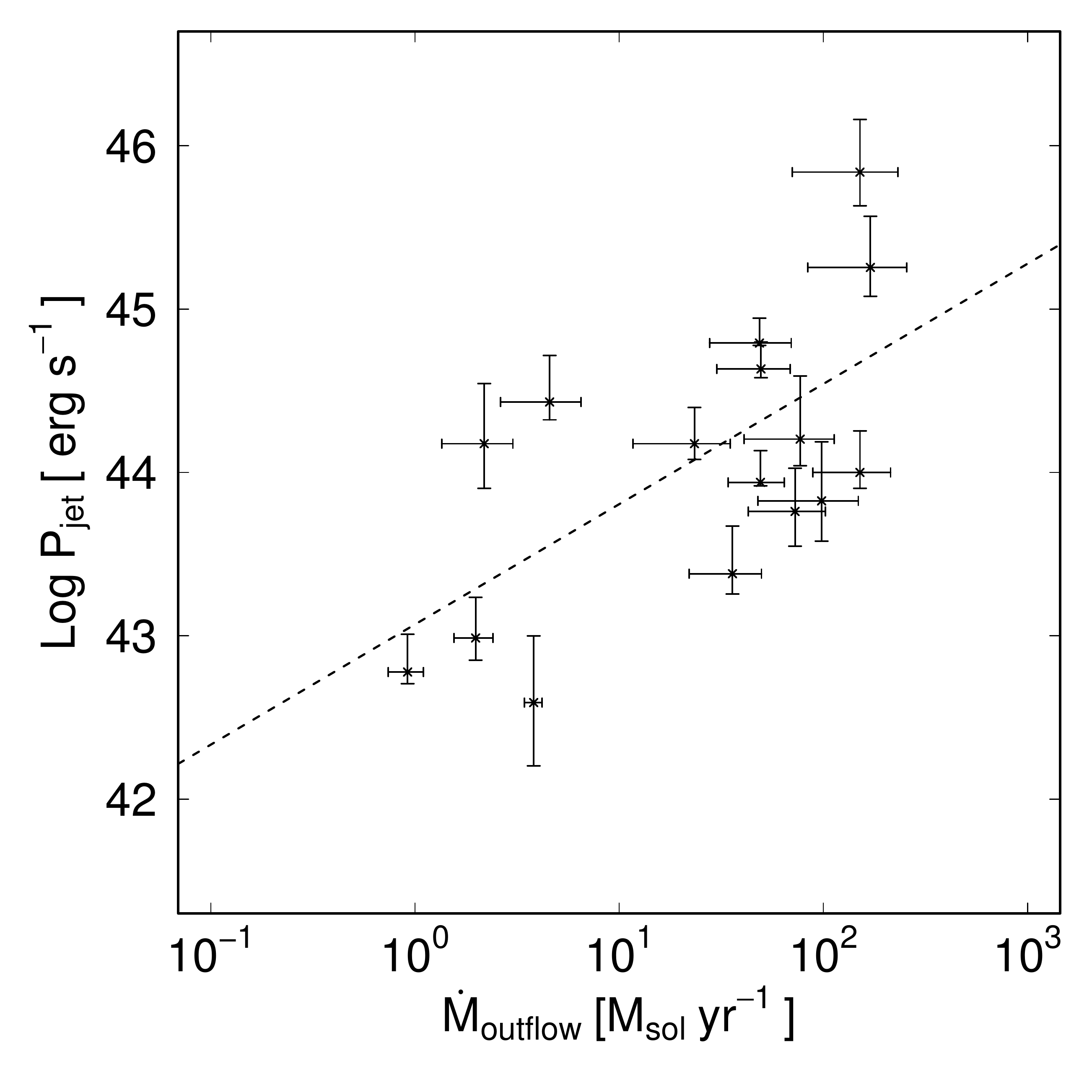}
\caption{Jet power vs. mass flow rate for systems in the high quality sample.  The dashed line is the best-fitting to the data.}
\label{fig:pjet-mdot}
\end{center}
\end{figure}

This conjecture is supported by recent ALMA discoveries of what appears to be outflowing molecular gas in Abell 1835 (included in this sample), and perhaps in Abell 1664 \citep{rm14,mr14}.  Similar molecular flows are present in the Perseus cluster \citep{lim08, sal11}. The molecular gas flows are surprisingly large, with masses of $10^{9}~\rm M_\odot$ to $10^{10}~\rm M_\odot$.  The hot and cold outflows may be physically related to each other.   As the rising bubbles lift the metal-rich hot gas, the cooler $\lae 1$ keV gas may condense into molecular clouds as it rises outward behind the bubble \citep{mr14,lb14}.  An alternative but more challenging possibility would be that molecular clouds are dragged upward in the wakes of the more massive hot outflows \citep{mr14}.  The outflowing hot and cold gas, perhaps associated with a circulating flow \citep{mat03,mat04,mcc08} may be delaying or removing gas otherwise destined to fuel star formation in the BCG.  Figure~\ref{fig:outflow_hist} shows that the total outflowing gas masses lie between $10^8$ M$_\odot$ to $10^{11}$ M$_\odot$ across our sample.  We have calculated the associated flow rates for each cluster assuming the gas is traveling at the bubble buoyancy speeds but no greater than the sound speed of the ICM.  The hot outflow rates lie between a few solar masses per year for jet powers of $\sim 10^{43}~\rm erg ~s^{-1}$ to more than $100~\rm M_\odot~yr^{-1}$ for jet powers exceeding $\sim 10^{45}~\rm erg ~s^{-1}$.  A crude relationship between jet power and outflow rate is shown in Figure~\ref{fig:pjet-mdot}.  The dashed line indicates the least-squares linear regression of the form,
\begin{equation} \label{eqn:mdot}
\dot{M}_{\rm outflow} = (22 \pm 16) \times P_{\rm jet}^{(1.4 \pm 0.4)}~(\rm M_\odot ~yr^{-1}).
\end{equation}
Here, jet power is in units of $10^{44}$ erg s$^{-1}$.  The scatter about the mean is large, with a standard deviation of approximately 1.47 dex.  Comparing the outflow rate to the enthalpy doesn't not improve the scatter.

\begin{figure}
\begin{center}
\includegraphics[totalheight=0.45\textwidth]{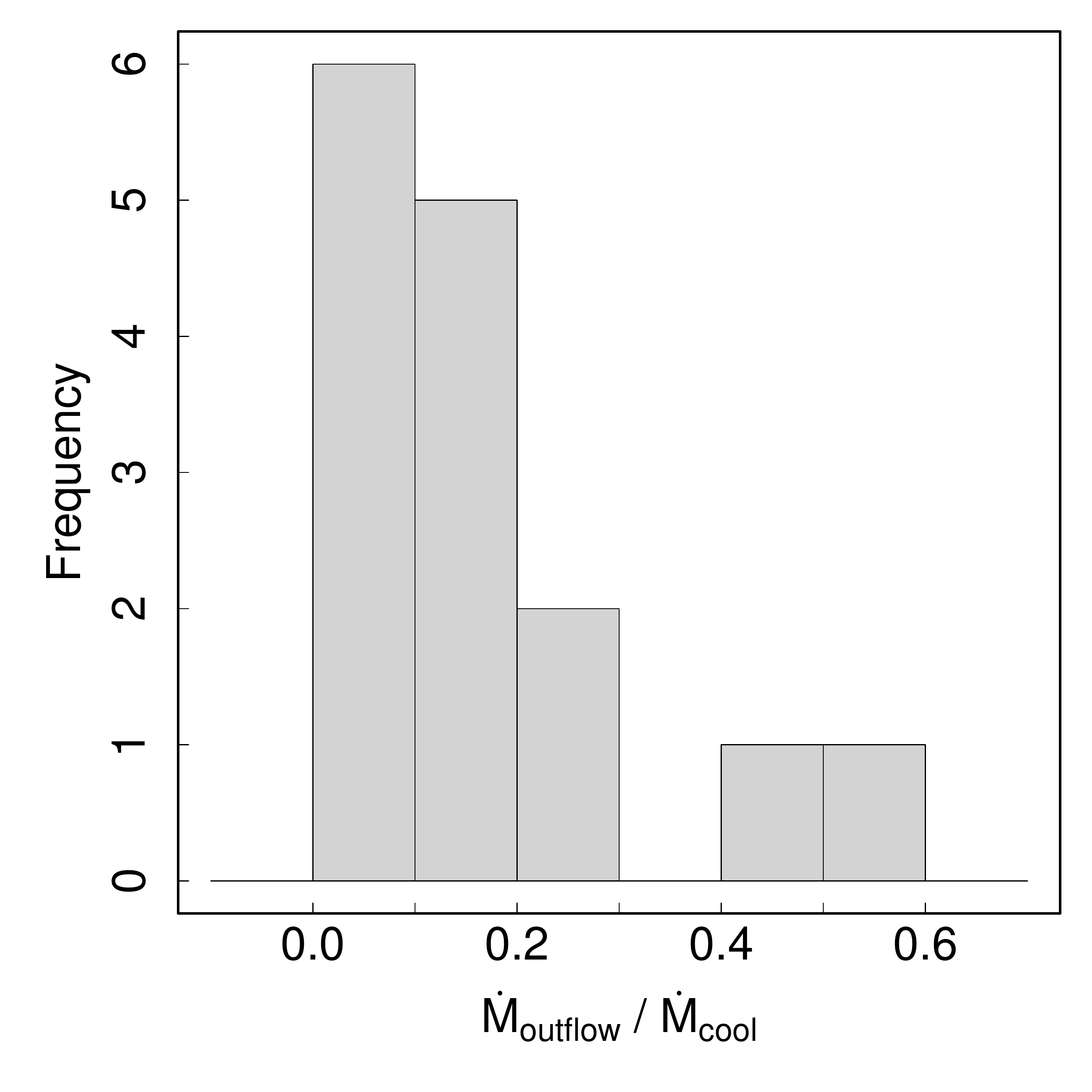}
\caption{The fraction of outflowing material to the expected total cooling rate of the cluster.  The redistribution of material by AGN alone is insufficient to balance a cooling flow.}
\label{fig:flowfrac_hist}
\end{center}
\end{figure}

We compare the outflow rates to the nominal cooling rates of the gas calculated as,
\begin{equation} \label{eqn:mcool}
\dot{M}_{\rm cool} = \frac{2}{5} \frac{\mu m_{\rm p} L_{\rm X}}{k_{\rm b} T_{\rm X}},
\end{equation}
where $L_{\rm X}$ is the cooling luminosity from \citet{raf06} and $T_{\rm X}$ is the average gas temperature within the cooling radius.  In Figure~\ref{fig:flowfrac_hist} we present a histogram of the ratios of outflow rates to the cooling rate.  While the uncertainties are admittedly large (as high as 50\%), the figure shows that outflows are able to remove only 10\% to 20\% on average of the gas that is expected to cool.  The higher fractions in the histogram are not associated with with higher jet powers, and likely reflect uncertainty in the outflowing gas masses.  This figure shows that hot outflows alone would be unable to remove all of the gas that is expected to cool and to accrete onto the central galaxy.  Instead, they may be one part of many processes which regulate the rate of cooling gas which fuel star formation and the nuclear black hole.

\section{Summary}

The 16 clusters of the High Quality sample we have presented here reveal higher metallicities along projected trajectories of their X-ray cavities.  We interpret this effect as the signature of massive, hot outflows of metal-enriched gas produced within the BCG.  At the same time the metallicity maps reveal a great deal of complexity, indicating that other processes, including returning flows of gas, may be taking place.  The outer iron radius is found to scale with jet power as $P_{\rm jet}^{0.45}$ and with cavity  enthalpy as $E_{\rm cav}^{0.33}$.  The jet power relation shows slightly lower scatter.  

The mean hot outflow rates are typically tens of solar masses per year and upward of $100 ~\rm M_\odot~yr^{-1}$ in the extreme.  These flow rates are generally insufficient to balance the radiative cooling rate of a cluster.  Recirculation of cooling gas accounts for only 10\% to 20\% of the radiative cooling rate, so that heating must be more important.  Nevertheless, the outflow rates are comparable to star formation rates in BCGs, and may have a significant impact on fueling star formation and the AGN itself.
  
Using the entire sample, we found that over the life cycle of AGN activity, hot outflows may alone be responsible for the broadening of abundance peaks in cool-core clusters, and effectively transport metal-enriched gas from the BCG to great distances in cluster atmospheres.  On average, only $\sim 15\%$ of the energy released by the AGN is required to lift the gas from the centre to the observed altitudes. Standard conversions between free energy and rest mass indicate that the SMBHs masses must have increased between $10^7$ and $10^9$ M$_\odot$ while driving the outflows. This mass increase lies comfortably within the estimated SMBH masses of BCGs, and is therefore a plausible explanation for the extended abundance peaks in clusters.

\section*{Acknowledgements}

We would like to thank Helen Russell, Adrian Vantyghem, Michael Balogh, Mike Hudson, and Sebastian Heinz for helpful discussions throughout the preparation of this paper.  BRM acknowledges financial support from the Natural Sciences and Engineering Research Council of Canada's Discovery Grant and Accelerator Supplement programs, and from the Canadian Space Agency.


\begin{thebibliography}{}

\bibitem[Allen \& Fabian(1998)]{all98} Allen S.~W., 
Fabian A.~C.\ 1998, MNRAS, 297, L63

\bibitem[B{\^i}rzan et al.(2004)]{bir04} B{\^i}rzan L., Rafferty D.~A., McNamara B.~R., 
Wise M.~W., Nulsen P.~E.~J.\ 2004, ApJ, 607, 800

\bibitem[B{\"o}hringer et al.(2004)]{boh04} B{\"o}hringer H., Matsushita K., 
Churazov E., Finoguenov A., Ikebe Y.\ 2004, A\&A, 416, L21

\bibitem[Bourdin \& Mazzotta(2008)]{bou08} Bourdin H., 
Mazzotta P.\ 2008, A\&A, 479, 307

\bibitem[Bower et al.(2006)]{bow06} Bower R.~G., Benson A.~J., 
Malbon R., Helly J.~C., Frenk C.~S., Baugh C.~M., Cole S., 
Lacey C.~G.\ 2006, MNRAS, 370, 645 

\bibitem[Bregman et al.(2003)]{bre03} Bregman J.~N., 
Novicki M.~C., Krick J.~E., Arabadjis J.~S.\ 2003, ApJ, 597, 399

\bibitem[Brighenti \& Mathews(2006)]{bm06} Brighenti F., 
Mathews W.~G.\ 2006, ApJ, 643, 120 

\bibitem[Br{\"u}ggen \& Kaiser(2002)]{bru02} Br{\"u}ggen M., 
Kaiser C.~R.\ 2002, Nature, 418, 301

\bibitem[Buote(2000)]{buo00} Buote D.~A.\ 2000, MNRAS, 311, 
176

\bibitem[Cappellari \& Copin(2003)]{cap03} Cappellari M., 
Copin Y.\ 2003, MNRAS, 342, 345

\bibitem[Cappellaro, Evans, \& Turatto(1999)]{cap99} Cappellaro E., Evans R., 
Turatto M.\ 1999, A\&A, 351, 459

\bibitem[Churazov et al.(2001)]{chu01} Churazov, E., Br{\"u}ggen, M., Kaiser, 
C.~R., B{\"o}hringer, H., \& Forman, W.\ 2001, ApJ, 554, 261

\bibitem[Churazov et al.(2000)]{chu00} Churazov, E., Forman, W., 
Jones, C., B{\"o}hringer, H.\ 2000, A\&A, 356, 788

\bibitem[Croton et al.(2006)]{cro06} Croton D.~J. et al.\ 2006, 
MNRAS, 365, 11 

\bibitem[David \& Nulsen(2008)]{dav08} David L.~P., 
Nulsen P.~E.~J.\ 2008, ApJ, 689, 837

\bibitem[De Grandi \& Molendi(2001)]{deg01} De Grandi S., 
Molendi S.\ 2001, ApJ, 551, 153

\bibitem[De Grandi et al.(2004)]{deg04} De Grandi S., Ettori S., 
Longhetti M., Molendi S.\ 2004, A\&A, 419, 7

\bibitem[de Plaa et al.(2006)]{dep06} de Plaa J. et al.\ 2006, 
A\&A, 452, 397 

\bibitem[Dickey \& Lockman(1990)]{dic90} Dickey J.~M., 
Lockman F.~J.\ 1990, ARA\&A, 28, 215

\bibitem[Diehl \& Statler(2006)]{die06} Diehl S., 
Statler T.~S.\ 2006, MNRAS, 368, 497

\bibitem[Doria et al.(2012)]{dor12} Doria, A., Gitti, M., 
Ettori, S., et al.\ 2012, ApJ, 753, 47

\bibitem[Dunn \& Fabian(2008)]{df08} Dunn R.~J.~H., 
Fabian A.~C. 2008, MNRAS, 385, 757 

\bibitem[Dupke \& White(2000)]{dup00} Dupke R.~A., 
White R.~E., III 2000, ApJ, 537, 123

\bibitem[Fabian(2012)]{fab12} Fabian, A.~C.\ 2012, 
Annual Review of Astronomy and Astrophysics, 50, 455

\bibitem[Gaspari et al.(2011)]{gas11} Gaspari, M., Melioli, 
C., Brighenti, F., D'Ercole, A.\ 2011, MNRAS, 411, 349

\bibitem[Gitti et al.(2011)]{git11} Gitti M., Nulsen P.~E.~J., David L.~P., 
McNamara B.~R., Wise M.~W.\ 2011, ApJ, 732, 13 

\bibitem[Gitti et al.(2010)]{git10} Gitti, M., O'Sullivan, 
E., Giacintucci, S., et al.\ 2010, ApJ, 714, 758 

\bibitem[Gopal-Krishna \& Wiita(2001)]{gop01} Gopal-Krishna, 
Wiita P.~J.\ 2001, ApJ, 560, L115

\bibitem[Gopal-Krishna \& Wiita(2003)]{gop03} Gopal-Krishna, 
Wiitan P.~J.\ 2003, Radio Astronomy at the Fringe, 300, 293

\bibitem[Grevesse \& Sauval(1998)]{gre98} Grevesse N., 
Sauval A.~J.\ 1998, Space Science Reviews, 85, 161

\bibitem[Gu et al.(2007)]{gu07} Gu J., Xu H., Gu L., An T., Want Y., 
Zhang Z., Wu X.\ 2007, ApJ, 659, 275

\bibitem[Heinz et al.(2006)]{hbyl06} Heinz S., Br{\"u}ggen M., 
Young A., Levesque E.\ 2006, MNRAS, 373, L65 

\bibitem[King et al.(2002)]{kin02} King, A., Frank, J., Raine, D.~J.\ 2002,
Accretion Power in Astrophysics (3rd ed.; Cambridge: Cambridge Univ. Press)

\bibitem[Kirkpatrick et al.(2009a)]{cck09b} Kirkpatrick C.~C., 
Gitti M., Cavagnolo K.~W., McNamara B.~R., David L.~P., 
Nulsen P.~E.~J., Wise M.~W.\ 2009a, ApJ, 707, L69

\bibitem[Kirkpatrick et al.(2009b)]{cck09a} Kirkpatrick C.~C., et al.\ 2009b, 
ApJ, 697, 867

\bibitem[Kirkpatrick et al.(2011)]{cck11} Kirkpatrick C.~C., 
McNamara B.~R., Cavagnolo K.~W.\ 2011, ApJ, 731, L23

\bibitem[Li \& Bryan(2014)]{lb14} Li Y., Bryan G.~L.\ 2014, ApJ, 789, 153 

\bibitem[Lim, Ao, \& Dinh-V-Trung(2008)]{lim08} Lim J., Ao Y., 
Dinh-V-Trung 2008, ApJ, 672, 252 

\bibitem[McCarthy et al.(2008)]{mcc08} McCarthy I.~G., 
Babul A., Bower R.~G., Balogh M.~L.\ 2008, MNRAS, 386, 1309 

\bibitem[McNamara et al.(2009)]{bmc09} McNamara B.~R., 
Kazemzadeh F., Rafferty D.~A., B{\^i}rzan L., Nulsen P.~E.~J., 
Kirkpatrick C.~C., Wise M.~W.\ 2009, ApJ, 698, 594

\bibitem[McNamara \& Nulsen(2007)]{bmc07} McNamara B.~R., 
Nulsen P.~E.~J.\ 2007, ARA\&A, 45, 117

\bibitem[McNamara \& Nulsen(2012)]{bmc12} McNamara, B.~R., 
Nulsen, P.~E.~J.\ 2012, New Journal of Physics, 14, 055023

\bibitem[McNamara et al.(2006)]{bmc06} McNamara B.~R. et al.\ 2006, 
ApJ, 648, 164

\bibitem[McNamara et al.(2014)]{mr14} McNamara B.~R. et al.\ 2014, 
ApJ, 785, 44 

\bibitem[Mathews et al.(2003)]{mat03} Mathews W.~G., 
Brighenti F., Buote D.~A., Lewis A.~D.\ 2003, ApJ, 596, 159 

\bibitem[Mathews, Brighenti, \& Buote(2004)]{mat04} Mathews W.~G., 
Brighenti F., Buote D.~A.\ 2004, ApJ, 615, 662 

\bibitem[Mendygral, O'Neill, \& Jones(2011)]{moj11} Mendygral P.~J., 
O'Neill S.~M., Jones T.~W.\ 2011, ApJ, 730, 100 

\bibitem[Molendi \& Gastaldello(2001)]{mol01} Molendi S., 
Gastaldello F.\ 2001, A\&A, 375, L14

\bibitem[Moll et al.(2007)]{moll07} Moll R., et al.\ 2007, A\&A, 463, 
513

\bibitem[Morita et al.(2006)]{mor06} Morita, U., Ishisaki, 
Y., Yamasaki, N.~Y., et al.\ 2006, PASJ, 58, 719 

\bibitem[Morsony et al.(2010)]{mor10} Morsony B.~J., Heinz S., 
Br{\"u}ggen M., Ruszkowski M.\ 2010, MNRAS, 407, 1277 

\bibitem[Mushotzky \& Loewenstein(1997)]{mus97} Mushotzky R.~F., 
Loewenstein M.\ 1997, ApJ, 481, L63

\bibitem[Mushotzky et al.(1996)]{mus96} Mushotzky R., Loewenstein M., 
Arnaud K.~A., Tamura T., Fukazawa Y., Matsushita K., Kikuchi K., 
Hatsukade I.\ 1996, ApJ, 466, 686

\bibitem[Nulsen et al.(2002)]{nuls02} Nulsen P.~E.~J., David L.~P., 
McNamara B.~R., Jones C., Forman W.~R., Wise M.~W.\ 2002, ApJ, 568, 163 

\bibitem[Omma et al.(2004)]{omm04} Omma H., Binney J., 
Bryan G., Slyz A.\ 2004, MNRAS, 348, 1105

\bibitem[Pope et al.(2010)]{pop10} Pope E.~C.~D., Babul A., Pavlovski G., 
Bower R.~G., Dotter A.\ 2010, MNRAS, 406, 2023

\bibitem[Rafferty et al.(2006)]{raf06} Rafferty D.~A., McNamara B.~R., 
Nulsen P.~E.~J., Wise M.~W.\ 2006, ApJ, 652, 216

\bibitem[Rafferty et al.(2013)]{raf13} Rafferty D.~A., B{\^i}rzan L., 
Nulsen P.~E.~J., McNamara B.~R., Brandt W.~N., Wise M.~W., 
R{\"o}ttgering H.~J.~A.\ 2013, MNRAS, 428, 58 

\bibitem[Rasera et al.(2008)]{ras08} Rasera Y., Lynch B., 
Srivastava K., Chandran B.\ 2008, ApJ, 689, 825

\bibitem[Rebusco et al.(2005)]{reb05} Rebusco P., Churazov E., 
B{\"o}hringer H., Forman W.\ 2005, MNRAS, 359, 1041

\bibitem[Rebusco et al.(2006)]{reb06} Rebusco P., Churazov E., 
B{\"o}hringer H., Forman W.\ 2006, MNRAS, 372, 1840

\bibitem[Reynolds, Casper, \& Heinz(2008)]{rey08} Reynolds C.~S., 
Casper E.~A., Heinz S.\ 2008, ApJ, 679, 1181

\bibitem[Roediger et al.(2007)]{roe07} Roediger E., Br{\"u}ggen M., 
Rebusco P., B{\"o}hringer H., Churazov E.\ 2007, MNRAS, 375, 15

\bibitem[Ruszkowski \& Oh(2010)]{ro10} Ruszkowski M., 
Oh S.~P.\ 2010, ApJ, 713, 1332 

\bibitem[Russell et al.(2014)]{rm14} Russell H.~R. et al.\ 2014, ApJ, 784, 
78 

\bibitem[Salom{\'e} et al.(2011)]{sal11} Salom{\'e} P., Combes, F., 
Revaz, Y., Downs D., Edge A.~C., Fabian A.~C.\ 2011, A\&A, 531, A85 

\bibitem[Sanders et al.(2005)]{san05} Sanders, J.~S., Fabian, 
A.~C., Dunn, R.~J.~H.\ 2005, MNRAS, 360, 133

\bibitem[Sijacki \& Springel(2006)]{ss06} Sijacki D., Springel V.\ 2006, 
MNRAS, 366, 397 

\bibitem[Simionescu et al.(2009)]{sim09} Simionescu A., Werner N., 
B{\"o}hringer H., Kaastra J.~S., Finoguenov A., Br{\"u}ggen M., 
Nulsen P.~E.~J.\ 2009, A\&A, 493, 409

\bibitem[Simionescu et al.(2008)]{sim08} Simionescu A., Werner N., 
Finoguenov A., B{\"o}hringer H., Br{\"u}ggen M.\ 2008, A\&A, 482, 97

\bibitem[Simionescu et al.(2012)]{sim12} Simionescu A. et al.\ 2012, 
ApJ, 757, 182 

\bibitem[Takahashi et al.(2012)]{tak12} Takahashi T. et al.\ 2012, 
Proc. SPIE, 8443,  

\bibitem[Tamura et al.(2001)]{tam01} Tamura T., Bleeker J.~A.~M., 
Kaastra J.~S., Ferrigno C., Molendi S.\ 2001, A\&A, 379, 107 

\bibitem[Tamura et al.(2004)]{tam04} Tamura T., Kaastra J.~S., den Herder J.~W.~A., 
Bleeker J.~A.~M., Peterson J.~R.\ 2004, A\&A, 420, 135

\bibitem[Vernaleo \& Reynolds(2007)]{vr07} Vernaleo J.~C., 
Reynolds C.~S.\ 2007, ApJ, 671, 171 

\end{thebibliography}
\end{document}